\pdfoutput=1
\documentclass[aps,prl,twocolumn,reprint,amsmath,amssymb,floatfix]{revtex4-2}
\usepackage[T1]{fontenc}
\usepackage{lmodern}
\usepackage{graphicx}
\usepackage{bm}
\usepackage[colorlinks=true,linkcolor=blue,citecolor=blue,urlcolor=blue]{hyperref}

\begin{document}

\title{Boundary-Sculpted Arctic Surface in the Cubic Ice Model}

\author{Benjamin Canals}
\email[Contact author: ]{benjamin.canals@neel.cnrs.fr}
\affiliation{Universit\'e Grenoble Alpes, CNRS, Grenoble INP, Institut NEEL, 38000 Grenoble, France}

\date{November 12, 2026}

\begin{abstract}
In square ice, domain-wall boundaries freeze the corners behind an arctic curve; strings of reversed arrows, unable
to pass each other, give it Tracy-Widom fluctuations. In 3D, they pass around one another. Monte Carlo simulations
calibrated on exact 2D results show that boundaries still sculpt an arctic surface in cubic ice, enclosing, by a lower
bound, 3\% of the box at the largest size. If local equilibrium holds up to it, it lies where the local field reaches
the bulk Kasteleyn threshold. Its fluctuations are not Tracy-Widom along lines, and much closer than in 2D to
independent strings.
\end{abstract}

\maketitle

\begin{figure*}
\includegraphics[width=\textwidth]{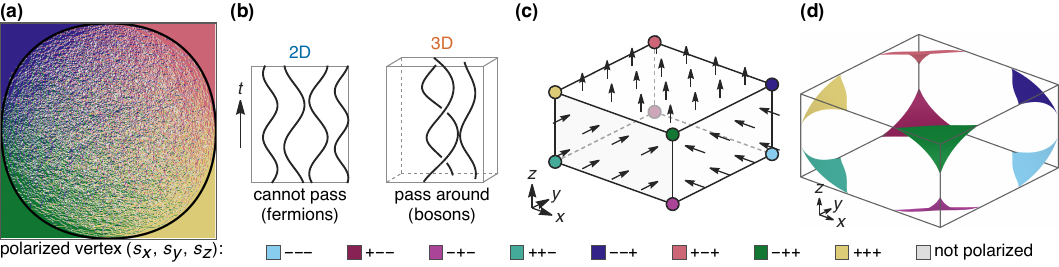}
\caption{\label{fig:region}Frozen regions sculpted by boundaries, and their strings. (a) Square ice with domain-wall
boundaries ($L=384$), one equilibrium sample: vertices in a polarized state are colored by their signs
$(s_x,s_y,s_z)$, with $s_z=+$ (one-layer box, vertical edges frozen upward); others are gray; black line: exact
arctic curve. (b) Strings as world lines along their direction $t$
(schematic). (c) Cubic ice: an $L\times L\times L/2$ box (here $L=4$) with frozen pendant edges (arrows, visible faces only); dots: corner states. (d) Region already
frozen at $L=160$: isosurface of the disorder, $\hat v=10^{-2}$, seen as in (c).}
\end{figure*}

\textit{Introduction}---Boundary conditions usually affect only a surface layer, about a correlation length
thick~\cite{Diehl1997}. In locally constrained systems, such as dimer and ice models, suitably chosen boundaries
can instead change the free energy per site~\cite{CohnKenyonPropp2001,KorepinZinnJustin2000} and fix the state of a
finite fraction of the bulk~\cite{Shah2026}.
In domino tilings of the Aztec diamond~\cite{JockuschProppShor1998,CohnElkiesPropp1996} and in
square ice with domain-wall boundaries~\cite{ColomoPronko2010JSP,Aggarwal2020}, the corners freeze into ordered
states, separated from a disordered liquid by a sharp arctic curve [Fig.~\ref{fig:region}(a)]. Its fluctuations are
universal, those of the largest eigenvalue of a large random matrix: the Tracy-Widom
distribution~\cite{TracyWidom1994} and, along the curve, the Airy process~\cite{PrahoferSpohn2002}, proven for
dominoes~\cite{Johansson2005} and, in recent preprints that prove a conjecture~\cite{ColomoPronko2024,ColomoPronko2026}, for
square ice~\cite{Li2026,Liao2026}.

Both the edge and its fluctuations come from the way a frozen corner melts. The corner is polarized: all arrows along
each axis point the same way, and no local move preserves the ice rule. Only strings can undo it, directed paths of
reversed arrows that cross it from side to side and preserve the ice rule~\cite{Jaubert2008,Morris2009}; the
liquid begins where they appear. Strings shape the edge in two ways. Their onset is a Kasteleyn transition, where the entropy of
strings overtakes their cost~\cite{Kasteleyn1963}, and its universality class sets how the string density rises across the edge.
In two dimensions (2D), moreover, strings cannot pass each other [Fig.~\ref{fig:region}(b)]: ordered like the world
lines of fermions, effectively free where dilute~\cite{Stephan2019}, they give the edge its Airy and
Tracy-Widom statistics~\cite{FerrariSpohn2003,Johansson2005,Li2026,Liao2026}. In three dimensions (3D), strings pass around each other
and share vertices, though never edges~\cite{NelsonSeung1989,Jaubert2008}: they are no longer ordered.

Whether
boundaries can then freeze a region of positive volume is open even for dimers~\cite[Sec.~1.4]{KenyonWolfram2025}.
Frozen regions appear in simulations of 3D tilings~\cite{RandallYngve2000,LindeMooreNordahl2001,Widom2002} and
dimers~\cite[Sec.~1.3]{ChandgotiaSheffieldWolfram2023}, but how their edge fluctuates is not known, to our knowledge. We ask what an arctic edge inherits from the statistics of its strings. Our
hypothesis: the class of the Kasteleyn transition sets the density profile across the edge, and with it how fast the
edge fluctuations grow; the ordering of the strings sets their law.

We examine this hypothesis on cubic ice, the 3D analog of square ice, with Monte Carlo simulations calibrated on exact 2D
results. Boundaries still freeze a region of cubic ice: near each corner, the string density falls almost exponentially
with the box size, as in 2D, and a lower bound puts 3\% of the box in it at the largest size, a fraction that still grows.
In local equilibrium, each part of the box behaves like the bulk in a field that favors the nearest corner state and
grows toward that corner; the region is frozen where this local field exceeds the Kasteleyn threshold, a transition
that is logarithmic in 3D~\cite{Bhattacharjee1983,Jaubert2008,Jaubert2009,PowellChalker2008}. The edge fluctuations grow mostly as
the mean density profile dictates, but they are not Tracy-Widom, and much closer than in 2D to those of independent strings.

\textit{Model}---We place Ising spins $s_e=\pm1$ on the edges $e$ of the simple cubic lattice, seen as arrows, and
keep the configurations obeying the ice rule, three arrows in and three out at every vertex, all with equal
weight: this is cubic ice. In its 2D analog, square ice with domain-wall boundaries~\cite{Korepin1982}, arrows point in
on two opposite sides and out on the other two. In 3D, analogous boundaries freeze the pendant edges of an
$L_x\times L_y\times L_z$ box, which connect its boundary vertices to the outside, inward on the sides and outward
on the top and bottom [Fig.~\ref{fig:region}(c)]. The ice rule conserves the flux, so what enters through the sides
must leave through the top and bottom:
\begin{equation}
L_yL_z+L_xL_z=L_xL_y\;\Rightarrow\;L_z=L/2
\label{eq:flux}
\end{equation}
for a square base, $L_x=L_y=L$. Each face fixes the arrows normal to it, so the three faces at each corner select a different one of the
eight polarized vertex types.

\textit{Deciding existence}---We sample the uniform measure with a loop algorithm~\cite{BarkemaNewman1998},
up to $L=160$ in 3D and 384 in 2D (Appendix~\ref{app:checks}). In 3D, strings do not cut
space, so, unlike in 2D, the cluster of corner-state vertices percolates between them into the liquid. We use
instead local fields: the string density $\rho$ at each site, the probability that an arrow is reversed relative
to the state of the nearest corner, and the disorder $1-\langle s_e\rangle^2$, equal to
$4\rho(1-\rho)$ on each edge, estimated by $\hat v$ (Appendix~\ref{app:definitions}). Whether a site is frozen shows when the box
doubles, through the rate function
\begin{equation}
I_L(\mathbf r)=\frac{\ln\rho_L(\mathbf r)-\ln\bar\rho_{2L}(\mathbf r)}{L},
\label{eq:rate}
\end{equation}
where $\bar\rho_{2L}$ is averaged over the matching $2\times2\times2$ block. In a frozen region, the probability of a
string at a site decays exponentially with $L$, and $I_L$ stays positive, the larger the deeper the site; in the
liquid, $\rho$ tends to a finite value and $I_L$ to zero [Fig.~\ref{fig:existence}(a)]. Five criteria on these fields test existence (S2 of the Supplemental Material~\cite{SM}). For existence and
each test below, criteria and analysis code were fixed before the data they judge were read (exceptions: Appendix~\ref{app:checks}), and checked on exact 2D results; the
local chemical potential and the comparisons with parameter-free predictions are descriptive.

\begin{figure}
\includegraphics[width=\columnwidth]{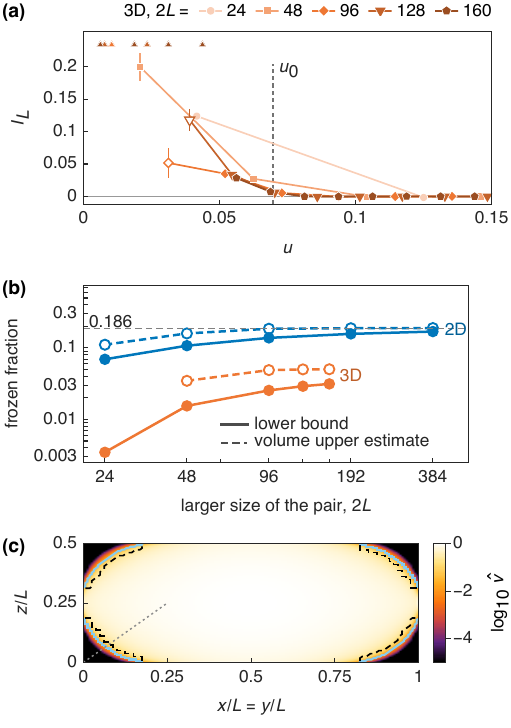}
\caption{\label{fig:existence}Existence of the frozen region. (a) Rate function $I_L$ [Eq.~\eqref{eq:rate}] along the
diagonal of cubic ice, for pairs $L\to2L$, with $1\sigma$ errors; open: $\bar\rho_{2L}$ unresolved (S2); triangles at
the top: no string seen at $2L$ (counted as frozen); dashed: zero $u_0$ of $\mu_{\rm loc}$ at $L=160$ [Fig.~\ref{fig:mechanism}(c)].
(b) Frozen fraction against $2L$: lower bound and volume upper estimate (Appendix~\ref{app:definitions}); gray: exact
2D value. (c)
Disorder $\hat v$ in the plane $x=y$ at $L=160$ (black: $\hat v\le10^{-5}$); contours: lower bound (light) and volume upper estimate (dashed) at $2L=160$; dotted: the diagonal of (a).}
\end{figure}

\textit{A frozen region that scales with the box}---Cubic ice passes the existence test up to $L=160$: near each
corner, $\rho$ falls almost exponentially with $L$, as in 2D at the same sizes (Appendix~\ref{app:checks}). At
$2L=160$, the sites with $\hat v<10^{-2}$ fill 3.1\% of the box, against 17\% in 2D at $2L=384$
[Fig.~\ref{fig:existence}(b)]; this lower bound, which in 2D stays below the exact value, still grows with $L$, and an upper
estimate of its volume (Appendix~\ref{app:definitions}) is about 5\% from $2L=128$ on. The region is thin, mostly within three layers of the faces,
one piece at each corner [Figs.~\ref{fig:region}(d) and~\ref{fig:existence}(c)], but not a surface layer: along the body diagonal
$x=y=z=uL$, it reaches 3 sites deep at $L=48$ and 11 at $L=160$, and its edge stays at $u\simeq0.07$, against
$1-\sqrt3/2=0.134$ in 2D~\cite{ColomoPronko2010JSP,Aggarwal2020} [Fig.~\ref{fig:existence}(a)]. What bounds the region depends on how a frozen corner melts.

\textit{Melting a frozen corner}---In a uniform field favoring a polarized state, with a weight $e^{H_as_e}$ per edge along axis $a$, a string pays $e^{-2|H_a|}$ per step along $a$, and strings
proliferate once their chemical potential, $\mu=\ln\sum_a e^{-2|H_a|}$, becomes positive. This is the Kasteleyn
transition, studied in pyrochlore spin ice in a $[100]$ field~\cite{Jaubert2008,Jaubert2009}. In cubic ice, a string
chooses at each step among three directions, so $\mu=\ln3-2H$ for a field $(H,H,H)$, which vanishes at
$H_K=\frac12\ln3$. How $\rho$ rises above the threshold depends on the statistics of
the strings. In 2D, where they are fermions, $\rho\propto\mu^{1/2}$~\cite{PokrovskyTalapov1979,MoessnerSondhi2003,Fennell2007}. In 3D, seen along
their direction, which acts as time, strings are world lines of particles in two transverse dimensions: bosons with a contact
repulsion, as they pass each other but never share an edge~\cite{Jaubert2008,PowellChalker2008}. Such a repulsion is marginal in $2+1$
dimensions, the upper critical dimension~\cite{PowellChalker2008}: in a dilute Bose gas, the effective coupling decays as
$1/\ln(1/n)$ at density $n$, so that $n\propto\mu\ln(1/\mu)$~\cite{Schick1971,FisherHohenberg1988}. For spin ice, Jaubert et al.\ derived an equation of state
with this logarithm and free coefficients, in reduced temperature, compatible with their
simulations~\cite{Jaubert2008,Jaubert2009}; in $\mu$,
\begin{equation}
\rho=A\mu\,[1+B\ln(1/\mu)].
\label{eq:eos}
\end{equation}
We test it first in the bulk, where $\mu$ is known exactly.

\begin{figure*}
\includegraphics[width=\textwidth]{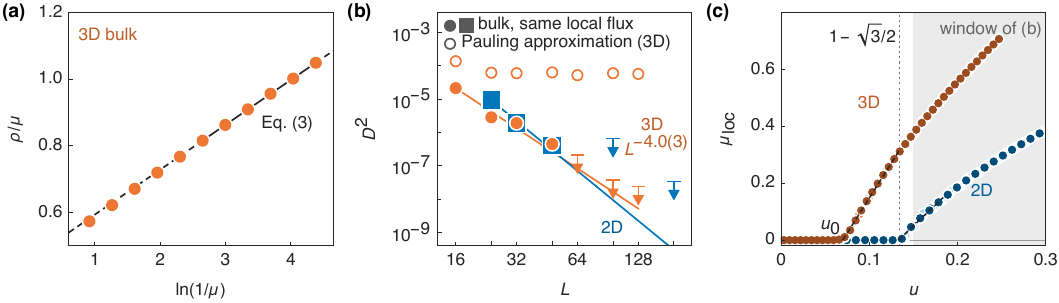}
\caption{\label{fig:mechanism}The surface as a Kasteleyn transition in space. (a) Bulk string density on $L^3$ tori in a
field along $(1,1,1)$; line: Eq.~\eqref{eq:eos}, fitted for $0.0125\le\mu\le0.0707$
(dashed outside). (b) Even part $D^2$ of the distance between the vertex frequencies along the diagonal of cubic ice
or square ice and those of the bulk at the same flux; lines: power-law fits, $L^{-5.0(4)}$ in 2D; arrows: upper limits $D^2+2\sigma$, where $D^2<2\sigma$.
(c) Local chemical potential along the diagonal of cubic ice ($L=64$, 96, 160) and square ice ($L=96$, 192, 384), from
light to dark; dashed: linear fits at $L=160$ and 384; dotted: exact 2D edge; gray: window of (b) in 3D. The zero
$u_0$ locates the edge by construction.}
\end{figure*}

\textit{The bulk equation of state}---We sample $L^3$ tori, $L=32$ to 256, in a uniform field along $(1,1,1)$, with
a two-defect worm~\cite{ProkofevSvistunov2001} (S1). For
$0.0125\le\mu\le0.0707$, Eq.~\eqref{eq:eos} describes $\rho$ to 0.07\%, where a power law and the mean-field form $\propto\mu(1+\eta\mu)$, with as
many parameters, fail: $\rho/\mu$ is linear in $\ln(1/\mu)$ [Fig.~\ref{fig:mechanism}(a)], up to resolved corrections (S3). Without free parameters,
dilute bosons whose mass is set by the diffusion of one string give $AB=\sqrt3/(4\pi)=0.138$, a value not obtained
before, to our knowledge; the fit gives 0.135, 2\% lower, and every fit of $\rho$ stays below, within the corrections
of several percent expected at these $\mu$ (S3). In the 2D control, $\rho$ follows the square root, with the
free-fermion amplitude $\sqrt2/\pi$. But the box has no field: does it behave locally like this bulk?

\textit{The surface is a transition in space}---If each region of the box behaves like the bulk at the same local
flux $\mathbf b(\mathbf r)$, the mean of the arrows, it sees the field that gives the bulk this flux. We test this local
equilibrium, unproven for ice, along the diagonal, comparing the frequencies of the twenty vertex types with
those of tori at the same flux. By symmetry, the deviation linear in the flux gradient is odd, and the even part should vanish as $L^{-2}$
(Appendix~\ref{app:definitions}). Summed over the twenty types, its square
$D^2$ falls as $L^{-4.0(3)}$, resolved up to $L=48$ and bounded beyond [Fig.~\ref{fig:mechanism}(b)], while that of the
Pauling approximation~\cite{Pauling1935}, with arrows independent apart from the ice rule, stays at a finite distance. Read through the bulk equation of state, the
measured $\rho$ gives a local chemical potential $\mu_{\rm loc}(u)$ [Fig.~\ref{fig:mechanism}(c)]. It falls
linearly toward the edge, at a slope independent of $L$, as a field set on the scale of the box should; its
zero, which locates the edge by construction, lies at $u_0=0.069$ to 0.070 at all sizes. On square ice, the same
reading returns the exact edge to 1\% at these sizes (S4). If local equilibrium holds up to the edge, the surface of the frozen region is where the local
field reaches the bulk Kasteleyn threshold. Local equilibrium is, however, tested only where strings are
dense, $\mu_{\rm loc}\gtrsim0.37$ [gray in Fig.~\ref{fig:mechanism}(c)]: the linear fit and the dilute layer next to the
edge, where the logarithm dominates, lie below this window, and only the zero is calibrated in 2D. The edge itself is sharp, about one site thick
(S5). If this transition differs from that of 2D, do its fluctuations differ as well?

\textit{What independent strings would do}---Along a line from a corner, let $\ell$ be the number of consecutive
vertices in the corner state. If the strings met along the line were independent,
$\ell\ge D$ would only require that none crosses its first $D$ vertices:
\begin{equation}
P(\ell\ge D)=\exp[-\Lambda(D)],
\label{eq:null}
\end{equation}
where $\Lambda(D)$, the mean number of string segments met, follows from the mean string density along the same line,
without free parameters (Appendix~\ref{app:definitions}). Near the edge, if $\rho\propto(d/L)^\beta$ at a distance $d$, the
line meets $\Lambda\sim w^{1+\beta}/L^\beta$ strings within $w$ sites beyond the edge; $\ell$ thus fluctuates over the
$w$ where $\Lambda\sim1$, and $\sigma_\ell\propto L^{\beta/(1+\beta)}$. This is $L^{1/3}$ for the square root of 2D,
the exponent of the Tracy-Widom edge, and $(L/\ln L)^{1/2}$ for the profile $(d/L)\ln(L/d)$ that Eq.~\eqref{eq:eos} and a
linear $\mu_{\rm loc}$ give in 3D. The class of the transition would set the growth. Equation~\eqref{eq:null} also sets
the shape of the law, $\exp(-cw^{1+\beta})$, a Weibull law with a long tail at large $\ell$, of positive skewness.
Ordered strings lengthen $\ell$ only by retreating together, which costs more~\cite{MajumdarSchehr2014}: the long tail
moves to small $\ell$, where the first string advances alone, like the largest eigenvalue in the Tracy-Widom law, and
$\ell$ has negative skewness. The ordering sets the law. Which law does the edge follow?

\begin{figure*}
\includegraphics[width=\textwidth]{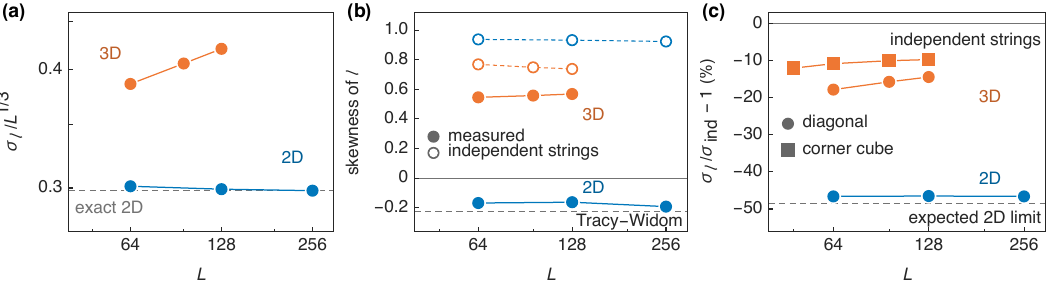}
\caption{\label{fig:fluctuations}Edge fluctuations along the diagonal, through the frozen length $\ell$ from a corner. (a) Width $\sigma_\ell/L^{1/3}$. (b) Skewness of $\ell$, measured and for independent strings [Eq.~\eqref{eq:null}]. (c) Deviation of $\sigma_\ell$ from the independent-string prediction $\sigma_{\rm ind}$, along the diagonal (disks)
and for the corner cube (squares). Widths carry Sheppard's correction~\cite{Sheppard1897}; errors are smaller than the
symbols.}
\end{figure*}

\textit{Ordered strings in 2D, unordered in 3D}---In 2D, along the diagonal, $\ell$ is also the side of the largest frozen
square at the corner, whose law is known~\cite{ColomoPronko2026,Li2026,Liao2026,LybergKorepinViti2023,PrahoferSpohn2024}; we recover its width and nearly its
skewness [Fig.~\ref{fig:fluctuations}(a,b), Table~\ref{tab:compare}]. Independent strings
give the right exponent, 0.325 against 0.326 measured, but a skewness of the wrong sign, and the measured width lies
47\% below theirs from $L=64$ to 256, near the exact 48.5\% by which the Tracy-Widom width falls below that of the largest
point of a Poisson process with the Airy density [Fig.~\ref{fig:fluctuations}(c)]: ordering halves the width. In 3D, the
skewness of $\ell$ is positive along every line instead, as for independent strings, and $\sigma_\ell$ grows faster than
$L^{1/3}$: the edge is not Tracy-Widom. Its width lies 14\% below the independent-string prediction at $L=128$, a gap that
shrinks with $L$. A line meets only the strings that touch it, which might make ordered strings look independent~\cite{Bohigas2009,BothnerBuckingham2018}; but
the corner cube, frozen only if no string enters it, sees every string of its region, and deviates even less, by 10\%, with a width below one site, at which square ice still deviates by 48\% (S6);
its width grows near $L^{1/3}$ and its skewness is negative, as for independent strings, since $\Lambda$ grows faster over a
cube than along a line: positive skewness and faster
growth belong to lines. Lines closer to the faces deviate more, up to 36\% at $L=128$, but there the model misses the exponent, even in 2D (S6). Along the diagonal,
the model gives most of the growth, an exponent 0.40 against 0.46 measured, the rest coming from the shrinking gap.
If $\sigma_\ell/\sigma_{\rm ind}$ tends to any nonzero limit, the exponent must come down to that of the model, itself rising slowly; at these sizes,
the rise of the model's exponent still offsets the waning share of the gap (S6, S7).

\begin{table}
\caption{\label{tab:compare}The edge of square ice (2D) and cubic ice (3D). Exact: theorems~\cite{Aggarwal2020,ColomoPronko2026,Li2026,Liao2026},
except the bulk density, a standard prediction~\cite{PokrovskyTalapov1979}, and the last row (S6). 2D and 3D: measured at the largest sizes ($\ell$ along the diagonal; 3D: effective exponent from $L=64$ to
128; in parentheses: corner cube).}
\begin{ruledtabular}
\begin{tabular}{llll}
 & Exact (2D) & 2D & 3D \\
\colrule
Strings & cannot pass & & pass around \\
Bulk density $\rho$ & $\sqrt{2\mu}/\pi$ & $0.4501\sqrt\mu$ & $\propto\mu\ln(1/\mu)$ \\
Frozen fraction & 0.1862 & 0.170--0.190 & 0.031--0.050 \\
Edge: $y_c$, $u_0$ & 0.1340 & 0.133 & 0.070 \\
$\sigma_\ell$ & $0.2980\,L^{1/3}$ & $0.2981(10)\,L^{1/3}$ & $\propto L^{0.457(2)}$ \\
Skewness of $\ell$ & $-0.224$ & $-0.19$ & 0.57 \\
$\sigma_\ell/\sigma_{\rm ind}-1$ & $\to-48.5\%$ & $-46.6\%$ & $-14\%$ ($-10\%$) \\
\end{tabular}
\end{ruledtabular}
\end{table}

\textit{What the edge inherits}---Up to $L=160$, boundaries alone sculpt an arctic surface in a three-dimensional ice, and, by descriptive comparisons, the
hypothesis holds in part (Table~\ref{tab:compare}). The class of the transition sets the density profile on tori with
planar edges, except close to them (S7), and by extrapolation in the box. Along the diagonal, the measured profile
accounts for all the growth of the edge fluctuations in 2D and most of it in 3D, an exponent of 0.40 against 0.46
measured. The ordering of the strings sets their law. In 2D, it reverses the skewness and halves the width.
In 3D, strings merely avoid sharing edges, a contact repulsion that produces the logarithm of Eq.~\eqref{eq:eos}; being
marginal, it may weaken near the arctic surface, where strings are dilute and are found much closer to independent than in 2D. What the 3D edge keeps from 2D is a Kasteleyn transition; what it loses is the order of its strings,
and with it the Tracy-Widom law.

\textit{Outlook}---These results rest on sizes up to $L=160$ and one geometry: the limit shape is not converged, and
a frozen fraction vanishing slowly at larger sizes, against which no control was calibrated, cannot be excluded. Our local tests could decide whether frozen regions exist in other 3D constrained
systems, such as dimers~\cite[Sec.~1.4]{KenyonWolfram2025} or artificial spin ices~\cite{Berchialla2024}. A variational principle in the flux, proven for 3D
dimers~\cite{ChandgotiaSheffieldWolfram2023}, should give the limit shape, and larger sizes the asymptotic law of the
edge.

\begin{acknowledgments}
The author thanks Rafik Ballou and Loren Coquille for their valuable suggestions. The text was drafted and revised
with Claude (Anthropic), each passage being proposed to and
approved by the author; the uses of AI tools in the research are stated in Appendix~\ref{app:checks}. The author is fully
responsible for the content.
\end{acknowledgments}

\textit{Data availability}---The data and code that support the findings of this article are openly available~\cite{DepotDonnees,DepotCode}.

\bibliography{letter}

\clearpage
\section*{End Matter}
\setcounter{secnumdepth}{1}%
\appendix

\section{Sampling and checks}
\label{app:checks}
\textit{Sampling}---The samplers obey detailed balance and match exact enumeration of small systems (S1 of Ref.~\cite{SM}). A sweep is one loop update per edge. Each run has eight replicas with distinct seeds, and errors are
jackknife estimates over replicas~\cite{Efron1982} (for the bulk equation of state on tori, from autocorrelation times). Runs of cubic ice last 1000
sweeps of thermalization (2000 at $L=128$ and 160) and, at $L=64$ to 160, 3100 to 15\,800 sweeps of measurement per replica, from one ice
state built by a maximum-flow algorithm~\cite{Dinic1970}. The longest time measured, that of the frozen volume of one
corner, is 25 to 83 sweeps from $L=48$ to 128 (extrapolated to 110 to 130 at 160), 15 to 40 times shorter than
thermalization. Runs whose replicas each start from their own disordered ice state agree edge by edge with those
from the common state, at $L=64$ and 80, and at $L=128$ with only 1280 sweeps of thermalization: if the relaxation time
grows as $L^2$ or slower, this is at most the thermalization of $L=160$ in units of that time (S1).

\textit{Tests and controls}---Five amendments to the existence protocol were made after some or all of the runs at
$L=64$ to 160 had been read (the estimator of the decay criterion E3, the bracket criterion E5, two quality controls, and the edge estimator); where there was an
original choice, it is reported next to the new one (S2), and none changes a verdict used here, with one reserve for E3
(Existence criteria). The edge criteria (S5) were set on these runs and judge runs at $L=64$, 96, and 128 with the same
parameters and other seeds, of which the vertex-type profiles and the frozen depth along the diagonal had been seen, but
not the head exponent and thickness they test. Tori in a sawtooth field, which have planar edges, and whose preset
verdict is ``not established'' because the 2D check of their local-equilibrium criterion fails (S7), enter only
descriptive comparisons. The existence protocol was set on square ice and on cubic ice at $L=12$, 24, 48, and 96, after a
rule based on the corner cluster had returned ``indeterminate'', and confirmed on new sizes (S2 of Ref.~\cite{SM}). The whole chain is run on square ice, whose frozen fraction, edge, and
fluctuations are known exactly (Table~\ref{tab:compare}): along the diagonal, $(Ly_c-\ell-1)/(\lambda L^{1/3})$ tends
to the Tracy-Widom distribution~\cite{TracyWidom1994}, with $y_c=1-\sqrt3/2$ and
$\lambda=2^{-4/3}3^{-1/6}$~\cite{ColomoPronko2026,Li2026,Liao2026}, whose moments~\cite{Bornemann2010} give
the width and skewness of $\ell$. Existence is also checked on negative controls, a one-layer box with
no frozen edge and square or cubic ice whose boundary arrows are frozen only near the corners: none returns ``exists'' (S2).
Applied to the 2D bulk, the test of Eq.~\eqref{eq:eos} does not find the logarithm (S3). Runs of cubic ice at $L=48$ to 128 with other
seeds but the same initial state give the same existence verdict and agree edge by edge (S1).

\textit{AI tools}---The code and the analyses were developed with Gemini (Google) and Claude (Anthropic) (S1), each
step being proposed to and approved by the author. The code and analyses are checked against results they do not use: exact enumeration of small
systems (S1), exact 2D results (S2, S3, S5), independent seeds, and a melted start (S1); the derivation of $AB$, by its 2D
counterpart, which gives the free-fermion amplitude (S3).

\textit{Existence criteria}---They test, as defined in S2, the nesting of frozen sites from $L$ to $2L$ (E1), their
anchoring to the corners (E2), the exponential decay of $\rho$ (E3), the growth of the frozen volume (E4), and the
narrowing of its bracket (E5). For the decay,
the slope E3 of $\ln\rho_L-\ln\bar\rho_{2L}$ against $-\ln\rho_L$ over the sites of a corner octant with
$10^{-6}\le\rho_L<10^{-2}$ where $\bar\rho_{2L}$ is resolved, 1 for
$\rho\sim e^{-cL}$ and, for a power law, 0 or, if its exponent varies from site to site, $\ln2/\ln L$ (0.16 at
$L=80$), is $0.856(34)(34)$ (statistical and systematic errors) from $L=80$ to 160; the criterion asks that E3 minus two
statistical errors be at least 0.5. This estimator was chosen after these data had been read and its value known. The
original one, over all sites with $\bar\rho_{2L}>0$, gives 0.946(10) but passes only with the error of a least-squares
fit, which treats the sites as independent: with a jackknife error over replicas, 0.95(24), it would fail, as it would on
square ice, where the frozen region exists; the new one, applied unchanged to square ice, gives 0.851(34) there (S2). At $48\to96$, the largest pair common to the 2D and 3D series, it is 0.80(4) in 3D and 0.82(6) in 2D, where
freezing is exponential. At $80\to160$, E3 does not rest on the face layers, which hold 43\% of its sites: on the sites at least one, two, or
three layers from the faces, it is 0.90 to 0.94; on the 14 sites at least four layers in, few and weakly frozen, it is
0.75 (S2).

\section{Definitions}
\label{app:definitions}
\textit{Geometry}---The diagonal $x=y=z=uL$ starts at a corner and reaches the center of the opposite face at $u=1/2$.
The box is split into eight octants, one per corner. Face layers are the sites of the outermost layer.

\textit{Disorder and string density}---Replica $r$ gives the time average $m_r(e)$ of each arrow. With
$S_e=\sum_r m_r(e)$ and $T_e=\sum_r m_r(e)^2$ over $R$ replicas, $\hat v_e=1-(S_e^2-T_e)/[R(R-1)]$ estimates
$1-\langle s_e\rangle^2$ without bias. The string density of an edge is $\rho_e=(1-s^c_e\langle s_e\rangle)/2$, where
$s^c_e$ is its arrow in the corner state of its octant. Site values average over a site's free edges,
those not frozen by the boundaries, and over its symmetry images (16 in 3D).

\textit{Rate function}---Its error $\sigma$ comes from those of $\rho_L$ and $\bar\rho_{2L}$. A site with no
string seen at $L$ or $2L$ counts as frozen. We call $\bar\rho_{2L}$ resolved where its relative jackknife error is below
0.5.

\textit{Frozen fraction}---The down-set of a set $\mathcal A$ of sites holds the sites $\mathbf r$ such that the box between
the corner of their octant and $\mathbf r$ lies in $\mathcal A$. The lower bound is the
volume fraction of the down-set of $\{\hat v<10^{-2}\}$ at $2L$. The volume upper estimate is that of the down-set of
$\{I_L>2\sigma\}$, with each face site given the value of its inner neighbor, since the rate function is biased in the
outermost layer, where the matching block at $2L$ mixes two layers (S2).
It is not a bound: in 2D it stays below the exact value up to $2L=96$.

\textit{Local equilibrium}---The symmetry $\Theta$, inversion combined with arrow reversal, keeps the flux
$\mathbf b$ and reverses its gradient, so that the vertex frequencies read
\begin{equation}
P_v(\mathbf r)=P_v^{\rm bulk}[\mathbf b(\mathbf r)]+O(L^{-1})_{\rm odd}+O(L^{-2})_{\rm even}.
\label{eq:le}
\end{equation}
The odd part obeys the exact identity $P({\rm source}_a)-P({\rm sink}_a)=\frac12\langle s_{+a}-s_{-a}\rangle$, where
${\rm source}_a$ (${\rm sink}_a$) holds the vertices whose two arrows along $a$ point out (in) and $s_{\pm a}$ are the
arrows on either side along $a$. $D^2=\sum_v(\Delta_v^2-\sigma_v^2)$, summed over the vertex types and averaged over the
$2\times2\times2$ cells of the diagonal with $0.15\le u\le0.45$ ($0.2\le u\le0.45$ in 2D), where $\Delta_v$ is the
$\Theta$-even part of $P_v-P_v^{\rm bulk}$ and $\sigma_v$ its error, has no noise bias. It is resolved up to $L=48$, and
bounded beyond.

\textit{Local chemical potential}---$\mu_{\rm loc}(u)=X_B^{-1}(\rho(u))$, where $X_B(\mu)$ interpolates the bulk
string density of the tori along $(1,1,1)$, with Eq.~\eqref{eq:eos} below their smallest $\mu$ (in 2D, the 2D tori,
with $\rho=a\sqrt\mu+b\mu$). The linear fits $\mu_{\rm loc}=g(u-u_0)$ use the sites where $0.03\le\rho\le0.2$.

\textit{Independent strings}---$\Lambda(D)=\sum_{k<D}n_k-\sum_{k<D-1}f_k$, where $n_k$, half the mean number of
reversed arrows at the $k$th vertex of the line, counts string passages, and $f_k$, the mean reversal of the line edge
between vertices $k$ and $k+1$, merges a string running along the line ($f_k=0$ on the diagonal). As strings are
directed relative to the corner state, the count does not depend on their pairing at shared vertices. For a box
anchored at the corner, $\Lambda$ counts the passages through its vertices minus the string steps along its inner
edges. $\sigma_{\rm ind}$ is the standard deviation of the law of Eq.~\eqref{eq:null}.

\end{document}


\title{Supplemental Material for ``Boundary-Sculpted Arctic Surface in the Cubic Ice Model''}

\author{Benjamin Canals}
\email[Contact author: ]{benjamin.canals@neel.cnrs.fr}
\affiliation{Universit\'e Grenoble Alpes, CNRS, Grenoble INP, Institut NEEL, 38000 Grenoble, France}

\date{November 12, 2026}

\maketitle

This Supplemental Material gives the details to which the Letter refers: the model and sampling (S1), the existence of
the frozen region (S2), the bulk equation of state (S3), local equilibrium and the local chemical potential (S4), the
edge and its fluctuations along lines (S5), independent strings (S6), and tori with a sawtooth field (S7). Each section
opens with its question and its answer, gives the evidence, and ends with the criteria and checks behind it. The words
verdict and descriptive are defined in S2 (``How the tests were run''), and the runs are listed in Table~\ref{tab:runs}. The criteria and the main values are in
Tables~\ref{tab:criteria} and~\ref{tab:e3} (existence), \ref{tab:b1} and~\ref{tab:ab} (equation of state), \ref{tab:le} (local
equilibrium), \ref{tab:edgepos} (edge), \ref{tab:lines} and~\ref{tab:probes} (lines and boxes), and \ref{tab:tori}
(tori).
Equations, figures, tables, and appendices of the Letter are cited as Eq.~(1), Fig.~1, Table~I, and
Appendix~A; those of this Material carry the prefix S.

Table~\ref{tab:notation} gathers the main symbols; a letter that serves again within a single
section, such as the forms K, P, and M of S3, is defined where it appears. Sizes are those of the box, $L\times L\times
L/2$ for the three-dimensional (3D) systems and $L\times L$ for the two-dimensional (2D) ones, or of
the torus, and a pair of sizes is written $L\to2L$.

\begin{table*}
\caption{\label{tab:notation}Main symbols.}
\begin{ruledtabular}
\begin{tabular}{lp{6.3cm}lp{6.3cm}}
$u$ & position on the body diagonal $x=y=z=uL$ (2D: $x=y=uL$) & $\mu$, $\mu_{\rm loc}$ & log-weight per string step,
$\ln\sum_ae^{-2|H_a|}$; its local value, read through Eq.~(3) \\
$\tilde u$ & $(x-y)\bmod L$ on sawtooth tori, in lattice units & $H_a$, $H_K$ & field along axis $a$; Kasteleyn threshold \\
$d$ & distance to the edge: $u-u_0$ or $u-u_c$ (S4, S5), in sites (S6, S7) & $c$ & direction of the field $(H,H,cH)$ \\
$u_c$, $u_0$, $y_c$ & edge on the diagonal; zero of $\mu_{\rm loc}$; exact edge of square ice, $1-\sqrt3/2$ (Appendix~A, S4 to S6) & $X_B$, $g$ & bulk string density as a function of $\mu$;
slope of $\mu_{\rm loc}$ in $u$ (S4) \\
$s_e$, $m_r(e)$ & arrow of edge $e$, $+1$ toward increasing coordinates; its time average in replica $r$ &
$\mathbf b$, $\Theta$ & flux, the local mean of the arrows (S1); inversion combined with arrow reversal \\
$\rho$, $\bar\rho_{2L}$ & string density: fraction of arrows reversed with respect to the state of the corner of its octant
(on tori, $(+,+,+)$); its average over the matching block at $2L$ & $P_v$, $D^2$ & frequency of vertex type $v$; squared
distance of the $\Theta$-even frequencies to the bulk \\
$\hat v$ & disorder, unbiased estimate of $1-\langle s_e\rangle^2$ & $\ell$, $\sigma_\ell$ & frozen length along a line
from a corner; its standard deviation, with Sheppard's correction \\
$I_L$ & rate function, Eq.~(2) & $D$, $\Lambda(D)$ & depth along a line; mean number of string segments within it \\
$D_c$ & down-set of corner $c$ (S2) & $n_k$ & string passages at the $k$th vertex of a line \\
$k$ & depth in layers from the faces (S2) & $\sigma_{\rm ind}$, $\delta$ & width of the independent-string model;
the gap $\delta=\sigma_\ell/\sigma_{\rm ind}-1$ \\
$K$ & distance of a line from the two faces it runs along & W1, W2 & windows of $\mu$ (S3) \\
$\tau_{\rm int}$ & integrated autocorrelation time & sweep & $E$ worms in boxes; on tori, worm attempts that reverse
about $E$ edges \\
$R$ & number of replicas of a run (8) & $m$ & mass of a string (S3) \\
$D^*_1$, $D^*_2$ & thickness of the edge (S5) & $p_a$ & probability that a string steps along axis $a$ (S3) \\
$\lambda$ & Tracy-Widom scale of square ice, $2^{-4/3}3^{-1/6}$ (Appendix~A, S4 to S6) & & \\
\multicolumn{4}{l}{Criteria: E1--E5 (S2), B1--B2 (S3), D0--D3 (S4), F1--F2 and H0--H2 (S5), C0--C2 (S7); $c_i$, $\eta$,
$\epsilon$: coefficients of fits.} \\
\end{tabular}
\end{ruledtabular}
\end{table*}

\section{Model and sampling}
\label{sec:model}

Which ensembles are sampled, and are the samplers exact? We sample the ice configurations of boxes whose pendant edges
are frozen, with a loop algorithm, and of tori in a field, with a two-defect worm; both match exact enumeration on small
systems.

\textit{Lattice and measure}---Each vertex of an $L_x\times L_y\times L_z$ box of the simple cubic lattice has six
edges. Those that leave the box, the pendant edges, end on vertices of their own, free of any constraint. The box thus
has
\[
E=(L_x+1)L_yL_z+L_x(L_y+1)L_z+L_xL_y(L_z+1)
\]
edges, 6\,195\,200 for cubic ice at $L=160$. The arrow $s_e=+1$ points toward increasing coordinates. Only pendant edges
can be frozen. We sample the uniform measure on the configurations that obey the ice rule, three arrows in and three
out, at every vertex of the box and the frozen edges: there is no energy, and the problem is one of counting. (This is
the ice rule of a six-coordinated lattice; on the cubic lattice, it defines a twenty-vertex model, whose phase diagram
with general weights was studied in Ref.~\cite{ShtengelChayes2005}; here all weights are equal. The name cubic ice
refers to this lattice, not to ice Ic, the cubic form of water ice, whose rule, two in and two out on the diamond lattice,
is that of pyrochlore spin ice. On the planar triangular lattice, the same rule gives a 2D twenty-vertex model, whose
arctic curves with domain-wall boundaries were obtained by the tangent method and seen in
simulations~\cite{DebinDiFrancescoGuitter2020}; there, as in square ice, strings that meet at a vertex can always be
drawn without crossing, and stay ordered.)

\textit{Cubic ice}---The pendant edges point inward on the four side faces and outward on the top and bottom faces
[Fig.~1(c)], the analog of the domain-wall boundaries of square ice~\cite{Korepin1982}. All are frozen: there is no
defect. The flux through a surface is the number of arrows that cross it one way minus the number that cross it the
other way. The ice rule makes the flux out of every vertex vanish; summed over any set of vertices, the inner edges cancel, so
that the net flux out of the set vanishes too: what enters through part of its boundary leaves through the rest. For
the whole box, this is Eq.~(1). For a layer of vertices, it fixes the sum of the arrows of
every plane of edges, in every configuration. A layer normal to $x$ receives $L$ arrows through its two faces normal to
$y$ and sends $2L$ out through the top and bottom, so that the polarization of the $k$th plane of $x$ edges, the average
of its arrows, counted from the pendant edges ($k=0$, all $+1$), is $1-2k/L$, and likewise along $y$; a horizontal
layer receives $4L$ arrows through its sides, and the $k$th plane of $z$ edges has the polarization $4k/L-1$. The
boundaries thus impose a flux whose plane averages turn from $+1$ to $-1$ across the box. Its local value, the flux
$\mathbf b(\mathbf r)$, the mean of the arrows near $\mathbf r$ along each axis, is not fixed: it is the coarse-grained,
divergence-free field of the Coulomb phase of ice~\cite{Henley2010}, to which S4 associates, region by region, the
uniform field that gives the bulk the same flux. Each
face fixes the arrows normal to it; at a corner, the three pendant arrows match exactly one of the eight polarized
vertex types, a different one at each corner, which we call the state of the corner: at $(0,0,0)$, they point in along
$x$ and $y$ and out along $z$, as in $(s_x,s_y,s_z)=(+,+,-)$. The corner vertex itself is not fixed: its three inner
edges carry one arrow in and two out, in three ways, only one of them polarized. That the region near each corner
freezes into its state as $L$ grows is a result (S2), not a constraint. The symmetry group,
generated by the three reflections and the exchange $x\leftrightarrow y$, has 16 elements, over which site values are
averaged. Exact enumeration gives 2 configurations for $L=2$ and 476\,004 for $L=4$ ($4\times4\times2$).

\begin{figure*}
\includegraphics[width=\textwidth]{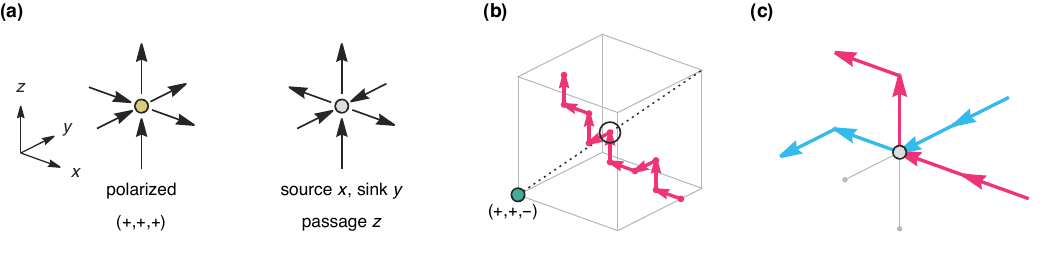}
\caption{\label{fig:strings}Vertex types and strings (schematic). (a) Two of the twenty vertex types of cubic ice: a
polarized type, here $(+,+,+)$, which passes along all three axes, and an unpolarized one, with a source along $x$, a
sink along $y$, and a passage along $z$. (b) A string, relative to the corner $(0,0,0)$ in the state $(+,+,-)$: its
reversed arrows step along $-x$, $-y$, or $+z$, so it meets the diagonal of the corner (dotted) at most once (circle;
S6). (c) Two
strings crossing the same vertex, which they enter and leave by distinct edges.}
\end{figure*}

\textit{Vertex types and strings}---Along each axis, the two arrows of a vertex either pass through it, in the $+$ or
$-$ direction, or both point out (a source along that axis) or both in (a sink). Eight of the twenty types of an ice
vertex pass along all three axes: these are the polarized types, one per state of a frozen corner. Each of the twelve
others has a source along one axis, a sink along another, and a passage along the third [Fig.~\ref{fig:strings}(a)].
Relative to the polarized state of a corner, the arrows that differ form strings. Take the corner $(0,0,0)$, in the
state $(+,+,-)$: a reversed arrow points toward the corner along $x$ and $y$, and away from it along $z$. Every vertex
keeps three arrows in and three out, so it has as many reversed arrows in as out, and reversed arrows join into
directed paths that step along $-x$, $-y$, or $+z$ [Fig.~\ref{fig:strings}(b)]. Two strings can cross the same vertex,
entering by two reversed arrows and leaving by two others, but they never share an edge [Fig.~\ref{fig:strings}(c)];
a configuration does not say which incoming string continues as which outgoing one. These strings run across a frozen
region from side to side, like the strings of spins reversed against a $[100]$ field in pyrochlore spin ice, whose
onset is a Kasteleyn transition~\cite{Jaubert2008}; neutron scattering revealed such strings in Dy$_2$Ti$_2$O$_7$ in this
field~\cite{Morris2009}. There, at the temperatures of the experiment, a string can also end inside the sample, on a
monopole, a tetrahedron that breaks the ice rule~\cite{Castelnovo2008}, and Morris et al.\ call these strings Dirac
strings. Here the ice rule holds at every vertex, and a string ends only on the boundary of the box.

\begin{figure}
\includegraphics[width=\columnwidth]{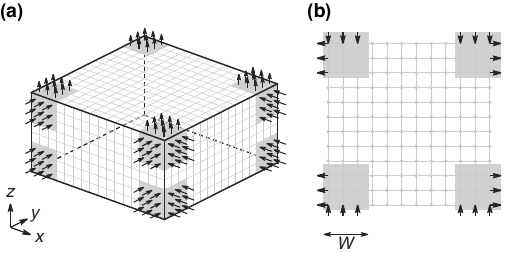}
\caption{\label{fig:band}Banded boxes (schematic). (a) A banded box, drawn for $L=16$ and $W=3$: the pendant edges of
cubic ice [Fig.~1(c)], inward on the sides and outward on the top and bottom, stay frozen (arrows) only at the vertices of
the eight corner cubes of width $W$, which meet the visible faces in the gray squares; the others are free (not drawn).
(b) A banded square, $L=12$ and $W=3$ ($x$ to the right): the pendant edges in the plane, out on the left and right
faces (normal to $x$) and in on the top and bottom ones (normal to $y$) as in square ice, stay frozen only at the
vertices of the four corner squares; the vertical pendant edges,
not drawn, are frozen at every vertex (dots). The width $W$ runs between two limits: at $W=0$, every pendant edge
would be free (in 2D, every one in the plane, the vertical ones staying frozen, unlike in the free box), and from $W=L/2$
on, every one is frozen, which gives back cubic ice in (a) and square ice in (b).}
\end{figure}

\textit{Square ice and controls}---Square ice with domain-wall boundaries is the $L\times L\times1$ box whose vertical
pendant edges are frozen, one in and one out at every vertex, the others pointing out on the faces normal to $x$ and in
on those normal to $y$. What remains is the six-vertex model at $\Delta=1/2$, whose configurations are the
alternating-sign matrices~\cite{MillsRobbinsRumsey1983,Kuperberg1996} (42 for $L=4$, as enumerated). The free box is an
$L\times L\times1$ box with no frozen edge at all; its string density is measured with respect to the state $(+,+,+)$.
Banded boxes keep the frozen pendant edges of cubic ice (of square ice in 2D) only at the vertices inside the eight
corner cubes (four corner squares) of width $W$, and leave the others free (in 2D, the vertical pendant edges stay frozen
at every vertex; Fig.~\ref{fig:band}). $W=0$ would leave every pendant edge free (in 2D, every one in the plane), and
$W\ge L/2$ recovers cubic or square ice; the corner regions of a box stay
separate cubes only for $W<L/4$, the box being $L/2$ high, and the runs have $W=4$ to 6 at $L=48$ to 96.

\begin{figure}
\includegraphics[width=\columnwidth]{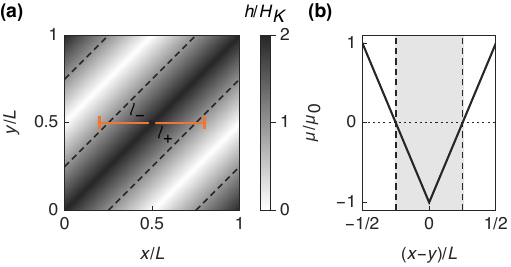}
\caption{\label{fig:field}Field of the sawtooth tori (schematic). (a) Reduced field $h_e/H_K$ in a plane of the torus
(the same in every plane $z$; the 2D torus is one such plane): largest on the diagonal $x=y$, zero on $x-y=\pm L/2$.
Dashed: the two planar edges, $\tilde u=L/4$ and $3L/4$, each cut in two by the periodic boundaries; they have no
corners and separate the polarized region around $x=y$ (dark; its continuation across the boundaries fills the top-left
and bottom-right corners) from the liquid region. Along a line parallel to $x$ from the diagonal, $\ell_+$ and $\ell_-$
count the vertices in the state $(+,+,+)$ up to the first string, on either side (S7); on average, the first string
lies a few sites beyond the edge.
(b) The local log-weight per string step, $\mu/\mu_0=1-h_e/H_K$, against $(x-y)/L$ brought into $[-1/2,1/2]$; gray:
polarized region, $\mu<0$; dashed: the edges.}
\end{figure}

\textit{Tori}---$L\times L\times L_z$ tori have no boundary. In the 2D control, $L\times L\times1$ tori keep their
vertical edges frozen, one in and one out at every vertex. A field enters through a weight $\prod_e e^{h_es_e}$; in spin
ice, $h_e$ would be a Zeeman energy over the temperature, a reduced field. In a uniform field, $h_e=H_a$ for an edge
along axis $a$ (S3, S4). On sawtooth tori (S4 to S7), the field is the same on the three axes and depends only on
$\tilde u=(x-y)\bmod L$, taken at the midpoint of the edge (Fig.~\ref{fig:field}):
\[
h_e=H_{\max}-(H_{\max}-H_{\min})\,\frac{2\min(\tilde u_e,L-\tilde u_e)}{L},
\]
with $H_{\min}=0$ and $H_{\max}=2H_K$ ($\ln3$ in 3D, $\ln2$ in 2D). A field that varies linearly cannot be periodic: this
one falls from $H_{\max}$ at $\tilde u=0$ to $H_{\min}$ at $\tilde u=L/2$ and rises back, a profile we call a sawtooth,
although in the strict sense it is a triangle wave. It crosses the Kasteleyn threshold $H_K$ at $\tilde u=L/4$ and
$3L/4$, where two planar edges, without corners, separate a polarized region centered at $\tilde u=0$ from a liquid
region centered at $\tilde u=L/2$. For $0\le\tilde u\le L/2$, the log-weight per string step of S3 is
$\mu=\mu_0(4\tilde u/L-1)$, where $\mu_0=\ln3$ ($\ln2$ in 2D) is its value at zero field. The edges are planes of
constant $\tilde u$, normal to $(1,-1,0)$, through which the polarized state $(+,+,+)$ carries no net flux. This is why
$\tilde u$ runs along $(1,-1,0)$: on a torus, the ice rule keeps the flux the same through every plane of constant
$\tilde u$, and with $\tilde u$ along an axis or along $(1,1,0)$, the polarized region would carry the largest possible
flux through these planes, which would freeze, everywhere, all the arrows that cross them.

\textit{The loop algorithm}---One update of the Markov chain on a box, which we call a worm, starts from a random vertex
and walks forward along free outgoing edges, chosen uniformly, until the path closes on itself; the closed loop is
reversed, and the part of the path before the loop is discarded. This is the loop algorithm of Barkema and
Newman~\cite{BarkemaNewman1998}. A
vertex keeps the same number of free outgoing edges in every ice state, so a loop and its reverse are proposed with the
same probability (the path before the loop is unchanged by the reversal), and the chain obeys detailed balance. When
pendant edges are free (free box, banded boxes), the forward walk can leave the box. The worm then walks backward from
its start, along free incoming edges chosen uniformly, until it leaves the box too, which gives an open string, or meets
its own path, which closes a loop. A loop closed by either walk on itself is accepted, as above. An open string, or a
loop closed by the backward walk on the forward one, is accepted with probability $\min(1,R)$, the ratio of the
probabilities of proposing the reverse move and the move. With $o(v)$ and $i(v)$ the numbers of free outgoing and
incoming edges of $v$, both constant, $R=\prod o(v)/i(v)$ over the vertices of the forward walk times $\prod i(v)/o(v)$
over those of the backward walk, the start (and the junction) excluded; $R\ne1$ only where frozen pendant edges make
$o\ne i$. Cubic and square ice have no free pendant edge, and this test never acts. Two ice states
with the same frozen edges differ on free edges that, at every vertex, point in as often as out: they split into
directed loops and, where pendant edges are free, open paths between them. A worm proposes and reverses each of these
with nonzero probability, so the chain goes from any ice state to any other: the uniform measure is its only stationary
law, and time averages converge to it whatever the initial state. A worm reverses
5.5 edges on average (5.65 to 5.46 from $L=64$ to 160), and a sweep is one worm per edge. Random numbers come from
xoshiro256++~\cite{BlackmanVigna2021}, replica seeds from SplitMix64~\cite{SteeleLeaFlood2014}.

\textit{Tori: the two-defect worm}---On a torus, the number of strings winding around it changes only through
system-spanning strings. A loop worm proposes a given winding string of $n\ge L$ steps with probability $3^{-n}$, and a
field at $H_K$ accepts it with $e^{-2H_Kn}=3^{-n}$: the number of strings does not equilibrate in practice. We use instead a two-defect
worm~\cite{ProkofevSvistunov2001}. Reversing a free outgoing edge $e=(t\to v)$ of a random vertex $t$ creates a tail at
$t$, with four arrows in, and a head at $v$, with four out: a pair of defects, the analog of a pair of monopoles in spin
ice~\cite{Castelnovo2008}. The opening is accepted with probability $\min(1,w_eA(t)/A(v))$, where $w_e=e^{-2h_es_e}$
and $A(v)$, defined at every vertex, is the number of free outgoing edges that a head at $v$ would have: 4 minus the
frozen outgoing edges of $v$, that is 4 on 3D tori and 3 on 2D ones (before the opening, the tail $t$ has one free
outgoing edge fewer). The head then moves one edge at a time,
from $v$ to $v'$, each step accepted with probability $\min(1,w_eA(v)/A(v'))$, backward steps included, and the worm
closes when the head returns to the tail, with probability $\min(1,w_eA(v)/A(t))$. Since $A(v)/A(v')$ is the ratio of the
probabilities of proposing the reverse step and the step, each step of the head obeys detailed balance in the space
extended by the two defects. So do the opening, proposed with probability $1/[N(A(t)-1)]$ for $N$ vertices, and the
closing, its reverse, proposed with $1/A(v)$, provided a state with defects carries, besides its field weight, the factor
$A(t)/[N(A(t)-1)]$: $4/(3N)$ on 3D tori, $3/(2N)$ on 2D ones. This factor weights only the states with defects, as would a fugacity of
the pair, which multiplies the ratio in the acceptance of the opening and divides that in the acceptance of the closing
(1 in all the runs of Table~\ref{tab:runs}; 0.3, 1, and 3 in the exactness tests, Checks): such a weight changes how
often worms open, not the law of the ice states, which alone are measured. Barkema and Newman forbid backward steps, which is exact without field only.
The head can trace any loop, winding ones included, so this chain too goes from any ice state to any other.
Configurations are measured only without defects: once per attempt to open a worm, after it has closed, rejected
attempts included. Measuring only after the worms
that open would weight configurations wrongly as soon as $h\neq0$, since the time spent without defects then depends on
the configuration: in the polarized state at $H_K$, a worm opens with probability $1/3$ only. A sweep is a fixed number
of attempts, chosen to reverse about $E$ edges.

\textit{Initial states}---Each run needs one ice state compatible with the frozen edges; a maximum flow builds it in the
same way for every boundary condition used here, and for the melted start (Checks). We orient the free edges arbitrarily
and count, at each vertex, half the excess of incoming over outgoing arrows: the net number of its incoming free edges
to reverse.
Reversing a free edge moves one unit of this excess from one end of the edge to the other, so the sets of reversals
that balance every vertex are the integer flows that carry one unit along each reversed edge, from the vertices with
too many outgoing arrows to those with too many incoming ones, and meet every excess. Such a flow exists if and only if
an ice state does, and Dinic's algorithm~\cite{Dinic1970}, exact on integer capacities, finds one; the program stops if
the flow does not meet every excess, and checks the ice rule at every vertex before the run. Since the chain reaches
every ice state, the initial state need not be typical: thermalization erases its memory (Durations), which two checks
test, the symmetry between corners (S2) and the melted start (Checks). Starting from all free edges at $+1$, the state
of the corner $(+,+,+)$, gives the common state from which all replicas of a run start. On tori, runs start from the
polarized state $(+,+,+)$.

\textit{Replicas and errors}---Each run has eight replicas with distinct seeds and their own thermalization; afterwards,
in boxes, each replica runs blocks of 1024 worms until the reversals of all replicas together reach a set total, so that
their lengths differ (by 0.5 to 22\% in the runs of cubic ice from $L=48$ on), while on tori each replica measures a
fixed number of sweeps. A replica's length thus depends on its own trajectory only weakly, through the time its worms
take and its share of the total; any resulting bias of its averages is at most of order $\tau_{\rm int}/T$, here under a
tenth of their statistical error. The jackknife variance of an equal-weight mean of replica averages is unbiased
whatever their lengths, and weights proportional to length would lower it by less than 1\%. Arrow averages are integrated over every worm of the measurement, at a cost
proportional to the number of reversals, and stored per replica in double precision, as $1-|m_r(e)|$ falls below
$10^{-6}$ near the frozen state. Errors are jackknife estimates over replicas~\cite{Efron1982} (for the equation of state of the tori, from autocorrelation times, S3); quantities that combine two sizes add the variances of both. For a fraction of sites below
a threshold, the jackknife error often vanishes, since removing one replica moves no site across the threshold: the
fraction then carries the range obtained by shifting $\hat v$ by $\pm2\sigma$ at every site. With eight replicas, an
error is itself uncertain by about $1/\sqrt{2(R-1)}=27\%$, and a deviation of $5\sigma$ has a one-sided Student
probability of $8\times10^{-4}$, that of 3.2 Gaussian $\sigma$. Fits are weighted by these errors. When the points
scatter about the fitted form more than their errors allow ($\chi^2/\mathrm{ndf}>1$, with ndf the number of degrees of freedom), the errors of the fitted
parameters are multiplied by $\sqrt{\chi^2/\mathrm{ndf}}$, the scale factor of the Particle Data Group~\cite{PDG2024},
so that the excess scatter enters them; they are never reduced when $\chi^2/\mathrm{ndf}<1$, which happens often by
chance with errors that are themselves uncertain. The rule is conservative: with errors estimated on eight replicas,
$\chi^2/\mathrm{ndf}$ is about 1.4 on average even for the right form. It does not correct a wrong form: where the form
itself is in doubt, the fitted value is given with its variants over forms and windows (S3 to S5), or only as a
description, as the exponent of a rejected power law (S7). The verdicts of S2 to S5 clear their thresholds by wide margins: E3
by 10 errors, D1 by 13, F2 by 7.5, H1 and H2 by 11 or more.

\textit{Durations}---Successive configurations of the chain are correlated. The integrated autocorrelation time
$\tau_{\rm int}$ of an observable, computed with Sokal's automatic window~\cite{Sokal1997,MadrasSokal1988}, is the
number of sweeps over which its values stay correlated: an average over $N$ sweeps has the error of $N/2\tau_{\rm int}$
independent values. We call an observable slow when its $\tau_{\rm int}$ is long, and a mode slow when it takes many
sweeps to relax; the chain forgets its initial state at the pace of its slowest mode, which an observable reveals only
if it couples to it. The slowest observables measured are the volumes of the frozen region of each sample, the
down-sets of S2, which follow its edge. The volume of the cluster of corner-state vertices connected to a corner (S2),
which contains the down-set of that corner, is dominated by fast local fluctuations: its $\tau_{\rm int}$, 0.6 to 2
sweeps in cubic ice, does not reveal the slow modes of the edge, although a drift of the edge would shift its mean. The
vertex-type runs, a second set of runs of cubic ice with other seeds (Table~\ref{tab:runs}), record, from $L=48$ on, the
down-set of every sample. There, the volume of a single corner has the longest time measured, 24.7, 34.4, 60.1, and
82.9 sweeps at $L=48$, 64, 96, and 128, growing roughly as $L^{1.25}$; that of the four corners together is shorter, 18 to
64 sweeps. Below, the lengths of the runs are also expressed in units of this single-corner time. At $L=160$, where no run records the down-set, it
extrapolates to about 110 sweeps as $L^{1.25}$, or 130 as $L^2$. The latter is the growth expected of long-wavelength
modulations of the flux: each worm reverses a short loop, which changes the flux only locally and keeps it free of
divergence, so a modulation over a length $\lambda$ should relax by diffusion, in a time $\propto\lambda^2$~\cite{Henley2010}. These modes
were not measured: the runs keep the arrows only as time averages. A mode that relaxes from the start in $\tau$ sweeps
shifts averages over $T$ sweeps, taken after $T_{\rm th}$ sweeps of thermalization, by about
$(\tau/T)\,e^{-T_{\rm th}/\tau}$ of its initial amplitude. At $L=160$ ($T_{\rm th}=2000$, $T\simeq15\,500$), this is less than
$10^{-3}$ for $\tau$ up to about 550 sweeps, four to five times the extrapolated single-corner time, but nearly
$10^{-2}$ for $\tau=1000$, enough to move sites across the threshold $\hat v=10^{-2}$. Runs of cubic ice last 1000 sweeps of thermalization
(2000 at $L=128$ and 160) and, at $L=64$ to 160, 3100 to 15\,800 sweeps of measurement per replica: at $L=128$ and 160,
about 11\,000 and 15\,500, or 120 to 140 single-corner times, after a thermalization of 24 and 15 to 18 such times. In the
vertex-type runs, thermalization lasts 17 to 40 single-corner times, and the measurement 87 to 122; along lines, the
autocorrelation time of the mean $\ell$ is at most 16 sweeps at $L=128$. On tori started from the polarized state, at
$L=64$ and 128, the number of strings reaches equilibrium in 30 to 40 sweeps in a uniform field, and the observables of
sawtooth tori in 10 to 25 sweeps; their autocorrelation times are 1 to 4 sweeps. A sweep of a torus makes about $E$
reversals, and one of a box about $5.5E$ (The loop algorithm; Table~\ref{tab:notation}): durations in sweeps compare within boxes or within
tori, not between them.

\begin{table*}
\caption{\label{tab:runs}The runs; each has eight replicas. Records: what the runs store besides the time average of
every arrow in each replica (on tori, besides the string density and its fluctuation). Cubic ice is run in four main sets:
the existence runs ($L=12$, 24, 48, 64, 80, 96, 128, and 160), which record the corner-state clusters; the vertex-type
runs, with the parameters of the existence runs and other seeds at $L=48$ to 128 (S1, ``Reproducibility''), which also
record the vertex types and, from $L=48$ on, the down-set of every sample (``types, down-sets''); the runs for lines;
and the melted start; the first vertex-type runs at $L=16$ to 32, on which the window of S4 was set, are listed with
those of square ice (``design of S4''). $L$: range of sizes (the sizes used are given in the sections); tori in a field $(1,1,c)$ are
$L\times L\times1.2L$. Sweeps: thermalization (Therm.), then measurement per replica, from the shortest to the longest
replica (on tori, a fixed number). Time: wall time; 393 runs, about 118 hours in all. Used in: the sections that use the runs. Not listed:
the small systems of the exactness tests (S1).}
\begin{ruledtabular}
\begin{tabular}{llllllll}
System & $L$ & Records & Runs & Therm. & Measurement & Time & Used in \\
\colrule
Cubic ice, existence runs & 12 to 160 & clusters & 8 & 1000, 2000 & 923--15\,754 & 54.6 h & S1, S2, S4, S5 \\
Cubic ice, vertex-type runs & 16 to 128 & types, down-sets & 7 & 1000, 2000 & 2162--11\,692 & 8.6 h & S1, S2, S4--S6 \\
Cubic ice & 32 to 128 & lines & 5 & 1000, 2000 & 3188--11\,250 & 8.5 h & S1, S5, S6 \\
Cubic ice, melted start & 64, 80, 128 & down-sets & 3 & 1000, 1280 & 3251--11\,130 & 7.9 h & S1 \\
Square ice & 12 to 384 & clusters & 6 & 1000, 2000 & 688--22\,841 & 2.2 h & S2, S4--S6 \\
Square ice & 24 to 192 & vertex types & 5 & 1000 & 3337--22\,203 & 9 min & S4 \\
Square ice & 64, 128, 256 & lines & 3 & 1000, 2000 & 2197--13\,621 & 0.5 h & S5, S6 \\
Cubic and square ice, design of S4 & 16 to 32 & vertex types & 5 & 1000 & 1928--2219 & 1 min & S2, S4 \\
Free box & 12 to 96 & clusters & 4 & 1000 & 361--2404 & $<1$ min & S2 \\
Banded box, $W=4$ & 48, 64, 96 & clusters & 3 & 1000 & 1678--5397 & 0.8 h & S2 \\
Banded box, $W\simeq\sqrt{L/3}$ & 48, 64, 96 & clusters & 3 & 1000 & 1675--5418 & 0.8 h & S2 \\
Banded square, $W=4$, $\simeq\sqrt{L/3}$ & 48 to 384 & clusters & 8 & 1000, 2000 & 1305--22\,935 & 1.5 h & S2 \\
Banded square, $W=20$ & 48 to 384 & clusters & 4 & 30\,000 & 3307--135\,433 & 1.6 h & S2 \\
Banded square, $W\simeq20\sqrt{L/48}$ & 48 to 384 & clusters & 4 & 1000, 2000 & 1309--25\,201 & 1.0 h & S2 \\
Tori, field $(1,1,1)$ & 32 to 256 & --- & 21 & 500, 1000 & 2000--4000 & 7.4 h & S3, S4, S7 \\
Tori, field $(1,1,1)$, finite size & 16 to 128 & --- & 26 & 200 to 1000 & 500--3000 & 1.6 h & S1--S3 \\
Tori, from the polarized state & 64, 128 & --- & 4 & 0 & 600--1500 & 23 min & S1 \\
Tori, field $(1,1,c)$ & 30 to 240 & --- & 18 & 500, 1000 & 2000--4000 & 7.3 h & S3 \\
2D tori, uniform field & 256, 512 & --- & 14 & 1000 & 4000 & 0.5 h & S3, S4, S7 \\
Tori, field $(H,H,H_z)$ & 32, 64 & vertex types & 173 & 500 & 1000--4000 & 1.8 h & S4 \\
2D tori, field $(H,H,0)$ & 128, 256, 512 & vertex types & 48 & 1000 & 4000 & 1.0 h & S4 \\
Sawtooth tori & 32 to 256 & lines & 7 & 500 to 2000 & 1000--4000 & 8.1 h & S4--S7 \\
2D sawtooth tori & 64 to 1024 & lines & 8 & 2000 to 6000 & 20\,000 & 1.5 h & S4--S7 \\
Sawtooth tori, short runs & 32 to 512 & lines & 6 & 500, 1000 & 2000--5000 & 8 min & S2 \\
\end{tabular}
\end{ruledtabular}
\end{table*}

\subsection{Checks}

\textit{Exactness}---On cubic ice $4\times4\times2$, sixteen replicas of $1.9\times10^6$ worms reproduce the exact
mean arrows of the 64 free edges, the largest normalized deviation being $2.6\sigma$ (normalized deviations whose error is
estimated on $R$ replicas follow a Student law with $R-1$ degrees of freedom, of variance $(R-1)/(R-3)$), and visit the
476\,004 configurations uniformly. Recorded after every worm, each configuration appears about four times per replica,
too few for a Student law: the $\chi^2/\mathrm{ndf}$ of the frequencies, 1.19, is to be compared with that of
independent uniform draws with the same numbers, 1.186 with a spread of 0.003 (simulated), rather than with the Student
value $15/13=1.15$. A deliberate bias, a field $h=0.05$ on one row of edges, is detected at $33.5\sigma$ with
$2\times10^5$ worms per replica. On square ice $4\times4$, three tests of the frequencies of the 42 configurations, each
visited $4\times10^3$ to $5\times10^4$ times per replica, give $\chi^2/\mathrm{ndf}=0.90$, 1.35, and 1.75, for 1.15
expected with a spread of 0.28. Open strings are checked in the same way on the free box $2\times2\times1$, on banded
squares $3\times3$ and $4\times4$, and on small boxes with part of their pendant edges frozen (7 to 10\,016
configurations): $\chi^2/\mathrm{ndf}$ of 0.81 to 1.16, mean arrows within $2.9\sigma$; without its
Metropolis-Hastings test, the worm gives $\chi^2/\mathrm{ndf}=2978$ on such a $3\times3\times1$ box. The two-defect worm
was checked by exact enumeration with fields, in 15 cases on small tori and boxes of up to $5\times10^5$ states, with
1.5 to $3\times10^5$ measurements per replica: mean arrows within $4.4\sigma$, the largest of about 450 Student
variables with 15 degrees of freedom, whose median maximum is 3.9 (it exceeds 4.4 one time in five), and the joint law of
the winding numbers on a torus. Where computed, the frequencies of the configurations expected at least five times per
replica, the others pooled into one class (35 to 11\,623 classes), give $\chi^2/\mathrm{ndf}$ of 0.88 to 1.36. The
largest is at the smallest fugacity of the defect pair, 0.3 (1.19 and 1.16 at 1 and 3, on the same torus), where most
attempts to open a worm are rejected: the same configuration is then recorded several times in a row, which probably
raises this mean as small counts do. On ordered tori ($H>H_K$, $L=4$ to 8), the ratios of the probabilities of the
winding sectors, known exactly for any field, are reproduced: the squared normalized deviations of 46 ratios sum to
64.8, as expected ($64\pm19$) for errors estimated on seven degrees of freedom.

\textit{Melted start}---Runs at $L=64$ and 80 whose replicas each start from their own random orientation of the free
edges, completed by the same flow (34\% of the free edges then differ from the common state), were compared with the
runs from the common state. The fraction of sites with
$\hat v<10^{-2}$ differs by 0.4 to 0.6 combined errors; edge by edge, the normalized deviations of the mean arrows have
the variance of a Student law with 14 degrees of freedom (1.14 to 1.17, for 1.167) and its share above 3 (0.0080 to
0.0095, for 0.0096), on all edges and in the frozen region $|\langle s_e\rangle|>0.9$; and the corner symmetry control
holds on the melted runs. There is no memory of the initial state at these sizes. At $L=128$, a melted run thermalized for only 1280 sweeps,
$2000/1.25^2$, which gives it at most the ratio of thermalization to relaxation time of $L=160$ if that time grows as
$L^z$ with $z\le2$ (the same ratio for $z=2$), was compared in the same way with the two runs of $L=128$ from the
common state, with criteria fixed before its data were read: the fraction of sites with $\hat v<10^{-2}$ differs by 0.4
and 0.7 combined errors, the normalized deviations of the mean arrows have the variance 1.157 to 1.169 and a share above
3 of 0.0087 to 0.0095, and the corner control holds; reported without criterion, the volume of the down-set does not
drift over the measurement. At $L=160$, no run starts elsewhere, and only the corner control of S2 tests this memory
there: it passes narrowly ($p=0.0015$), three corners deviating by about 0.06\%, but not the corner $(+,+,+)$ of the
common state (Quality controls).

\textit{Reproducibility}---At $L=48$, 64, 96, and 128, the existence runs and the vertex-type runs of cubic ice have the
same parameters and initial state, and different seeds (Table~\ref{tab:runs}); they test the errors and the verdict,
not the memory of the start. Their mean arrows agree edge by edge: the normalized
deviations have the variance 1.158 to 1.172 and a share above 3 of 0.0090 to 0.0097, as for a Student law with 14
degrees of freedom (1.167 and 0.0096), so that the jackknife errors are right at these sizes. Both sets give
``exists'' by the criteria of S2. At the pairs they share, E3 is 0.924(48) and 0.808(47) at $48\to96$ and $64\to128$
on the vertex-type runs, against 0.798(44) and 0.880(29), within two combined errors, and the lower bounds agree within
one error at every size.

\textit{Code and AI tools}---The initial code, including the loop algorithm, was generated with Gemini 4 (Google) and
was audited and corrected before any run used here. The later code (the two-defect worm, the optimized loop algorithm,
the maximum-flow initial states, the measurements, the analysis criteria and estimators, the figure scripts), the
analyses, the derivation of $AB$ (S3), and the text were developed with Claude Opus 5.5 (Anthropic), each step being
proposed to and approved by the author. The checks compare the code and analyses with results they do not use: exact
enumeration of small boxes and tori, with and without a field, independent seeds, and a melted start (this section);
the bit-for-bit identity of the optimized loop algorithm with a reference version; and exact 2D results, the frozen
fraction (S2), the free-fermion amplitude (S3), and the Tracy-Widom moments (S5) of square ice.

\section{Existence of the frozen region}
\label{sec:existence}

Does cubic ice have a frozen region of positive volume? Up to $L=160$, the data say it does: criteria calibrated on
square ice, and set on cubic ice at $L=12$, 24, 48, and 96, return ``exists'' (Table~\ref{tab:criteria}). Near each
corner, the string density falls with the box size as fast as in square ice, where the fall is exponential; at $2L=160$,
the sites with $\hat v<10^{-2}$ fill 3.1\% of the box, a fraction that still grows. The region is thin, one piece at each
corner, mostly near the faces, but it is not a surface layer: its depth along the body diagonal keeps pace with the box,
from 3 sites at $L=48$ to 11 at 160. A region whose depth grows slightly more slowly than $L$, so that its share of the
volume would vanish at larger sizes, is not excluded (Negative controls; Growth at a fixed position).

\begin{table*}
\caption{\label{tab:criteria}Existence criteria at the last pair of each system. Square ice: $L=12$ to 384, last pair
$192\to384$; cubic ice: existence runs, $L=12$ to 160, $80\to160$; free box: $L=12$ to 96. E1: smallest fraction over
pairs and thresholds (number of cases). E3: slope, with its jackknife error over replicas and, in brackets, the resolved
sites; margin: slope minus two errors. E4: fraction of sites with $\hat v<10^{-2}$ at the three largest sizes. E5: lower
bound, then width of the bracket at the two last pairs. Upper bound: at the two last pairs.}
\begin{ruledtabular}
\begin{tabular}{llll}
Criterion (threshold) & Square ice & Cubic ice & Free box \\
\colrule
E1 ($\ge0.99$) & 1.000 (14) & 1.000 (11) & empty sets \\
E2 ($\le0.001$) & 0 & 0 & empty sets \\
E3 slope & $0.851(34)$ [399] & $0.856(34)$ [776] & no site \\
E3 margin ($\ge0.5$) & 0.78 & 0.79 & --- \\
E4 (non-decreasing, $>0$) & 0.138, 0.157, 0.170 & 0.0255, 0.0292, 0.0314 & 0 \\
E5 lower bound ($>0$) & 0.1697 & 0.0314 & 0 \\
E5 width (non-increasing) & 0.0314, 0.0202 & 0.0210, 0.0189 & 0 \\
Upper bound & 0.200, 0.194 & 0.111, 0.102 & 0.0017, 0 \\
Verdict & exists & exists & does not exist \\
\end{tabular}
\end{ruledtabular}
\end{table*}

\textit{Why local fields}---In square ice, strings cut the plane, and the cluster of corner-state vertices connected to a
corner converges to the frozen region: the share of its sites that are disordered ($\hat v>0.01$) falls from 46\% at
$L=96$ to 21\% at 384. In 3D, strings do not cut space: the cluster percolates between them into the liquid, and 91\%
of its sites are disordered at $L=96$; it does not converge to the frozen region. We therefore compare local fields site by site between $L$ and $2L$, and
replace the cluster by down-sets.

\textit{Local fields}---Replica $r$ gives the time average $m_r(e)$ of each arrow. With $S_e=\sum_r m_r(e)$ and
$T_e=\sum_r m_r(e)^2$ over $R$ replicas, the disorder $\hat v_e=1-(S_e^2-T_e)/[R(R-1)]$ estimates $1-\langle
s_e\rangle^2$ without bias. The box is split into eight octants, one per corner, and the string density of an edge,
$\rho_e=(1-s^c_e\langle s_e\rangle)/2$, is the probability that its arrow is reversed with respect to $s^c_e$, its
arrow in the polarized state of the corner of its octant; thus $1-\langle s_e\rangle^2=4\rho(1-\rho)$ on each edge,
which $\hat v_e$ estimates without bias: it exceeds the value of $4\rho_e(1-\rho_e)$ computed from the replica mean
$\bar m_e=S_e/R$ by the squared error of that mean, $\sum_r(m_r(e)-\bar m_e)^2/[R(R-1)]$, negligible on frozen sites. Site values average
over the free edges of the site, those not frozen by the boundaries, and over its symmetry images (16 in cubic ice, 4 in
square ice). Face layers are
the sites of the outermost layer of the box.

\textit{The rate function}---A site at $L$ corresponds to a $2\times2\times2$ block at $2L$, at the same reduced
position, over which $\bar\rho_{2L}$ is averaged, and
\[
I_L(\mathbf r)=\frac{\ln\rho_L(\mathbf r)-\ln\bar\rho_{2L}(\mathbf r)}{L}
\]
[Eq.~(2) of the Letter]. In a frozen region, the probability of a string at a site decays exponentially with $L$, and
$I_L$ stays positive, the larger the deeper the site; in the liquid, $\rho$ tends to a finite value and $I_L$ to zero.
Its error $\sigma$ is propagated from the jackknife errors of $\rho_L$ and $\bar\rho_{2L}$. A site where no string is
seen at $L$ or at $2L$ is censored and counted as frozen; in the data, every site with no string at $L$ has none at $2L$
either. In Fig.~2(a) of the Letter, open symbols mark the sites where $\bar\rho_{2L}$ is not resolved (relative jackknife
error above 0.5, the rule of E3, ``Error of E3''): their signal rests on one or two replicas, and the propagated error
understates their uncertainty.

\begin{figure}
\includegraphics[width=\columnwidth]{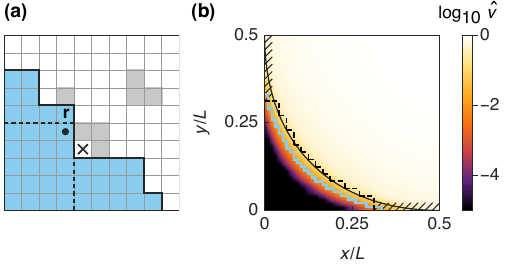}
\caption{\label{fig:downset}Down-sets and the bracket. (a) Schematic, in 2D near a corner (bottom left): a set
$\mathcal A$ of sites (gray and blue) and its down-set (blue, thick outline), the sites whose whole box from the corner
(dashed, for one site $\mathbf r$) lies in $\mathcal A$. A site outside $\mathcal A$ (cross) removes the sites beyond it
along both axes, and a protrusion and a part of $\mathcal A$ detached from the corner are not counted. (b) Square ice,
pair $48\to96$, the quarter of the box at the corner $(0,0)$: disorder $\hat v$ at $2L=96$; lower bound, the down-set
of $\{\hat v<10^{-2}\}$ at $2L$ (light blue); volume upper estimate (dashed); what the upper bound adds (hatched: face
sites counted as frozen); exact arctic curve (thin line). The sets follow the sites of $L$ or $2L$, hence the steps.
Their volume fractions are 0.1385, 0.1858, and 0.2170, for the exact 0.1862. In the data, $\{\hat v<10^{-2}\}$ is already a
down-set (E2), and the construction acts on $\{I_L>2\sigma\}$ only.}
\end{figure}

\textit{Down-sets and the bracket}---For a set $\mathcal A$ of sites and a corner, the down-set holds the sites
$\mathbf r$ of the corner's octant such that the whole box between the corner and $\mathbf r$ lies in $\mathcal A$; we
take the union over the corners [Fig.~\ref{fig:downset}(a)]. A single liquid site thus removes every site of its octant
beyond it along all three axes, and a part of $\mathcal A$ detached from the corner is not counted. Three volume
fractions of down-sets bracket the frozen fraction (Table~\ref{tab:bracketdefs}), drawn for square ice in
Fig.~\ref{fig:downset}(b); Fig.~2(b) of the Letter shows the lower bound and the volume upper estimate against $2L$ for
both systems. The lower bound counts the sites already frozen at $2L$, $\hat v<10^{-2}$; in the data, this set
is already a down-set (E2 below). It is a calibrated bound, not a strict one: with E1 (frozen sites stay frozen when $L$
doubles) and E4 (their set grows), it approaches the frozen fraction from below, as in 2D, where it stays below the exact
value at every size for thresholds up to 0.05. Its value of at least 3\% at $L=160$ is that of the calibrated threshold
$10^{-2}$; the stricter thresholds $10^{-3}$ and $10^{-4}$ give 0.022 and 0.017, still growing. The upper bound counts the
sites where $I_L$ is resolved above zero, $I_L>2\sigma$, censored sites included. The volume upper estimate is the same
with each face site given the value of its inner neighbor, since the rate function is biased in the outermost layer
(``The face layers and the upper side of the bracket''). It is an estimate, not a bound: in 2D, it stays below the exact
value only up to $2L=96$.

\begin{table}
\caption{\label{tab:bracketdefs}The three quantities that bracket the frozen fraction, each the volume fraction of the
down-set of a set of sites (from the corners), and their values at the last pair of cubic ice ($80\to160$) and of square
ice ($192\to384$; exact: 0.1862).}
\begin{ruledtabular}
\begin{tabular}{p{2.2cm}p{2.9cm}ll}
Quantity & Set of sites & 3D & 2D \\
\colrule
Lower bound & $\hat v<10^{-2}$ at $2L$ & 0.031 & 0.170 \\
Upper bound & $I_L>2\sigma$ & 0.102 & 0.194 \\
Volume upper estimate & $I_L>2\sigma$, each face site from its inner neighbor & 0.050 & 0.190 \\
\end{tabular}
\end{ruledtabular}
\end{table}

\begin{figure*}
\includegraphics[width=\textwidth]{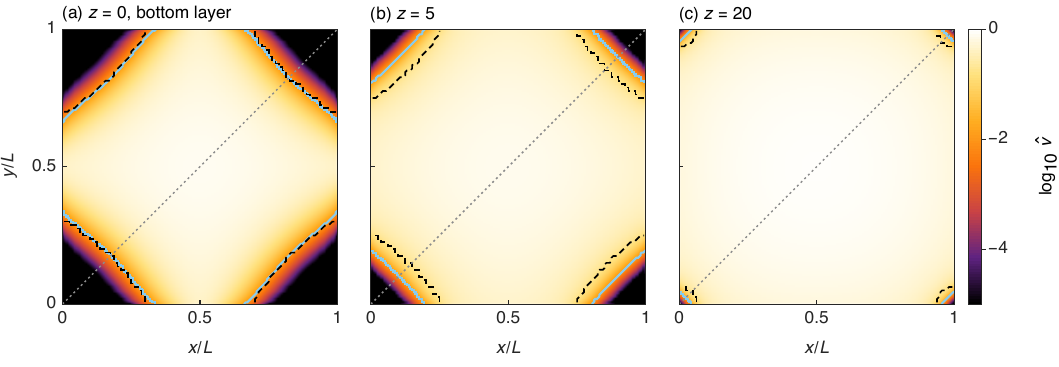}
\caption{\label{fig:slices}Disorder $\hat v$ of cubic ice at $L=160$ in three horizontal planes, on a logarithmic scale
(black: $\hat v\le10^{-5}$): (a) the outermost layer of sites, at the bottom face ($z=0$); (b) $z=5=L/32$; (c) $z=20=L/8$. On the
body diagonal, the region ends at $z\simeq0.07L\simeq11$, between (b) and (c). Contours: region of the lower bound (light
blue) and of the volume upper estimate (dashed, in steps since it follows the sites of $L=80$) at $2L=160$
(Table~\ref{tab:bracketdefs}); dotted: the plane diagonal $x=y$. The outermost layer behaves as a 2D layer, with frozen
corners of its own, where 24\% of its sites have $\hat v<10^{-2}$; inside the box, the frozen region shrinks fast away
from the faces, to 8\% of the sites at $z=5$ and 0.4\% at $z=20$, near the vertical edges of the box.}
\end{figure*}

\textit{Where the frozen region lies}---Figure~\ref{fig:slices} shows the disorder in three horizontal planes at $L=160$.
The outermost layer of sites behaves as a 2D layer, with frozen corners of its own; inside the box, the frozen region
shrinks fast away from the faces. Along the body diagonal $x=y=z=uL$, which starts at a corner and reaches the center
of the opposite face at $u=1/2$, the sites with $\hat v<10^{-2}$ reach 3 sites deep at $L=48$ and 11 at $L=160$, that
is $\lfloor0.07L\rfloor$ at every size from 48 to 160: the tip of the region stays at $u\simeq0.07$, where the local
chemical potential vanishes (S4). Away from the diagonal the region is thinner. At $2L=160$, 30\%, 22\%, and 16\% of the
lower bound lie in the first, second, and third layers from the faces, and 12\% deeper than $1/32$ of the box side from
every face; this deep part grows with $L$ (0.0025, 0.0033, and 0.0037 of the volume at $2L=96$, 128, and 160). Almost none lies
deeper than $1/16$ of the side ($4\times10^{-6}$ of the volume at $2L=160$, about eight sites), since the tip, at
$u\simeq0.07$, lies just beyond $u=1/16$; in square ice, 0.033 and 0.038 of the volume do at $2L=192$ and 384. Why the
3D region is so much smaller than in 2D is not established. A heuristic: a string chooses among three directions at
each step instead of two, so that the threshold $H_K=\frac12\ln3$ exceeds the $\frac12\ln2$ of 2D, and a site freezes
only where the local field is larger, closer to the corner.

\textit{Criteria}---Five criteria test existence. E1 (nesting): at least 99\% of the sites with $\hat v<\varepsilon$
at $L$ ($\varepsilon=10^{-3}$, $10^{-2}$, 0.1) have a matching block at $2L$ with smaller $\rho$ and $\hat v<\varepsilon$.
E2 (anchoring to the corners): at most 0.1\% of these sets lie outside their down-set. E3 (exponential decay): the slope
of $\ln\rho_L-\ln\bar\rho_{2L}$ against $-\ln\rho_L$, over the sites of one corner octant with $10^{-6}\le\rho_L<10^{-2}$
and $\bar\rho_{2L}$ resolved (relative jackknife error below 0.5), minus two errors, is at least 0.5. E4 (growth of the
frozen volume): the fraction of sites with $\hat v<10^{-2}$ does not decrease over the three largest sizes and is
resolved above zero at the largest. E5 (narrowing of the bracket): the lower bound is positive, and the width of the bracket, with
the volume upper estimate, does not increase from the second-largest to the largest pair. The verdict is ``exists'' if E1 to E5 hold, ``does not
exist'' if the fraction of sites with $\hat v<0.1$ is at most $10^{-3}$ at the two largest sizes and the upper bound at
most 0.01, and ``indeterminate'' otherwise.

\textit{What the criteria test}---E1 and E2 are expected for a region anchored at the corners and monotonic away from
them. E1 counts as decreasing the blocks where $\rho$ vanishes at both sizes, 15\% to 68\% of the sites retained in 2D
and 2\% to 26\% in 3D; these blocks must still have $\hat v<\varepsilon$ at $2L$, and since E1 is exactly 1 in every
case, it is 1 on the other sites as well. E3 asks how much $\ln\rho$ drops when $L$ doubles,
$\ln\rho_L-\ln\bar\rho_{2L}$, as a function of how frozen the site already is, $-\ln\rho_L$. If $\rho\propto
e^{-\alpha L^\theta}$ with $\alpha$ varying from site to site, the drop grows with depth at the rate $2^\theta-1$: the
slope is 1 for exponential freezing ($\theta=1$), as in a frozen region, and smaller for slower, stretched freezing. For
a power law $\rho\propto L^{-a}$, slower still, the drop $a\ln2$ does not depend on depth: the slope is 0 if only the
prefactor varies from site to site, and $\ln2/\ln L$ if $a$ does (0.16 at $L=80$). A slope of at least 0.5 after
subtracting two errors thus requires $\theta\ge0.585$: the threshold separates freezing close to exponential from
strongly stretched or algebraic freezing; at finite size, square ice calibrates it (``The verdict''). E3 thus tests how fast $\rho$ falls at a fixed reduced
position, not how the depth of the region grows (Negative controls). E4 is meant to
exclude a frozen layer of fixed or sub-linear depth, whose volume fraction would decrease, provided the decrease shows at
the sizes measured. E5 asks the two sides of the bracket to approach each other, as they must if both converge to the
frozen fraction. At these sizes, the volume upper estimate is flat (0.0502, then 0.0504), so the bracket narrows only
because the lower bound grows, from 0.0292 to 0.0314. That growth is already required by E4: since E2 finds no site with
$\hat v<10^{-2}$ outside its down-set, the fraction of E4 is the lower bound itself. What E5 adds is that the upper side
does not rise faster than the lower one: the width falls from 0.0210 to 0.0189 with all eight replicas, and does not
change in one half of them (``Sensitivity of the bracket'').

\textit{The verdict}---By these criteria, the frozen region exists in cubic ice and in square ice, and not in the free
box (Table~\ref{tab:criteria}). E1 and E2 hold without exception. E3 is $0.856(34)(34)$ at $80\to160$
(statistical and systematic errors, the latter from resolution thresholds 0.3 and 0.7; all-site estimator: 0.946(10),
Table~\ref{tab:e3}), 0.79 after subtracting two statistical errors (0.76 with the systematic error added in quadrature), against a threshold
of 0.5. The lower bound grows over the three largest sizes,
and the bracket narrows. In square ice, where freezing is exponential ($\theta=1$), E3 is 0.85 rather than 1 at these
sizes: the 2D control calibrates the finite-size slope of the same estimator, and the 3D values, 0.80 to 0.88 at every
pair from $48\to96$ on, match it (Table~\ref{tab:e3}). At $48\to96$, the largest pair common to the 2D and 3D series, E3
is 0.80(4) in 3D and 0.82(6) in 2D. Moving the upper bound of the window in $\rho_L$ to $3\times10^{-3}$ or $3\times10^{-2}$ gives 0.78(2)
to 0.90(5) at $80\to160$, and 0.80(3) to 0.90(5) at $192\to384$ in square ice; moving the lower bound to $10^{-5}$ or
$10^{-7}$ changes nothing, since no resolved site has $\rho_L$ between these bounds. Over these windows, the smallest
margin, slope minus two statistical errors, is 0.66, at $48\to96$ in cubic ice with the upper bound $3\times10^{-2}$. The vertex-type runs, with other seeds, give the same verdict (S1, ``Reproducibility'').
Descriptively, the steps of the lower bound shrink as $1/L$ would make them (ratio 0.61, against 0.60); depending on the
form assumed, $1/L$ or $1/\sqrt L$, its three largest values extrapolate to 0.040 or 0.052, below the upper bound
(0.102), the second just above the volume upper estimate (0.050; 0.048 to 0.052 over the variants of
Table~\ref{tab:bracket}). Three
points do not fix the form. On square ice, the same two forms through the three largest values give 0.179 and 0.201, on
either side of the exact 0.186.

\textit{E3 and the face layers}---Since the frozen region lies against the faces near the corners
(Fig.~\ref{fig:slices}), E3 might owe its value to face layers that freeze as 2D layers. It does not
(Table~\ref{tab:e3}). On the sites at least one, two, or three layers from the faces, the slope is 0.90 to 0.94 at
$80\to160$, above the resolved slope over all depths (0.856), and passes the threshold of E3 at all three depths; the outermost layer
alone has a lower slope, 0.80, as the boundary row does in 2D (0.69 to 0.76 at
large sizes, against 0.85 to 0.89 inside). The sites at least four layers from the faces are few and only weakly frozen (small $-\ln\rho_L$), and give lower
slopes (0.75 on 14 sites), as 2D does at $48\to96$ but not at larger sizes; a regression slope depends on the
population of sites, so the 2D control under the same masks is the reference. The lower bound behaves likewise: its share
in the outermost layer is 0.47, 0.36, and 0.30 at $2L=96$, 128, and 160, falling with effective exponents 0.88 and 0.86
(0.97 in square ice from $2L=192$ to 384), much faster than for a layer of fixed or logarithmic depth, and the part at least
three layers from every face grows from 0.0025 to 0.0064 and 0.0100 of the volume; with each face site replaced by its inner neighbor, the lower bound is 0.030 instead of 0.031 at $80\to160$.

\begin{table*}
\caption{\label{tab:e3}Freezing slope E3. All sites: all sites with $\bar\rho_{2L}>0$, error of a regression;
resolved: sites where $\bar\rho_{2L}$ is resolved, paired jackknife error (the estimator of the verdict); margin:
resolved slope minus two errors. Depth: resolved slope on the sites at least $k$ layers from the faces, or on the
outermost layer alone (face layer). In brackets: the number of sites in each regression, those of one corner octant
(quadrant in 2D) with $10^{-6}\le\rho_L<10^{-2}$ where $\bar\rho_{2L}$ is resolved (relative jackknife error below 0.5);
the counts at $k\ge1$ and in the face layer add up to that of the resolved slope. ---: no site. 2D: square ice; 3D:
cubic ice (existence runs).}
\begin{ruledtabular}
\begin{tabular}{lllllllll}
Pair & All sites & Resolved & Margin & $k\ge1$ & $k\ge2$ & $k\ge3$ & $k\ge4$ & Face layer \\
\colrule
2D, $48\to96$ & 0.688(49) & 0.818(62) [40] & 0.69 & 0.74(8) [30] & 0.68(11) [21] & 0.67(10) [13] & 0.68(7) [6] & 0.90(7) [10] \\
2D, $96\to192$ & 0.937(20) & 0.840(48) [136] & 0.74 & 0.86(5) [122] & 0.88(6) [107] & 0.89(7) [95] & 0.88(7) [81] & 0.69(7) [14] \\
2D, $192\to384$ & 0.883(17) & 0.851(34) [399] & 0.78 & 0.85(4) [374] & 0.86(4) [352] & 0.86(5) [330] & 0.86(5) [310] & 0.76(7) [25] \\
3D, $48\to96$ & 0.651(30) & 0.798(44) [149] & 0.71 & 0.71(10) [46] & 0.52(8) [7] & --- & --- & 0.80(6) [103] \\
3D, $64\to128$ & 0.982(17) & 0.880(29) [391] & 0.82 & 0.91(4) [187] & 0.88(7) [65] & 0.75(6) [13] & --- & 0.84(3) [204] \\
3D, $80\to160$ & 0.946(10) & 0.856(34) [776] & 0.79 & 0.90(5) [441] & 0.94(5) [214] & 0.90(5) [78] & 0.75(7) [14] & 0.80(3) [335] \\
\end{tabular}
\end{ruledtabular}
\end{table*}

\textit{The face layers and the upper side of the bracket}---At a face, the pendant edge normal to it is frozen, and the
outermost layer of sites behaves as a 2D layer, with frozen corners of its own: $\{\hat v<10^{-2}\}$ covers 15.4\% of the
top face at $L=80$ and 23.7\% at 160 (square ice: 0.157 at $L=192$, 0.186 in the limit). Comparing such a layer between $L$ and $2L$ says
nothing about the bulk, and the rate function is biased there: the block at $2L$ that matches a face site averages the
face layer of $2L$ with the next one, whose string density is lower. At the center of the top face, $\rho_L=0.400$ and
$\bar\rho_{2L}=0.367$ at $80\to160$, so that $I_L=1.1\times10^{-3}$ is resolved above zero (its error is
$4\times10^{-7}$) although the site is liquid ($\hat v=0.82$), while the next layer matches its block ($\rho=0.344$ at
both sizes). The upper bound thus contains the top and bottom layers whole and part of the side layers: 0.071 of its
0.102 at $80\to160$, 83\% of which is liquid. This share falls nearly as $1/L$, and it alone makes the upper bound
decrease: without it, the bound no longer decreases, in 2D as in 3D, where it stays near 0.050. The volume upper estimate
removes this artifact; it stays below the exact value in 2D only up to $2L=96$ (0.0004 below at $48\to96$). E5 therefore
asks the width of the bracket, with the volume upper estimate as its upper side, not to increase; a decrease of the volume upper estimate
would fail even in 2D, where the region exists.

\textit{Sensitivity of the bracket}---The bracket depends on two choices, the threshold of the lower bound and the
factor in the set $I_L>2\sigma$ that defines the upper bound and the volume upper estimate, and on eight replicas. We varied them at $64\to128$ and $80\to160$ (Table~\ref{tab:bracket}): $\hat v$ shifted by $s\sigma$ at every site, $s=-3$ to 3; the factor $k_\sigma$
of $I_L>k_\sigma\sigma$ set to 1, 2, and 3; and two groups of four replicas. The lower bound of at least 3\% at
$80\to160$ and its growth from $64\to128$ hold in every variant, as does the decrease of the upper bound, by only 0.0004 in one group of replicas.
A plateau of the volume upper estimate, within 2\% over the last step, holds with the choice of the criteria ($+0.4\%$) but
not with $k_\sigma=3$ ($+2.2\%$) nor in one group of four replicas ($+4.8\%$): what holds is 4.8\% to 5.2\% of the volume
from $2L=128$ on, an estimate rather than a bound. At $48\to96$, the upper bound and the volume upper estimate are fragile: with $k_\sigma=3$, an
unresolved site near a corner removes every site beyond it. Censored sites, with no string seen at $L$ or $2L$, counted
as frozen, make up 0.7\% of the volume at $80\to160$.

\begin{table}
\caption{\label{tab:bracket}Sensitivity of the bracket of cubic ice, $64\to128$ then $80\to160$. Lower bound: $\hat v$
shifted by $s\sigma$, range over $s=-3$ to 3; upper bound and volume upper estimate: factor $k_\sigma$ in $I_L>k_\sigma\sigma$ (criteria:
$k_\sigma=2$); groups: two halves of the replicas, $s=0$ or $k_\sigma=2$.}
\begin{ruledtabular}
\begin{tabular}{lll}
 & $64\to128$ & $80\to160$ \\
\colrule
Lower bound & 0.0292 & 0.0314 \\
\quad $s=-3$ to 3 & 0.0287--0.0297 & 0.0311--0.0319 \\
\quad groups & 0.0292, 0.0292 & 0.0314, 0.0315 \\
Upper bound, $k_\sigma=1$ & 0.117 & 0.103 \\
\quad $k_\sigma=2$ & 0.111 & 0.102 \\
\quad $k_\sigma=3$ & 0.105 & 0.094 \\
\quad groups & 0.1110, 0.1022 & 0.0931, 0.1018 \\
Volume upper, $k_\sigma=1$ & 0.0513 & 0.0516 \\
\quad $k_\sigma=2$ & 0.0502 & 0.0504 \\
\quad $k_\sigma=3$ & 0.0488 & 0.0499 \\
\quad groups & 0.0502, 0.0480 & 0.0494, 0.0503 \\
\end{tabular}
\end{ruledtabular}
\end{table}

\textit{Calibration on square ice}---The whole chain runs unchanged on square ice with domain-wall boundaries, $L=12$
to 384, whose frozen fraction is known exactly: in each corner, the arctic curve is an arc of the ellipse
$4x(1-x)+4y(1-y)+4xy=1$, in units of $L$ from that corner~\cite{ColomoPronko2010SIAM,Aggarwal2020}, and the four regions
between the corners and the curve cover $2-\pi/\sqrt3=0.1862$ of the square [Fig.~\ref{fig:downset}(b)]. The bracket
closes on it: 0.1697 and 0.194 at $192\to384$, with a volume upper estimate of 0.190 (Table~\ref{tab:bracketdefs}; the
Letter's Table~I gives 0.170--0.190). The edge
and the fluctuations of square ice calibrate the other tests (S4, S5).

\textit{Negative controls}---Square ice shows that the criteria find a frozen region where one exists; negative
controls ask whether they return ``exists'' where there is none. The free box concludes ``does not exist'' already with
the sizes 12 and 24: no site has $\hat v<10^{-2}$ at any size, and the upper bound vanishes at $48\to96$. The
alternative that matters for cubic ice, whose region is thin, is a region frozen at every size but whose depth grows
more slowly than $L$, so that its share of the volume vanishes. E4 rejects it only if that share already falls at the
sizes measured. E3 is not built to test it: if $-\ln\rho_L=(h_L-uL)/\xi$ near an edge at depth $h_L\propto L^{\omega}$, with
a decay length $\xi$ in sites, the drop $\ln\rho_L-\ln\rho_{2L}$ at the same reduced position equals
$-\ln\rho_L-h_L(2-2^{\omega})/\xi$, of slope 1 for every $\omega$; only the intercept depends on $\omega$. A profile
curved as in square ice, $-\ln\rho_L\propto(h_L-uL)^{3/2}/h_L^{1/2}$, makes the slope depend on $\omega$, but weakly near
$\omega=1$. Banded boxes and
squares (S1, Fig.~\ref{fig:band}) are meant to build it: their pendant edges are frozen only within $W$ of the corners, so that only
corners of size of order $W$ freeze, with $W$ fixed or growing as $\sqrt L$. A control tests E3 only if its corners
freeze at the threshold. Those with $W=4$ and $W\simeq\sqrt{L/3}$ do not, or barely (12 sites at $L=384$ for
$W\simeq\sqrt{L/3}$ in 2D; smallest $\rho$: 0.058 in 2D and 0.068 in 3D for $W=4$): ``does not exist'' for all four
series, but E3 has nothing to judge (Table~\ref{tab:controls}). A banded square with $W=20$, thermalized for 30\,000
sweeps, freezes 60 to 68 sites at the threshold at every size; E3 rejects it, with slopes $-0.19$, $-0.18$, and $-0.17$: for a frozen corner
of fixed size, the same reduced position moves away from it as $L$ doubles, and disorders. A banded square of sub-linear width,
$W\simeq20\sqrt{L/48}$, gave no false positive (slopes $+0.17$, $+0.07$, $-0.08$), but could not be equilibrated: its
corners stay asymmetric and drift, through a slow mode of the free boundary segments between large frozen corners, which
the corner control detects. None of these controls returns ``exists'', but a frozen region whose depth
grows more slowly than $L$ remains without a valid control; cubic ice, all of whose pendant edges are frozen, has no such
slow mode.

\textit{Growth at a fixed position}---Two readings of the existence runs, made after the data, bear on a sub-linear
depth without settling it. The intercept of the regression of E3 and its drop at $-\ln\rho_L=6$ drift as in square ice:
$-2.87(21)$, $-3.42(15)$, and $-3.33(17)$, and 1.92(7), 1.86(3), and 1.81(4), from $48\to96$ to $80\to160$, against
$-2.75(32)$, $-3.07(25)$, and $-3.29(19)$, and 2.15(11), 1.97(6), and 1.82(2), from $48\to96$ to $192\to384$; in the
linear profile above, with $\xi\simeq0.4$ site and a depth of $0.07L$, a depth growing as $L^{0.9}$ would add $-1$ to
$-2$ to the intercept. At a fixed reduced position inside the region ($u\le0.065$ on the body diagonal), $-\ln\rho_L$
grows linearly with $L$, at a rate that does not fall from $L=48$ to 160: at $u=0.06$, 0.020(5), 0.026(4), 0.021(3),
0.024(2), and 0.021(2) per unit of $L$ over the five successive pairs. In the same linear profile, a depth growing as
$L^{0.95}$, with the edge where it is at $L=160$, would lower this rate by a third over this range, and $L^{0.9}$ by
nearly two thirds, whereas the last rate is 1.03(28) times the first; in square ice, where the depth grows as $L$, the
rate at $u=0.12$ nevertheless falls by 28\% from $L=48$ to 384, toward its limit. These readings disfavor a depth
growing as $L^{0.9}$, but not one growing as $L^{0.95}$ or closer to $L$.

\begin{table}
\caption{\label{tab:controls}Negative controls: banded boxes of width $W$, in 2D and 3D for $W=4$ and
$W\simeq\sqrt{L/3}$, in 2D for the others ($L=48$ to 384 in 2D, 48 to 96 in 3D); free box, $L=12$ to 96; square ice as
the reference. E3: slopes at the pairs $48\to96$, $96\to192$, $192\to384$ (---: too few sites in the window). QC: quality
controls of the runs (the free box has no symmetry images and no corner orbit).}
\begin{ruledtabular}
\begin{tabular}{llll}
Series & Verdict & E3 & QC \\
\colrule
Free box & does not exist & --- & pass \\
$W=4$ & does not exist & --- & pass \\
$W\simeq\sqrt{L/3}$ & does not exist & --- & pass \\
$W=20$ & indeterminate & $-0.19$, $-0.18$, $-0.17$ & pass \\
$W\simeq20\sqrt{L/48}$ & indeterminate & $+0.17$, $+0.07$, $-0.08$ & fail \\
Square ice & exists & 0.82, 0.84, 0.85 & pass \\
\end{tabular}
\end{ruledtabular}
\end{table}

\subsection{How the tests were run, and checks}

\textit{How the tests were run}---We call \emph{verdict} the outcome of criteria and analysis code fixed before the data
they judge were read, and \emph{descriptive} a result compared without such criteria, often after the data were known.
For existence, a first rule, based on the corner-state cluster, returned ``indeterminate'' on cubic ice (``Why local
fields''). The present criteria, their thresholds, the quality controls, and the analysis code were then set on square ice
($L=12$ to 384), the free box (12 to 96), and the existence runs of cubic ice at $L=12$, 24, 48, and 96, on which they
find the region in square ice and in cubic ice, and not in the free box; they were fixed before the runs at $L=64$, 80,
128, and 160 were read. These runs form the new pairs $64\to128$ and $80\to160$, on which E3 and E5 are judged; E1, E2, and E4 also use sizes read before (E1 every pair, E2 every size, E4 the three
largest, of which 96 was read before). The new sizes were judged with the amendments listed next, none of which changes the verdict. The
criteria of the bulk equation of state (S3) were set on the tori in a field $(1,1,1)$ at $L=16$ to 128, before the
tori at 32 to 256 were read; along $(1,1,c)$, the same criteria judge tori read afterwards. Those of local equilibrium
(S4), with the even part and the window, were set on runs up to $L=32$, before the comparisons at $L=48$ to 128 were
made; the sizes 16 to 32 and the 2D calibration were then run anew, with other seeds. Those of the edge (S5) were set on
square ice and on the existence runs at $L=64$ to 160, and judge the vertex-type runs at 64, 96, and 128, of which the
vertex-type profile and the frozen depth of the down-sets along the diagonal (mean, width, and skewness) had been looked
at, but not the exponent of the head nor the thickness $D^*_2$. Those of the lines (S5) were
set on square ice and on cubic ice at $L=32$ and 48, and judge cubic ice at $L=64$ to 128. Those of the sawtooth tori
(S7) were set at the same time as those of the bulk, on short runs of sawtooth tori at $L=32$ to 512
(Table~\ref{tab:runs}), before the runs at $L=32$ to 192 (2D: 64 to 1024) were read; the 3D run at $L=256$, made later,
enters only descriptive comparisons (S6, S7). The
comparisons with independent strings (S6), the derivation of the coefficient $AB$ (S3) and the local chemical potential
(S4) are descriptive.

\textit{Amendments made after the data}---Five changes were made to the existence protocol after some or all of the
runs at $L=64$, 80, 128, and 160 had been read; where there was an original choice, it is reported next to the new one, and the
verdict ``exists'' is the same with both, with one reserve for E3. The images control (Quality controls), first applied to all cells, fails at
$L=48$, 80, and 96 (not at 64, 128, and 160) through one or two nearly frozen cells ($\hat v<0.01$), whose jackknife error
rests on rare excitations of one or two replicas, as for E3 (Error of E3); it was restricted to the cells with $\hat
v\ge0.01$ before the run at $L=160$ was read, as in square ice, where such cells see no string and have zero error, so
that they drop out of the control by themselves. These cells lie well below the threshold of the lower bound ($\hat
v\le2\times10^{-3}$, against $10^{-2}$), which they cannot change. Without the sites of the cells whose $\chi^2/\mathrm{ndf}$ exceeds 10 at either
size of a pair, nearly frozen or liquid (one to five resolved sites per pair in the window of E3), E3 is 0.796(44),
0.878(29), and 0.855(35) at $48\to96$, $64\to128$, and $80\to160$, against 0.798(44), 0.880(29), and 0.856(34). E3, first the slope over all sites with $\bar\rho_{2L}>0$ with the error of
the regression (the all-site estimator, 0.946(10) at $80\to160$), is now the slope over the resolved sites with a paired
jackknife error, 0.856(34) (Error of E3, Table~\ref{tab:e3}). The first form passes only through the error of the
regression, which treats the sites as independent: with a jackknife error over replicas, it would fail on square ice,
where the region exists, as on cubic ice (Error of E3). The second form, with its resolution threshold of 0.5, was chosen
after all the runs had been read and its value known; it was applied unchanged to square ice, where it gives 0.851(34),
and it rejects the banded square with $W=20$, run afterwards (slopes $-0.19$ to $-0.17$, Negative controls); thresholds
from 0.3 to 0.7 give the same verdict and its systematic error. E5, first the decrease of the upper bound, which holds
(0.111, then 0.102), but only through the face layers, is now the narrowing of the bracket (``The face layers and the
upper side of the bracket''). The corner control was added (Quality controls). The estimator of the edge from the rate
function, a reported value rather than a criterion, was redefined (S4); its first form, symmetric about the edge, gave
0.095(19)(13) in 3D, the same central value, since all fitted points of the diagonal lie below the fitted edge, where the
two forms coincide, and 0.1387(9)(25) in 2D. The criteria of S3, S4, and S5 were not changed after their test data were
read.
Elsewhere, one rule set after the data would alter a verdict, that of the sawtooth tori (S7); the original verdict is
kept, and the Letter uses neither.

\textit{Quality controls}---Each run must have at least eight replicas, a thermalization of at least ten and a
measurement of at least a hundred autocorrelation times of the cluster volume; all runs pass with wide margins (630 to
1640 and 1300 to 7700 in cubic ice), a weak test, since the autocorrelation time of this volume is set by fast local
fluctuations (S1, Durations). Two controls compare quantities that symmetry
makes equal. For the images, the $\chi^2/\mathrm{ndf}$ of the distinct symmetry images of each cell of the vertical
plane $x=y$ with $\hat v\ge0.01$, divided by its expected value $(R-1)/(R-3)=1.4$, must be at most 1.5; the runs of cubic
ice give 0.91 to 1.41. For the corners, the volume of each corner's down-set $D_c$ (Down-sets and the bracket), or else of its
corner-state cluster, is compared between corners, $\chi^2_S$ over the replica means, and between the two halves of the measurement,
$\chi^2_D$; both must have $p\ge10^{-3}$ under a simulated null law, since errors estimated on seven degrees of freedom
make each term the square of a Student variable (the 99.9\% quantile of $\chi^2_D$ for four corners is about 52, against
18.5 for a $\chi^2$ law). All runs of the verdicts pass; cubic ice at $L=160$ narrowly ($p=0.0015$ and 0.017), a $p$
value that arises about once in ten among some 70 tests. That run does not record the down-set, and the volume compared
is that of each corner's cluster, about 72\,600 sites; three corners deviate by about 40 sites, 0.06\%, the corner
$(+,-,-)$ above, $(+,-,+)$ and $(-,-,+)$ below, while the corner $(+,+,+)$ of the common initial state does not
($+13$ sites, 0.4 error). The corner control fails on the sub-linear banded square (Negative controls); at $L=48$,
the banded square with $W=20$ passes the images control only with a measurement a hundred times longer (Table~\ref{tab:runs}), since with only eight free pendant edges per side, the flux through each side changes by steps of two
every 1600 sweeps or so. A control that fails leaves the verdict in place but calls for new runs.

\textit{Error of E3}---The error of a regression assumes independent sites. A jackknife over replicas on all sites gives
0.95(24) at $80\to160$ and 0.88(26) on square ice at $192\to384$, where E3 would fail although the 2D
frozen region exists: at a few very frozen sites, $\bar\rho_{2L}$ rests on the rare excitations of one or two replicas.
E3 therefore uses the sites where $\bar\rho_{2L}$ is resolved, with a paired jackknife error: the slope is recomputed
without replica $r$ at both sizes at once, on the same sites, for $r=1$ to 8, which keeps the correlations between sites
that the error of a regression ignores. At $80\to160$, of the 959
sites in the range of $\rho_L$, 99 are censored ($\bar\rho_{2L}=0$) and 84 unresolved, leaving 776 (in 2D at
$192\to384$: 638, 114, 125, and 399). The censored sites, the most frozen, are left out rather than treated by a censored
regression. Table~\ref{tab:e3} gives both estimators.

\section{Bulk equation of state}
\label{sec:eos}

Does the string density of the bulk follow Eq.~(3)? It does. On tori in a uniform field, Eq.~(3) describes $\rho$ to
$7\times10^{-4}$ on the dilute window W1 with two parameters, and to $10^{-3}$ over a decade and a half of $\mu$ with a
correction factor $1+\eta\mu$, where a power law and the mean-field form, with as many parameters, fail. Its coefficient
$AB$ lies about 2\% below the value, without free parameters, of a dilute Bose gas in $2+1$ dimensions, within the
corrections to that value at these $\mu$. Square ice follows the square root of free fermions instead.

\textit{Tori and observables}---The equation of state is measured where the field is known: on tori in a uniform
field, sampled with the two-defect worm of S1, which lets the number of strings winding around the torus change. Along
$(1,1,1)$, the tori are cubes, $L=32$ to 256. The variable is the log-weight per string step, $\mu=\ln\sum_ae^{-2H_a}$ (all $H_a>0$ here, the field favoring
$(+,+,+)$; in general $|H_a|$),
the sum running over the axes: a step along axis $a$ reverses an arrow of weight $e^{-2H_a}$, and an isolated string
steps along $a$ with probability $p_a=e^{-2H_a-\mu}$. Along $(1,1,1)$, $\mu=\ln3-2H$, which vanishes at the Kasteleyn
threshold $H_K=\frac12\ln3$; in the 2D control, on $L\times L$ tori, $\mu=\ln2-2H$. This $\mu$ is the chemical potential of the strings, the variable in which Eq.~(3) is
derived below; another variable, such as $t=H_K/H-1$, with $\mu=\ln3\,t/(1+t)$, brings a factor $1-t+\dots$ that the
leading form lacks (Criteria and checks). The string density $\rho$ is the fraction of the free edges, all $3L^2L_z$ edges of a 3D
torus and the $2L^2$ horizontal ones in 2D, whose arrow is reversed with respect to the polarized state $(+,+,+)$ that
the field favors: $\rho=(1-\langle\mathcal M\rangle/E_f)/2$, with $\mathcal M=\sum_es_e$ over the $E_f$ free edges. The
temperature is contained in the reduced field (S1). Along $(1,1,1)$, the weight is $e^{H\mathcal M}$, so that
$d\langle\mathcal M\rangle/dH={\rm Var}(\mathcal M)$ exactly, at every size, and the response $\chi={\rm Var}(\mathcal
M)/E_f$ obeys $\chi=4\,d\rho/d\mu$, one factor 2 coming from $\rho$ and the other from $\mu$; dividing by $E_f$ makes it
intensive, like $\rho$. Relative to $(+,+,+)$, a reversed arrow points along $-x$, $-y$, or $-z$, so a string decreases
one coordinate at each step and increases none: a closed string must wind around the torus. Every reversed arrow thus
belongs to a winding string, $\mathcal M$ depends only on the windings, and $\chi$ measures the fluctuation of the number
of strings. Eleven values of $\mu$, from 0.4 down to 0.0125 by factors of
$\sqrt2$, are each run at one size, large enough for finite-size effects to be negligible (Criteria and checks). The
window W1, written $\mu\le0.1$ in the criteria, holds the six values $0.0125\le\mu\le0.0707$ in 3D, where strings are
dilute: the run at $\mu=0.1$ was computed at $\mu=0.1000003$ and falls just outside it (seven values in 2D). With this seventh point, criterion B1 gives the same result in both directions; along
$(1,1,1)$, the largest residual of Eq.~(3) becomes 0.17\% and $AB=0.1351(4)$. The window W2 holds all eleven.

\begin{figure*}
\includegraphics[width=\textwidth]{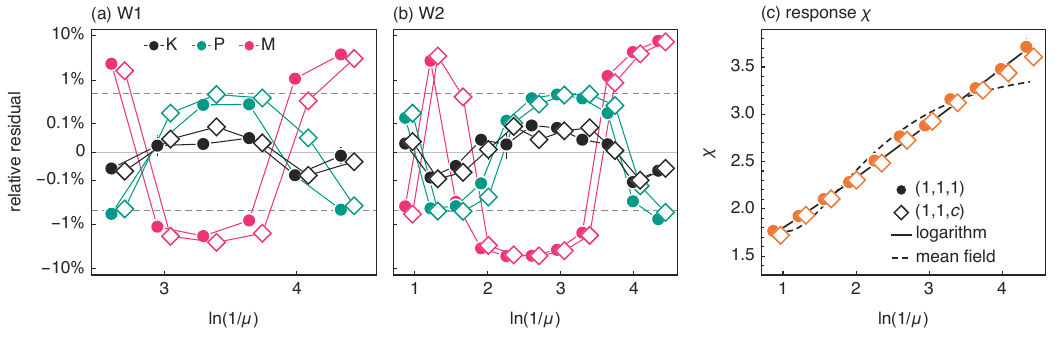}
\caption{\label{fig:eos}Bulk equation of state on tori, along $(1,1,1)$ (disks) and $(1,1,c)$, $c=0.843$ (diamonds),
shifted by $-0.05$ and $+0.05$ in $\ln(1/\mu)$ for legibility. (a), (b) Relative residuals of the fits of criterion B1,
with two parameters on W1 ($0.0125\le\mu\le0.0707$) and three on W2 ($0.0125\le\mu\le0.4$): Eq.~(3) (K), a power law
(P), and the mean-field form (M); the scale is linear within $\pm0.05\%$ and logarithmic beyond; dashed: the bound of
0.5\% on the largest residual of K. (c) Response $\chi=4\,d\rho/d\mu$: along $(1,1,1)$, $4/E_f$ times the variance of the number of reversed arrows; along
$(1,1,c)$, a covariance (S3, ``A second direction''); with the fits
of criterion B2 on W2 along $(1,1,1)$: $c_0+c_1\ln(1/\mu)+c_2\mu$ (solid) and the mean-field form $c_0+c_2\mu+c_3\mu^2$
(dashed).}
\end{figure*}

\textit{The logarithm}---Criterion B1 fits three forms with two parameters each to $\rho(\mu)$ on W1: Eq.~(3),
$A\mu[1+B\ln(1/\mu)]$, written K (for Kasteleyn); a power law $A\mu^\beta$, written P; and the mean-field form
$A\mu(1+\eta\mu)$, written M. These forms describe the dilute limit. W2 extends to $\mu=0.4$, where 23\% of the arrows
are reversed; there, each form gets one more parameter, the same for all, a factor $1+\eta\mu$ (for the mean-field form,
which already has a term in $\mu$, a term $\eta'\mu^2$), so that the comparison stays at equal numbers of parameters. This
factor is empirical: the corrections that Eq.~(3) leaves out are logarithmic (``Where the logarithm comes from''), and
$\eta$ absorbs them over the window, together with corrections of the lattice; along $(1,1,1)$, it is $-0.048(6)$, 2\% at
$\mu=0.4$. Equation~(3) describes $\rho$ to $7.4\times10^{-4}$ on W1 and, with this factor, $1.1\times10^{-3}$ on W2, as
the largest relative residual, while the power law misses by up to 0.8\% and the mean-field form by up to 7.5\% (Table~\ref{tab:b1},
Fig.~\ref{fig:eos}). The coefficient of the logarithm, $B=0.293(2)$ on W1, lies more than 130 errors above zero. By the
criteria (Criteria and checks), this logarithmic form wins over the other two along $(1,1,1)$ and along a second
direction, $(1,1,c)$, introduced below. Equation~(3) is only the leading form: its $\chi^2$, several times the number of
degrees of freedom, and the pattern of its residuals show corrections of a few $10^{-4}$, which is why B1 compares forms
rather than testing one alone. On W1 alone, a third parameter would not separate the logarithm from a power law: with
the factor $1+\eta\mu$, the power law fits about as well along $(1,1,1)$ ($\chi^2=17.8$ against 13.3) and better along
$(1,1,c)$ (7.6 against 15.5), for 3 degrees of freedom; the decade and a half of W2 separates them ($\chi^2=5430$ and 5140 against 129 and
122, for 8). A diagnostic free of any form says the same. With $\beta_{\rm eff}=d\ln\rho/d\ln\mu$,
the quantity $Q=1/(1-\beta_{\rm eff})$ is constant for a power law and equals $1/B+\ln(1/\mu)$ for Eq.~(3); it grows
linearly with $\ln(1/\mu)$, with a slope of $1.091(8)$ in both directions over the eleven values of W2. The correction
factor of the W2 fit accounts for the departure from 1: along $(1,1,1)$, the fitted form gives 1.10 at the same points
and with the same weights, and 1.000 without the factor, since a derivative turns $\eta\mu$ into a shift of $\beta_{\rm
eff}$ of up to 8\% of $1-\beta_{\rm eff}$, at $\mu=0.4$.

\textit{The response}---The identity $\chi=4\,d\rho/d\mu$, applied to Eq.~(3), gives
$\chi=4A(1-B)+4AB\ln(1/\mu)$, and applied to the mean-field form, $\chi=4A+8A\eta\mu$. Criterion B2 fits these two forms,
$c_0+c_1\ln(1/\mu)$ and $c_0+c_2\mu$, to $\chi$, an observable distinct from $\rho$, the fluctuation of $\mathcal M$
rather than its mean, though measured on the same runs. Its coefficient $c_1$ matches $4AB$ from the fit of $\rho$
within $1.2\sigma$ on W1 along both directions and on W2 along $(1,1,c)$, and lies $2.3\sigma$ above it on W2 along
$(1,1,1)$; the mean-field form fits $\chi$ worse, by $\Delta\chi^2=97$ on W2 (Table~\ref{tab:b1}). The response does not, however,
separate the logarithm from a power law in $\mu$, with $\chi^2$ of 3.7 for the logarithm against 3.4 for the power law on
W1, and 7.5 against 12.9 on W2: it is the equation of state that excludes the power law. The identity itself holds at the 11 points where neighboring
values of $\mu$ were run at the same size: the squared normalized deviations sum to 9.88 and 9.16 along the two
directions ($p=0.54$ and 0.61).

\textit{A second direction}---Nothing forces the local field of cubic ice to lie along $(1,1,1)$: no symmetry of the box
exchanges $z$ with $x$ or $y$ (S1). For dilute strings, each step along $a$ reverses one arrow along $a$, so the ratio
$r$ of the string densities along $z$ and $x$ is $p_z/p_x=e^{-2(H_z-H_x)}$. Reading the box near its edge as a bulk in a
field, as local equilibrium does (S4), $r$ gives the direction of the local field: on the body diagonal, in the dilute
cells next to the edge, $r=1.18$ to 1.26 from $L=64$ to 128, so strings step slightly more often along $z$. A field
$(H,H,cH)$ at the threshold, $2e^{-2H}+e^{-2cH}=1$, gives the ratio $r$ for $c=1-\ln r/\ln(2+r)$, that is, 0.80 to 0.86.
We take $r=6/5$, near the measured values: $c=1-\ln(6/5)/\ln(16/5)=0.84325$, and at the threshold a string steps along
$x$, $y$, and $z$ with probabilities $5/16$, $5/16$, and $3/8$. The tori are then $L\times L\times1.2L$, $L=30$ to 240,
with integer sides, so that a string winding once around them takes exactly these shares of steps; on a cube, it would
be forced to $1/3$ per axis, at a cost of 0.004 in $\mu$ per step, 30\% of the smallest $\mu$. The field couples to
$\mathcal M_x+\mathcal M_y+c\mathcal M_z$, and the identity becomes $\chi={\rm Cov}(\mathcal M,\mathcal M_x+\mathcal
M_y+c\mathcal M_z)/(\bar cE_f)=4\,d\rho/d\mu$, with $\bar c=\sum_ac_ae^{-2H_a}/\sum_ae^{-2H_a}$ and $c_a=(1,1,c)$.
Along $(1,1,c)$, the verdict is the same (Table~\ref{tab:b1}), and at the same $\mu$ the string densities along the two
directions differ by 0.27\% at most: near the threshold, the equation of state hardly depends on the direction of the
field, which lets S4 read the local field of cubic ice along $(1,1,1)$. Since $c$ is close to 1, this is a check in the
direction that matters for cubic ice, not a test far from $(1,1,1)$; the small change of $AB$ that the derivation
predicts is compared below.

\textit{The 2D control}---On 2D tori, the same procedure does not confirm the logarithmic class of Eq.~(3), that of the
3D Kasteleyn transition: the
power-law exponent is 0.540(2), near the 1/2 of Pokrovsky and Talapov~\cite{PokrovskyTalapov1979}, and $Q$ decreases
toward 2, its value for their square root, $\rho\propto\mu^{1/2}$ of free fermions, instead of growing as $\ln(1/\mu)$. This negative control is part of the verdict (Criteria and checks). Square ice in a
field is solvable by the Bethe ansatz~\cite{SutherlandYangYang1967,LiebWu1972}; outside the verdict, we compare the tori
only with its leading term, set by the mass of one string, and fit the next one, $b\mu$ (Against the fits,
Table~\ref{tab:ab}).

\begin{table*}
\caption{\label{tab:b1}Criteria B1 and B2 along $(1,1,1)$ and $(1,1,c)$, on the windows W1 and W2: largest relative
residual and $\chi^2$ of the three forms of B1 (degrees of freedom: 4 on W1, 8 on W2); $B$ of Eq.~(3); coefficient $c_1$
of the response, to compare with $4AB$ from the fit of $\rho$ (0.539 and 0.543); excess $\chi^2$ of the mean-field form
in B2.}
\begin{ruledtabular}
\begin{tabular}{lllll}
 & \multicolumn{2}{l}{$(1,1,1)$} & \multicolumn{2}{l}{$(1,1,c)$} \\
 & W1 & W2 & W1 & W2 \\
\colrule
K, residual & $7.4\times10^{-4}$ & $1.1\times10^{-3}$ & $8.4\times10^{-4}$ & $1.0\times10^{-3}$ \\
K, $\chi^2$ & 24.8 & 129 & 50.8 & 122 \\
P, residual ($\chi^2$) & 0.6\% (2050) & 0.8\% (5430) & 0.48\% (2520) & 0.55\% (5140) \\
M, residual ($\chi^2$) & 3.7\% ($7.0\times10^4$) & 7.5\% ($3.5\times10^5$) & 3.0\% ($7.6\times10^4$) & 7.1\% ($4.3\times10^5$) \\
$B$ & 0.2934(15) & 0.2912(22) & 0.2968(18) & 0.2924(22) \\
$c_1$ & 0.535(33) & 0.594(24) & 0.507(30) & 0.540(25) \\
B2, $\Delta\chi^2$ & 16.5 & 97 & 8.4 & 95.5 \\
\end{tabular}
\end{ruledtabular}
\end{table*}

\textit{Pyrochlore spin ice}---The same transition occurs in pyrochlore spin ice in a field along
$[100]$~\cite{Jaubert2008,Jaubert2009}. A field beyond the Kasteleyn threshold selects a single ice state, a vertex of the polytope of the mean
polarizations that the ice rule allows. In cubic ice, the moment of a vertex, half the sum of its six arrows, is one of
the eight $(\pm1,\pm1,\pm1)$ of the polarized types or one of the six $\pm\hat{\mathbf e}_a$ of the others: the polytope
is the cube $[-1,1]^3$, with vertices along $\langle111\rangle$. In pyrochlore, the six two-in, two-out states of a
tetrahedron have moments along $\pm\hat{\mathbf e}_a$: the polytope is the octahedron, with vertices along
$\langle100\rangle$. Hence a field along $(1,1,1)$ here and along $[100]$ there; in both cases, it makes the same angle
with every local spin axis. A string chooses among three directions at each step here, two there. For a field of norm
$h$ (the Zeeman energy of a unit moment, with $k_B=1$), every arrow has the projection $1/\sqrt3$ on the field in both
lattices, so that it carries the reduced field $h_e=h/(\sqrt3T)$ of S1, and a step costs $e^{-2h/(\sqrt3T)}$. The
Kasteleyn temperatures, below which the field holds the polarized state without strings, thus have the same form,
$T_K=2h/(\sqrt3\ln3)$ here, which is the threshold $H_K=\frac12\ln3$, and $2h/(\sqrt3\ln2)$ there; in pyrochlore, the Kasteleyn temperature has been computed for any field direction~\cite{PottsBenton2022}. In
both, strings are
directed lines with two transverse dimensions, in the same class [(iii) below]. Jaubert et al.\ obtained for spin
ice, by carrying over results for dilute bosons~\cite{FisherHohenberg1988}, the form of Eq.~(3) in the reduced
temperature $t=T/T_K-1$, which is the $H_K/H-1$ of the tori, with free coefficients, and found their simulations
compatible with it~\cite{Jaubert2008,Jaubert2009}. Powell and Chalker derived its critical theory, the dilute Bose gas in
$2+1$ dimensions, and carried over its thermodynamics~\cite{PowellChalker2008}; Powell later built a scaling theory of
the transition~\cite{Powell2013}. The coefficient $m/4\pi$ of the dilute Bose gas is universal; its evaluation for
strings, through their mass and transverse density, which gives $AB$ without free parameters below, is not in these
works, to our knowledge. On a 2D lattice, kagome ice, the plateau of pyrochlore spin ice in a $[111]$ field, has a Kasteleyn
transition when the field is tilted, the analog of that of square ice~\cite{MoessnerSondhi2003,Fennell2007}.

\textit{Where the logarithm comes from}---The following derivation was made after the fits; it enters no criterion,
and its agreement with them is descriptive. We write $\nu$ for the dimension of the lattice, and keep ``two dimensions''
for the transverse plane of the Bose gas. Near the threshold, strings are dilute, and each one is a directed walk: a step
reverses an arrow along axis $a$ with probability $p_a$ (Tori and observables) and changes the sum $x+y+z$ of the
coordinates by one, always in the same direction, whatever its axis. This sum, $\sqrt3$ times the position along
$(1,1,1)$, thus counts the steps and plays the role of time: seen along $(1,1,1)$, a string is the world line of a
particle moving in the transverse plane [Fig.~1(b)]. Four steps lead from this picture to Eq.~(3) and to its
coefficient.

(i) \emph{The mass.} Take first a single string. Each arrow it reverses along axis $a$ has the weight $e^{-2H_a}$, so a
string of $\tau$ steps along the axes $a_1,\dots,a_\tau$ has the weight $\prod_ie^{-2H_{a_i}}=e^{\mu\tau}\prod_ip_{a_i}$.
The factor $e^{\mu\tau}$ is the same for all strings of $\tau$ steps, and the product is the probability of the path for a
random walk whose steps $\hat{\mathbf e}_a$ are drawn independently with the probabilities $\mathbf p=(p_1,\dots,p_\nu)$.
A step thus has the mean $\mathbf p$ and the covariance $\Sigma={\rm diag}(\mathbf p)-\mathbf p\mathbf p^{\sf T}$, which
vanishes along $(1,\dots,1)$ for every $\mathbf p$, since every step changes the sum of the coordinates by exactly one:
all the randomness of the walk lies in the $\nu-1$ transverse directions. Along $(1,\dots,1)$, all $p_a$ equal $1/\nu$,
the mean step has no transverse part, and the transverse part of a step, $\hat{\mathbf e}_a-(1,\dots,1)/\nu$, has the
squared length $1-1/\nu$, shared equally among the transverse directions: each gets the variance $1/\nu$ per step. After
$\tau$ steps, the transverse displacement thus spreads as that of a free particle of mass $m$ in imaginary time $\tau$
($\hbar=1$), whose variance is $\tau/m$ per direction: $m=\nu$, that is, $m=3$ along $(1,1,1)$ and $m=2$ in 2D. Along
$(1,1,c)$, at the threshold, $\mathbf p=(5,5,6)/16$ (A second direction), and two things change. The mean step has the
transverse part $(-1,-1,2)/48$: strings drift toward $z$, which a shift of each slice removes without changing the density
in a slice or the distances between strings, hence without changing the equation of state. And the transverse diffusion
becomes anisotropic, with the variances $5/16$ along $(1,-1,0)$ and $45/128$ along $(1,1,-2)$; a linear map of the
transverse plane with unit determinant makes it isotropic and keeps areas, hence densities, and the mass is the inverse
of the geometric mean of the two variances, $m=32\sqrt2/15=3.017$. In general, an anisotropic diffusion counts through
its determinant, $m=(\nu\prod_ap_a)^{-1/(\nu-1)}$: the product of the nonzero eigenvalues of $\Sigma$ is the sum of its
$\nu$ principal minors of order $\nu-1$, each equal to $\prod_bp_b$ by the matrix determinant lemma.

(ii) \emph{The density.} Between neighboring slices of constant $x+y+z$, each site of a slice sends $\nu$ edges to the
next slice, and each reversed one carries a string: $\nu\rho$ strings cross per site of a slice. Since slices are
$1/\sqrt\nu$ apart and each site fills a unit volume, a site of a slice covers the transverse area $\sqrt\nu$, and the
strings have the transverse density $n_\perp=\sqrt\nu\,\rho$.

(iii) \emph{The statistics.} In 2D, strings cannot pass each other: where dilute, they are effectively free fermions in
one dimension (square ice is the six-vertex model at $\Delta=1/2$, not at its free-fermion point $\Delta=0$), with
$n_\perp=\sqrt{2m\mu}/\pi$, hence $\rho=(\sqrt2/\pi)\sqrt\mu=0.45016\sqrt\mu$. In 3D, they pass around each other: they
are bosons in two dimensions with contact repulsions, of two kinds. An edge carries a single arrow, so two strings cannot
share it, as free bosons could. And a vertex crossed by $k$ strings counts once, since a configuration does not say which
incoming string continues as which outgoing one (S1), whereas free bosons would give it the weight $k!$, one term for
each way of continuing them: for two strings entering by given edges and leaving by distinct ones, 3 configurations
against a weight of 6. Both are short-range repulsions; they change only the scattering length $r_0$ introduced below,
not $AB$. For bosons in two
dimensions, in $2+1$ dimensions with time, such a repulsion is marginal~\cite{PowellChalker2008}: at large scales, it
neither grows until it dominates, as in $1+1$ dimensions, where it makes the strings fermions, nor fades as a power of
the scale, as it would with more transverse dimensions; its effective strength decreases only logarithmically. Counting
encounters shows why. The transverse separation of two strings is a random walk in $\nu-1$ dimensions, at zero at time
$\tau'$ with a probability of order $\tau'^{-(\nu-1)/2}$, so that the mean number of their encounters up to time $\tau$
grows as $\sum_{\tau'\le\tau}\tau'^{-(\nu-1)/2}$. In 2D, where the separation walks on a line, it grows as $\sqrt\tau$.
In 3D, the separation walks in a plane and returns to zero infinitely often, but so rarely that the number of returns
grows only as $\ln\tau$. With three or more transverse dimensions, a case without counterpart here, a random walk
returns to its origin only a finite number of times. In the language of the Bose gas, imaginary time scales as the
square of a length, as for the walk of (i) (dynamic exponent 2), so that time counts as two dimensions, and the contact
repulsion is marginal when $(\nu-1)+2=4$: the gas in $2+1$ dimensions is at its upper critical
dimension~\cite{PowellChalker2008}. There, the chemical potential is $\mu\simeq(4\pi n_\perp/m)/\ln[1/(n_\perp r_0^2)]$
at leading order, where $r_0$ is the two-dimensional scattering length of the repulsion, of the order of its
range for a hard core and smaller for a softer one~\cite{Schick1971,FisherHohenberg1988,LiebYngvason2001}: the effective coupling, the factor of $n_\perp$, decays as $1/\ln(1/n_\perp)$.

(iv) \emph{The inversion.} Inverting this leading order alone would leave a term in $\ln\ln(1/\mu)$. The energy density
is known to the next order~\cite{Fournais2022}. With $Y=1/\ln[1/(n_\perp r_0^2)]$,
\begin{align*}
e&=\frac{2\pi}m\,n_\perp^2\,Y\bigl[1-Y|\ln Y|+c_FY+o(Y)\bigr],\\
\mu&=\frac{de}{dn_\perp}=\frac{4\pi n_\perp}m\,Y\bigl[1-Y|\ln Y|+(c_F+\tfrac12)Y+o(Y)\bigr],
\end{align*}
with $c_F=2\gamma_E+\frac12+\ln\pi$, where $\gamma_E$ is Euler's constant (the second line differentiates the expansion term by term, a standard step
assumed here). Hence
\begin{align*}
n_\perp&=\frac{m\mu}{4\pi}\Bigl[\frac1Y+\ln\frac1Y-c_F-\frac12+o(1)\Bigr]\\
&=\frac{m\mu}{4\pi}\Bigl[\ln\frac1\mu+C+o(1)\Bigr],
\end{align*}
since $1/Y=\mathcal L-\ln\mathcal L+o(1)$ with $\mathcal L=\ln[4\pi/(m\mu r_0^2)]$: the two terms in $\ln\mathcal L$ cancel, through
the sign of $-Y|\ln Y|$, and $C=\ln[4\pi/(mr_0^2)]-c_F-\frac12$. With $n_\perp=\sqrt3\,\rho$, this is Eq.~(3), with
\[
AB=\frac{m}{4\pi\sqrt3},
\]
independent of the strength of the repulsion, while $A$ depends on it through $C$. Along $(1,1,1)$, $AB=\sqrt3/(4\pi)=0.13783$;
along $(1,1,c)$ at the threshold, $p=(5,5,6)/16$ gives $m=32\sqrt2/15=3.017$ and $AB=0.13861$, 0.57\% higher. The terms
left out, $o(1)$ inside the bracket, vanish only slowly as $\mathcal L$ grows, and are not small at the $\mu$ of the tori
(Against the fits).

\textit{Against the fits}---Table~\ref{tab:ab} compares these values with the fits. On W1, $AB=0.1347(3)$ along
$(1,1,1)$ and 0.1358(4) along $(1,1,c)$, 2.3\% and 2.1\% below the predictions. These deviations exceed the statistical
errors, but not the terms that the derivation leaves out, which are not small at these $\mu$: read as its asymptotic value, the measured $C=1/B$ gives $r_0\simeq0.07$, well below
the lattice spacing [(iii)], and $\mathcal L=\ln(1/\mu)+6.7$, 9.4 to 11.1 on W1, so that the corrections kept in (iv),
$Y|\ln Y|$ and $(c_F+\frac12)Y$, are still about 0.25 and 0.4. As an illustration, keeping in the energy density of Fournais et al.~\cite{Fournais2022} only the terms
written in (iv), with $r_0$ set so as to give the measured $C=1/B$, and fitting Eq.~(3) on the six values of $\mu$ of W1, gives $AB$ 7\% below $m/(4\pi\sqrt3)$
with their constant $c_F$, and 1.5\% to 4\% above it with $c_F$ set to 1 or 0. Every fit of $\rho$ stays below the
prediction: adding $\epsilon/\ln(1/\mu)$ inside the bracket of Eq.~(3), or a factor $1+\eta\mu$, gives 0.131 to 0.133.
The response, less precise, gives 0.134(8) and 0.149(6) on W1 and W2 along $(1,1,1)$, on either side of the prediction.
$AB$ thus agrees with the value without free parameters within these corrections, without being a test at the 2\%
level. The two directions differ by $+0.78(40)\%$ on W1 and $+0.44(51)\%$ on W2, for the $+0.57\%$ of the anisotropy
alone. In 2D, $\rho=a\sqrt\mu+b\mu$ gives $a=0.45006(23)$ on W1, 0.02\% below $\sqrt2/\pi$. This derivation was made with
Claude Opus 5.5 (S1, ``Code and AI tools''); three checks bear on it: its 2D counterpart, with the same mass and density
steps, gives this free-fermion amplitude; the cancellation of the $\ln\ln$ terms in (iv) was checked numerically; and
the two directions differ as the anisotropy alone predicts.

\begin{table}
\caption{\label{tab:ab}Coefficient $AB$ of Eq.~(3) and 2D amplitude $a$ of $\rho=a\sqrt\mu+b\mu$, fitted on the values of
$\mu$ used by B1 (errors scaled as by the Particle Data Group, S1), against their values without free
parameters. W1: 6 points in 3D, $0.0125\le\mu\le0.0707$, and 7 in 2D, $\mu\le0.1$; W2: 11 points, $\mu\le0.4$.
Response: $AB=c_1/4$ (in 2D, $\chi=2a/\sqrt\mu+4b$).}
\begin{ruledtabular}
\begin{tabular}{lll}
3D & $(1,1,1)$ & $(1,1,c)$ \\
\colrule
Prediction, $m/(4\pi\sqrt3)$ & 0.13783 & 0.13861 \\
W1, Eq.~(3) & 0.1347(3) & 0.1358(4) \\
W2, Eq.~(3)$\times(1+\eta\mu)$ & 0.1344(5) & 0.1350(5) \\
W1, $+\epsilon/\ln(1/\mu)$ & 0.1320(17) & 0.1311(18) \\
W1, $\times(1+\eta\mu)$ & 0.1328(12) & 0.1324(13) \\
W1 without the smallest $\mu$ & 0.1348(5) & 0.1360(7) \\
W1 without the largest $\mu$ & 0.1343(4) & 0.1350(4) \\
Response, W1 & 0.134(8) & 0.127(8) \\
Response, W2 & 0.149(6) & 0.135(6) \\
\colrule
2D & $a$ & \\
\colrule
Prediction, $\sqrt2/\pi$ & 0.45016 & \\
W1 & 0.45006(23) & \\
W1, $+\eta\mu^{3/2}$ & 0.4490(9) & \\
W2, $+\eta\mu^{3/2}$ & 0.45012(37) & \\
$\mu\le0.05$ & 0.44954(47) & \\
Response, W1 & 0.463(11) & \\
\end{tabular}
\end{ruledtabular}
\end{table}

\subsection{Criteria and checks}

\textit{Sizes and durations}---Finite-size effects depend mainly on the length of a winding string in units of
$1/\mu$, $3L\mu$ on a cube: on the tori at $L=16$ to 128 run for this purpose (Table~\ref{tab:runs}), $\rho$ lies 0.9\% below its limit at $3L\mu=2.4$ at $L=32$ and 64, and within
0.1\% of it from $3L\mu=4.8$ on. The sizes used give at least 6.8 in both directions ($L=32$ or 30 for $\mu\ge0.1$, up to
256 or 240 at $\mu=0.0125$); the next smaller size, where this length is still at least 4.8, is a required variant. Each
run measures 4000 sweeps per replica (2000 at the largest size), after 500 to 1000 sweeps of thermalization; the number
of strings decorrelates in 1 to 4 sweeps, and thermalization from the polarized state takes 30 to 40 sweeps (S1). Errors
come from the integrated autocorrelation time of each series~\cite{Sokal1997}; the jackknife over replicas, with seven
degrees of freedom, is a variant.

\textit{Criteria}---B1 fits the three forms on W1 with two parameters each; on W2, each gets a factor $1+\eta\mu$ (the
mean-field form, a term $\eta'\mu^2$). The verdict ``Kasteleyn class confirmed'', meaning that the logarithmic form
of Eq.~(3), that of the 3D Kasteleyn transition and of the dilute Bose gas in $2+1$ dimensions, wins over the other two, requires that, in both windows, K
describe $\rho$ to $5\times10^{-3}$ (largest relative residual), beat P and M by $\Delta\chi^2\ge100$, and give $B>0$ at
$5\sigma$; it is refuted if P or M wins in the same way. B2 fits $\chi=c_0+c_1\ln(1/\mu)$ against $c_0+c_2\mu$ (on W2, one
more term each), and requires $c_1>0$ at $5\sigma$ in both windows and the mean-field form worse by $\Delta\chi^2\ge25$ on
W2. The verdict also requires the quality controls, the required variants (next smaller size, jackknife errors) and the
negative control: applied to the 2D tori, the procedure must not confirm the logarithmic class. The quality controls check
that the runs are complete, that the two halves of the measurement agree, that each run lasts at least 100
autocorrelation times, and that the identity $\chi=4\,d\rho/d\mu$ holds between neighboring values of $\mu$; a run whose
ratio of the two error estimates leaves $[0.38,1.70]$, the 99\% band for seven degrees of freedom and the sign of a hidden
slow mode, is flagged and takes the larger error.

\textit{Three choices}---Three of the criteria answer what the tori at $L=16$ to 128 showed. First, B2 asks for its
$\Delta\chi^2$ on W2 only: on those tori, the curvature that separates $\ln(1/\mu)$ from a straight line hardly showed
over W1 ($\Delta\chi^2=7.8$, and 12 to 88 expected at the precision of the tori at $L=32$ to 256, against a threshold of
25); on the latter, the mean-field form of $\chi$ is indeed worse on W1 by only $\Delta\chi^2=16.5$ (Table~\ref{tab:b1}),
and requiring W1 as well would have left the verdict ``indeterminate''. Second, the variable $t=H_K/H-1$ is reported
outside the verdict, since $\mu=\ln3\,t/(1+t)$ brings a factor $1/(1+t)$ that the forms lack: on those tori, Eq.~(3) in
$t$ missed by 1.2\% on W1. On the tori of Table~\ref{tab:b1}, it still fits best in $t$, by $\Delta\chi^2>8000$, but
less well than in $\mu$ ($\chi^2=1250$ and 1520 for 4 degrees of freedom on W1, against 25 and 51); on W2, where $t$
reaches 0.57 and the factor $1+\eta t$ no longer follows $1/(1+t)$, its largest residual, 1.0\% and 0.8\%, exceeds the
bound of 0.5\%, and B1 is indeed indeterminate in $t$. Third, the next smaller size counts only where $3L\mu\ge4.8$. Along $(1,1,c)$, the same criteria and
thresholds apply; only the reading changes ($\mu$, $\chi$ through the covariance of ``A second direction'', the variable
$t$, the sizes of the tuned tori, and the length $3.2L\mu$ of a winding string).

\section{Local equilibrium and the local chemical potential}
\label{sec:le}

Does each region of the box behave like the bulk at the same flux? Along the body diagonal, where strings are dense,
$u\ge0.15$ or $\mu_{\rm loc}\gtrsim0.37$, it does for the frequencies of the vertex types: beyond the first-order effect
of the flux gradient, their deviation from the bulk falls below the resolution as $L$ grows, consistently with $L^{-4}$,
while that of the Pauling approximation stays at a finite distance. Read through the bulk equation of state, the string
density then gives a local chemical potential that falls linearly toward the edge, at a slope independent of $L$, and
vanishes at $u_0=0.069$ to 0.070; on square ice, the same zero lands within 1\% of the exact edge at the sizes reached.
Below $u=0.15$, local equilibrium is assumed, not tested (Table~\ref{tab:diagonal}).

\begin{table*}
\caption{\label{tab:diagonal}Along the body diagonal of cubic ice, from the corner: what is measured in each range of
$u$, and its status ($L=64$ to 160). The local chemical potential is read along $(1,1,1)$; the local field points near
$(1,1,c)$ at the edge, where the two directions give the same $\rho$ within 0.3\%, and the ratio of the string densities
along $z$ and $x$ grows to 1.5--1.7 at $u=0.15$ to 0.2, where $\mu_{\rm loc}$ is only indicative.}
\begin{ruledtabular}
\begin{tabular}{lp{2.6cm}p{6.8cm}p{4.6cm}}
Range of $u$ & $\mu_{\rm loc}$ & What is measured & Status \\
\colrule
$u<u_0\simeq0.07$ & $<0$ (extrapolated; the reading gives 0) & frozen region: $\hat v<10^{-2}$, $\lfloor0.07L\rfloor$ sites deep; $I_L>0$ & verdict (S2) \\
$u\simeq u_0$ & 0 & edge, about one site thick, with a tail on the frozen side & verdict F2 (S5); tail descriptive, measured on tori (S7) \\
0.076 to 0.084 & 0.03 to 0.07 & dilute regime of W1 (S3), where the logarithm dominates; start of the linear fit &
local equilibrium assumed, not tested \\
0.076 to 0.14 & 0.03 to 0.33 & linear fit of $\mu_{\rm loc}$, slope independent of $L$; zero $u_0$ by construction &
descriptive; zero calibrated in 2D (1\%) \\
0.10 to 0.15 & 0.15 to 0.37 & even part of the vertex frequencies converges & reported without criterion (S4) \\
0.15 to 0.45 & $\gtrsim0.37$ & window of the test: $D^2$ resolved up to $L=48$, then bounded & verdict (S4), one-vertex statistics \\
\end{tabular}
\end{ruledtabular}
\end{table*}

\textit{The idea}---Cubic ice has no field: its boundaries impose the flux. They fix the polarization of every plane
of edges (S1; along $z$, it grows by $4/L$ per plane), so that the liquid is a Coulomb phase~\cite{Henley2010} whose
flux $\mathbf b(\mathbf r)$ varies on the scale of the box. The frozen regions are where
$\mathbf b$ saturates at a vertex $(\pm1,\pm1,\pm1)$ of the cube of fluxes, where the entropy per site has a finite
slope, the Kasteleyn field, as for dimers~\cite{KenyonWolfram2025}. In the multinomial limit of dimers, where this slope
diverges, no frozen region forms~\cite[Theorem~1.3]{KenyonWolfram2025}. On a face of the cube, where a single component
of $\mathbf b$ saturates, the entropy stays positive, each layer normal to that axis being then an independent square
ice, as on the faces of the octahedron of fluxes of 3D dimers~\cite[Sec.~4.2]{ChandgotiaSheffieldWolfram2023}. We expect,
without proof for ice, that a face is reached only in an infinite field: there, the strings of each layer have a finite
density, and a small fraction of steps along the third axis, placed among their steps, brings an entropy of mixing that
outweighs, for a small enough fraction, a cost proportional to it, as at the boundary of the polytope in the multinomial
limit of dimers. If so, no region of the box is polarized along a single axis in local equilibrium; such a region, which
the site values, averaged over the three axes, would count as unfrozen, would in any case not enter the lower bound of
the frozen fraction. If each region behaves like the bulk at the same flux
$\mathbf b(\mathbf r)$, it sees the field that gives the bulk this flux, and the frozen region ends where this local field
crosses the Kasteleyn threshold. Such local equilibrium is standard in physics but not proven for ice. We test it on the
frequencies of the vertex types along the body diagonal, against tori in a uniform field.

\textit{Vertex types and the bulk reference}---The twenty vertex types are those of S1 [Fig.~\ref{fig:strings}(a)]:
eight polarized, which pass along all three axes, and twelve with a source, a sink, and a passage. The flux $\mathbf b$ of
a type is the mean of its arrows along each axis: $\pm1$ for a passage, 0 for a source or a sink. We count the types in
cells of $2\times2\times2$ vertices along the diagonal, with the eight corners brought to the frame of the corner
$(0,0,0)$ and the counts symmetrized under $x\leftrightarrow y$. The bulk reference is a map of tori in fields
$(H,H,H_z)$, which have the flux $(b_{xy},b_{xy},b_z)$ of the diagonal: 173 tori, 144 at $L=32$ and 29 at $L=64$, placed
along the path of the measured flux and on one row on either side of it, wide enough to cover the profiles of all sizes;
in 2D, 48 tori with $L=128$ to 512. The bulk frequencies at the flux of each cell come from a local regression in flux
space (tricube weights; second order over 12 neighbors in 3D, third order over 6 in 2D; as a variant, third order over 24
and second order over 6).

\textit{The symmetry $\Theta$}---Inversion combined with arrow reversal, $\Theta$, keeps the arrow vectors and moves them
to the opposite point: it keeps the flux and reverses its gradient. It is a symmetry of the uniform measure on the ice states of a torus and of the tori in a uniform field, since
inversion and reversal each change the sign of every $s_e$, so that their product leaves the field term
$\sum_aH_a\sum_{e\parallel a}s_e$ unchanged. At a vertex, it fixes the passages
and exchanges sources and sinks: it fixes the polarized types and exchanges (source along $a$, sink along $b$) with
(source along $b$, sink along $a$), the passage being kept. In the bulk, $P_v=P_{\Theta v}$. In a gradient, the
deviation from the bulk splits into an odd part, $(P_v-P_{\Theta v})/2$, and an even part. Since $\Theta$ reverses
$\nabla\mathbf b$ and keeps $\mathbf b$ and its second derivatives, a correction of first order in the gradient is odd,
and the even part deviates from the bulk only at second order, $O(L^{-2})$ [Eq.~(B1)]. Its squared distance to the bulk,
$D^2$, should thus decay as $L^{-4}$.

\begin{figure}
\includegraphics[width=\columnwidth]{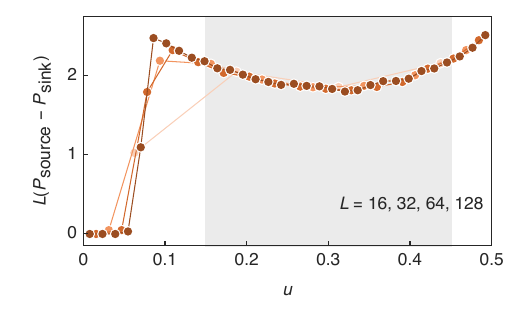}
\caption{\label{fig:le}Odd part of the vertex frequencies under $\Theta$ along the body diagonal $x=y=z=uL$ of cubic ice,
$L[P({\rm source}_z)-P({\rm sink}_z)]=\frac L2\langle s_{+z}-s_{-z}\rangle$, per cell of $2\times2\times2$ vertices, for
$L=16$, 32, 64, and 128 (light to dark; 24, 48, and 96 collapse as well); gray: window of the test of local equilibrium,
$0.15\le u\le0.45$. The curves collapse, as the derivative of a flux profile set on the scale of the box does, near 2,
the exact mean of this quantity over any horizontal plane (S1). The even part is in Fig.~3(b), the local chemical
potential in Fig.~3(c).}
\end{figure}

\textit{The odd part, exactly}---Let $s_{+a}$ and $s_{-a}$ be the arrows of the two edges of a vertex along axis $a$,
on the side of increasing and of decreasing coordinate. A source along $a$ has $(s_{+a},s_{-a})=(+1,-1)$, a sink
$(-1,+1)$, a passage equal arrows. At every vertex, the indicator of a source minus that of a sink is thus
$(s_{+a}-s_{-a})/2$, and
\[
P({\rm source}_a)-P({\rm sink}_a)=\tfrac12\langle s_{+a}-s_{-a}\rangle,
\]
the discrete derivative of the flux $b_a$ along $a$, of order $1/L$ (Appendix~B). At every ice vertex there are as many
sources as sinks, so the sum over the three axes vanishes, and this combination is odd under $\Theta$. On the diagonal,
after the symmetry $x\leftrightarrow y$, the odd part has two components, $O_+$ and $O_-$ (Table~\ref{tab:theta}): the
identity along $z$ fixes their sum, and those along $x$ and $y$ add nothing, each being then $-1/2$ of that along $z$.
Their difference is fixed by no identity and is not tested here. Figure~\ref{fig:le} shows $L[P({\rm source}_z)-P({\rm
sink}_z)]$. Since the boundaries fix the polarization of every plane (S1), this quantity averages exactly $(L/2)(4/L)=2$
over every horizontal plane. Along the diagonal, outside the frozen region, it collapses from $L=16$ to 128, with a mean
of 1.92 to 2.02 over the window of the test, as the derivative of a flux profile $\mathbf b(\mathbf r/L)$ does; the
component along $x$ is $-1/2$ of it in every cell, to rounding, as the identities require. In 2D, the symmetry
$x\leftrightarrow y$ coincides with $\Theta$ on the two unpolarized types, and the odd part vanishes on the diagonal.

\begin{table}
\caption{\label{tab:theta}The twelve unpolarized vertex types on the body diagonal, written (source, sink; passage), and
their partners under $\Theta$, which exchanges source and sink and keeps the passage. The symmetry $x\leftrightarrow y$
of the diagonal joins the types of each row; the eight polarized types are fixed by $\Theta$. $O_\pm$ is the frequency of
the types with a source along $z$ and a passage in the $\pm$ direction, minus that of their partners. The identity along $z$
fixes $O_++O_-=P({\rm source}_z)-P({\rm sink}_z)=\frac12\langle s_{+z}-s_{-z}\rangle$; $O_+-O_-$ is not fixed.}
\begin{ruledtabular}
\begin{tabular}{lll}
Types & Partners under $\Theta$ & Odd part \\
\colrule
$(x,y;\pm z)$, $(y,x;\pm z)$ & each other & vanishes \\
$(z,x;+y)$, $(z,y;+x)$ & $(x,z;+y)$, $(y,z;+x)$ & $O_+$ \\
$(z,x;-y)$, $(z,y;-x)$ & $(x,z;-y)$, $(y,z;-x)$ & $O_-$ \\
\end{tabular}
\end{ruledtabular}
\end{table}

\textit{Results}---All criteria hold (Table~\ref{tab:le}, Fig.~3(b); Criteria and checks). $D^2$ is
resolved up to $L=48$ only; from $L=64$ on, $D^2+2\sigma$ stays below 1\% of the Pauling distance $D^2_{\rm VP}$ and bounds
it, down to $2.4\times10^{-8}$ at $L=128$, against a threshold of $5.7\times10^{-7}$. A fit from $L=16$ to 128, by weighted least squares on $D^2$ itself, so that values compatible with zero enter, gives
$D^2\propto L^{-q}$ with $q=4.00(31)$, consistent with the expected 4; the errors are scaled as by the Particle Data
Group (S1): the excess $\chi^2$ of the fit, 22.6 for 5 degrees of freedom, comes from $L=24$ and 32, with 3 and 5 cells
in the window, and the factor, 2.1, doubles the error of $q$ (1.1 in 2D). The criteria
require only $q>0$ at $5\sigma$; over windows of sizes, $q$ ranges from 3.3(4) to 4.7(1.3) (Criteria and checks). The Pauling
approximation~\cite{Pauling1935}, extended here to a nonzero flux, gives each vertex type a frequency
$\propto e^{\boldsymbol\lambda\cdot\mathbf b_v}$, with $\boldsymbol\lambda$ set by the flux, as if the arrows were
independent apart from the ice rule: it has the right flux, not the right correlations, and the measured frequencies
stay at a finite distance from it, near $6\times10^{-5}$ from $L=24$ on ($q=0.19(12)$), close to its own distance to
the bulk, $D^2_{\rm VP}$. Square ice gives $q=4.97(41)$, which
rests on $L=24$, three cells (without it, $q=3.9(7)$), and $3.5\times10^{-8}$ at $L=192$ for a threshold of
$1.2\times10^{-5}$. For the twenty frequencies without symmetrization, odd part included, $D^2$ decays as
$L^{-2.03(2)}$, the square of a first-order term. The layer $0.10\le u<0.15$, between the window and the edge,
reported without criterion, converges as well.

\begin{table}
\caption{\label{tab:le}Local equilibrium: distance $D^2$ of the even part of the vertex frequencies to the bulk at the
same flux, summed over the types and averaged over the cells of the window ($0.15\le u\le0.45$ in 3D, $0.2\le u\le0.45$
in 2D), with its jackknife error; edge layer: $0.10\le u<0.15$, without criterion (---: no cell); Pauling: the same
distance $D^2$, with the Pauling approximation at the measured flux of each cell as the reference in place of the bulk;
$D^2_{\rm VP}$: distance between the Pauling approximation and the bulk at the same cells, whose 1\% is the threshold of
D2. 3D: the vertex-type runs (Table~\ref{tab:runs}).}
\begin{ruledtabular}
\begin{tabular}{rrllll}
 & & \multicolumn{2}{l}{$D^2$ ($10^{-8}$)} & \multicolumn{2}{l}{($10^{-5}$)} \\
$L$ & Cells & Window & Edge layer & Pauling & $D^2_{\rm VP}$ \\
\colrule
16 & 3 & 2160(42) & --- & 13.8 & 7.79 \\
24 & 3 & 287(40) & 3890(170) & 6.24 & 4.52 \\
32 & 5 & 192(24) & --- & 6.05 & 4.84 \\
48 & 7 & 45(11) & 347(61) & 6.38 & 5.92 \\
64 & 9 & 8.7(6.3) & 84.9(9.9) & 5.27 & 5.05 \\
96 & 15 & $-1.1(2.5)$ & 5.5(7.7) & 6.05 & 5.93 \\
128 & 19 & 1.14(64) & 2.4(1.8) & 5.75 & 5.71 \\
\colrule
\multicolumn{6}{l}{Square ice} \\
24 & 3 & 943(55) & & 111 & 104 \\
32 & 4 & 193(30) & & 119 & 114 \\
48 & 6 & 40(10) & & 128 & 124 \\
96 & 12 & 11(28) & & 123 & 124 \\
192 & 24 & $-5.1(4.3)$ & & 119 & 119 \\
\end{tabular}
\end{ruledtabular}
\end{table}

\textit{What is established}---Local equilibrium holds for one-vertex statistics along the body diagonal, for
$0.15\le u\le0.45$, the layer $0.10\le u<0.15$ converging as well. The test does not reach other lines: in the plane $x=y$, symmetry keeps $b_x=b_y$, but the map covers only a band around
the flux of the diagonal; elsewhere it would need three field parameters. Nor does it reach two-vertex statistics, the
layer $u<0.10$ next to the edge, or whether the local field derives from a potential, as a variational principle would
require.

\textit{The local chemical potential}---Reading the measured string density through the bulk equation of state gives
the local chemical potential, $\mu_{\rm loc}(u)=X_B^{-1}(\rho(u))$ (Appendix~B). $X_B(\mu)$ interpolates the string density
of the tori along $(1,1,1)$ (S3), including tori deeper in the liquid: it interpolates $\rho/\mu$ in $\ln(1/\mu)$ to
third order, uses Eq.~(3), fitted on W1, below the smallest $\mu$, and ends at $\rho=1/2$ for $\mu=\ln3$. In 2D, it does
the same with the 2D tori, with $\rho=a\sqrt\mu+b\mu$ below their smallest $\mu$ and $\rho=1/2$ at $\mu=\ln2$. Near the
edge, where $0.03\le\rho\le0.2$, $\mu_{\rm loc}$ is nearly linear in $u$ [Fig.~3(c)]: $\mu_{\rm loc}\simeq g(u-u_0)$. In cubic ice, the
eight runs from $L=64$ to 160 (existence runs at 64, 80, 96, 128, and 160; vertex-type runs at 64, 96, and 128) give
$g=4.83$ to 4.95 and $u_0=0.0692$ to 0.0702, without trend in $L$,
though beyond their fit errors (at most 0.0003). On square ice, the same reading gives $u_0=0.1349(2)$, 0.1337(1), and
0.1328(1) at $L=96$, 192, and 384, for the exact edge $1-\sqrt3/2=0.1340$, with $g=3.086(21)$, 3.036(7), and 2.951(4). In
2D, both drift beyond their errors: $u_0$ crosses the exact edge, from $+0.7\%$ to $-0.9\%$, and a power law through the
three points extrapolates to 0.130, 3\% below it. The 1\% thus holds at these sizes, and a bias of a few percent of the
linear extrapolation, from its curvature or from finite size, cannot be excluded, in 3D as in 2D. At the edge itself, the 2D slope is known. Matching the Airy density
$\sqrt{|s|}/\pi$ of the strings that cross the diagonal, in the variable $s$ of S6, with $n_k=2\rho$, to the bulk law
$\rho\simeq(\sqrt2/\pi)\sqrt\mu$ gives $\rho\simeq0.838\sqrt{u-y_c}$ and $d\mu_{\rm loc}/du=1/(8\lambda^3)=2\sqrt3\simeq3.46$
at the exact edge $u=y_c=1-\sqrt3/2$, where $\lambda=2^{-4/3}3^{-1/6}$ is the Tracy-Widom scale of S5. The fitted slopes
lie 11\% to 15\% below it, and further as $L$ grows: over $0.03\le\rho\le0.2$, the fit measures a chord of a concave
$\mu_{\rm loc}$, whose slope lies below that at the edge and whose zero lies on the frozen side of the edge; at the
smaller sizes, the rounding of the profile near the edge moves the zero the other way, so that $u_0$ crosses the exact
edge as $L$ grows. The zero also depends on the window: with the four windows of S5, $0.08\le\rho\le0.25$,
$0.05\le\rho\le0.25$, $0.03\le\rho\le0.2$, and $0.02\le\rho\le0.15$, it ranges from 0.129 to 0.134 in 2D at $L=384$,
the last, nearest the edge, landing within 0.2\% of the exact edge, and from 0.063 to 0.071 in 3D at $L=160$, again
largest for the window nearest the edge.

\textit{What the reading shows}---Since $\mu_{\rm loc}$ vanishes where $\rho$ does, its zero locates the edge by
construction. What the reading shows is that the local field reaches the threshold linearly, with a slope in $u$ that
does not depend on $L$, as a field set on the scale of the box does: across one site, $\mu_{\rm loc}$ changes only by
$g/L$, a gradient that vanishes as $L$ grows, as local equilibrium requires. On the frozen side, $\rho$ vanishes for any
field beyond the threshold: there, the field is known only through the extrapolation of this line. With Eq.~(3), it then gives the string density at a distance $d=u-u_0$ from the
edge, $\rho\simeq X_B(g\,d)\propto d\ln(1/d)$, the profile used in S5 and S6. On square ice, where the edge is known, the
zero lands within 1\% of it. What 2D calibrates is the position of this zero, not the shape of the profile that the linear fit
gives next to the edge, which misses the 2D exponents of the head (S5).

\textit{Two reserves}---The first concerns the direction of the local field. Near the edge, it points near $(1,1,c)$,
where $\rho$ differs from its value along $(1,1,1)$ by less than 0.3\% at the same $\mu$ (S3); farther in, the ratio of
the string densities along $z$ and $x$ grows to 1.5--1.7 at $u=0.15$ to 0.2, and $\mu_{\rm loc}$ read along $(1,1,1)$ is
only indicative. Restricting the fit to $\rho\le0.1$, where the field points near $(1,1,c)$, moves $u_0$ by $+2.9\%$ at
$L=160$, a third of a site (a fit over four sites), as the same restriction does in 2D ($+2.4\%$ at $L=384$, 1.2
sites), where the direction is fixed by
symmetry: the change comes from the curvature of $\mu_{\rm loc}$, not from the direction. The second concerns the dilute layer. The window of the test starts at $\mu_{\rm loc}\simeq0.37$
($u=0.15$), and the layer $0.10\le u<0.15$ covers $\mu_{\rm loc}\simeq0.15$ to 0.37. The linear fit, where
$0.03\le\rho\le0.2$, covers $\mu_{\rm loc}\simeq0.03$ to 0.33, or $u\simeq0.076$ to 0.14: it lies entirely below the
window of the test, partly in the layer that converges without criterion, partly below $u=0.10$, where local
equilibrium is not tested, down into the dilute regime $\mu\lesssim0.07$ of the window W1 of S3, where the logarithm
dominates. What is assumed below $u=0.15$ is local equilibrium itself, that is, the reading of $\rho$ through $X_B$; the
line and its zero depend on it. On sawtooth tori, where the local field is exact, the same reading finds the
threshold $\tilde u=L/4$ within 0.15 to 0.26 site, on the liquid side, from $L=64$ to 256 (0.9\% to 0.4\% of $L/4$), and
the slope $4\ln3$ within 2.1\%; on 2D tori, the zero lies 0.09 to 0.65 site on the liquid side from $L=64$ to 1024 (0.2
site at 1024), and the slope comes within 3\% of $4\ln2$ from $L=384$ on (0.5\% at 1024). On the same tori, the
profile that local equilibrium and Eq.~(3) predict reproduces how the measured one changes with $L$, not its level next
to the edge (S7). In the box at $L=160$, the dilute regime $\mu\le0.07$ spans about two sites of the diagonal; the box
has the field gradient of a sawtooth torus of side 350, 2.2 times its own, on which this regime lies within the layer
where the density departs from local equilibrium by more than 1\% (S7), so that the logarithm of Eq.~(3) cannot be
tested there at these sizes. Table~\ref{tab:diagonal} sums up this chain along the diagonal.

\textit{Three estimates of the edge}---The edge on the diagonal is estimated in three ways (Table~\ref{tab:edgepos}):
by the rate function of S2, the point where $I_L$ vanishes, fitted jointly over the pairs from $L=48$ on with the
one-sided form $I_L=c_1\max(u_c-u,0)^{3/2}+c_2\ln2/L$, flat on the liquid side (exponents 1 and 2 among the variants of the
systematic error); by the zero $u_0$ of $\mu_{\rm loc}$; and by the head of the
profile, fitted with the form $d\ln(1/d)$ (S5). They agree within their spread, near the exact edge in 2D and at about
0.07 in 3D, where the rate function gives a higher value within its error: as $L$ grows, the point where $I_L$ vanishes
approaches $u_0$ from above [Fig.~2(a)].

\begin{table}
\caption{\label{tab:edgepos}Three estimates of the position $u_c$ of the edge on the diagonal: the rate function, fitted
jointly over pairs (S2), with statistical and systematic errors; the zero $u_0$ of the linear $\mu_{\rm loc}$; and the
head of the profile, fitted with the form $d\ln(1/d)$ (S5). 3D: cubic ice, $L=64$ to 160 (rate function: pairs from
$L=48$); 2D: square ice, $L=96$ to 384, exact edge $1-\sqrt3/2=0.1340$.}
\begin{ruledtabular}
\begin{tabular}{lll}
Estimate & 3D & 2D \\
\colrule
Rate function & $0.095(23)(18)$ & $0.1371(17)(79)$ \\
Zero of $\mu_{\rm loc}$ & 0.069--0.070 & 0.133--0.135 \\
Head, $d\ln(1/d)$ & 0.072--0.074 & 0.134--0.136 \\
\end{tabular}
\end{ruledtabular}
\end{table}

\subsection{Criteria and checks}

\textit{Why the even part, this window and these sizes}---Without the symmetrization, the exact odd term of first
order in the gradient dominates: the squared distance $D^2$ of the twenty raw frequencies decays only as $L^{-2.03(2)}$
(Results), the square of a first-order term, whereas the even part should deviate from the bulk only at second order,
$D^2\propto L^{-4}$. The window
starts about 0.07 in $u$ beyond the edge, at $u=0.15$ in 3D, where the edge lies at 0.07, and at 0.2 in 2D, where it lies
at 0.134: next to the edge, the deviation is about ten times larger where it is resolved ($L=24$ to 64; edge layer,
Table~\ref{tab:le}, reported without criterion), so that a few cells there would dominate $D^2$. In 2D, the window also stays where the map of the bulk has converged. If
local equilibrium holds, the even deviation expected at $L=48$, 64, 96, and 128, estimated before the comparison from the
trend at $L\le32$, stands at 6, 3.5, 1.7, and 0.9 standard errors: the largest sizes bound the deviation rather than
measure its exponent.

\textit{Criteria}---$D^2=\sum_v(\Delta_v^2-\sigma_v^2)$, summed over the types and averaged over the cells of the window,
where $\Delta_v$ is the even part of $P_v-P_v^{\rm bulk}$ and $\sigma_v$ combines its jackknife error, with the flux
recomputed in each jackknife subsample, and the error of the map, has no noise bias (Appendix~B). D1 requires the fit $D^2\propto L^{-q}$
over all sizes, 16 to 128 in 3D and 24 to 192 in 2D, with errors scaled as by the Particle Data Group (S1), to
give $q>0$ at $5\sigma$. D2 requires $D^2+2\sigma$ at the largest size to be at most 1\% of $D^2_{\rm VP}$, the distance
between the Pauling approximation and the bulk at the same cells. D3 requires the Pauling approximation to fail D2 at
$5\sigma$, or to fail D1: otherwise the test would not distinguish the bulk from any model with the right flux. D0
requires square ice to satisfy D1 and D2.

\textit{Descriptive checks}---Not part of the criteria: D1 holds at $5\sigma$ without $L=16$ ($q=3.3(4)$) and without 16 and 24 ($4.1(4)$); on 48 to
128 alone, $q=4.7(1.3)$, as the power estimate implies. At $L=16$, the cell nearest the edge carries two thirds of
$D^2$, and the last cell, which reaches the face layer at this size only, about a fifth. The fluctuation of the map, absent from the jackknife error of
$D^2$, makes up at most 2.2\% of the variance of the $\Delta_v$, up to a quarter of that of $D^2$ at $L=128$, and leaves
$q=4.01(32)$; the variant regression changes $D^2$ by less than a tenth of its error. The local chemical potential is
descriptive, judged by no criterion; $X_B$ has no parameter fitted on the box, being built from the bulk tori alone (S3),
and is continuous at its junction with Eq.~(3) to $10^{-4}$ in 3D ($1.3\times10^{-3}$ in 2D).

\section{The edge and its fluctuations along lines}
\label{sec:edge}

What is the edge like, and how does it fluctuate? Along the body diagonal, it is sharp, about one site thick, with a head
that rises faster than the square root of 2D. Along lines from the corners, its fluctuations are not Tracy-Widom: the
frozen length has a positive skewness, and its width grows faster than $L^{1/3}$. On square ice, the same chain recovers
the exact Tracy-Widom law.

\textit{The profile across the edge}---Along the body diagonal, the string density $\rho(u)$, averaged over the 16
images of each site with jackknife errors, describes the edge by three quantities. The first is its shape on the
liquid side next to the edge, which we call the head of the profile (sites where $0.03\le\rho\le0.2$): at a distance
$d=u-u_c$ from the edge, local equilibrium with Eq.~(3) gives $\rho\propto d\ln(1/d)$ in 3D
(S4), against $\sqrt d$ in 2D. The second is its amplitude at the edge. The third is its thickness on the frozen side,
which says whether the edge is sharp but not which class it belongs to. Both are set by one string. Near the edge, the
local chemical potential is linear, $\mu_{\rm loc}=g(u-u_0)$, with $g\simeq4.9$ in cubic ice (S4, ``The local chemical
potential''); since one site of the diagonal is a step $1/L$ in $u$, it changes by $g/L$ from one site to the next, and a
string near the edge wanders over the Airy length, of order $(L/g)^{1/3}$ up to a factor of order one, in any dimension
and for any statistics (in 2D, where a string moves by $\pm\frac12$ diagonal site per step, it is $(L/8g)^{1/3}$, the
Tracy-Widom scale $\lambda L^{1/3}$ for the exact slope $g=2\sqrt3$ at the edge, S4); this rounds the mean profile over that length and sets its amplitude,
$\rho^*=X_B(\mu^*)$ with $\mu^*\propto L^{-2/3}$, that is $\rho^*\propto L^{-2/3}$ up to logarithms in 3D and
$L^{-1/3}$ in 2D. The thickness is expected to grow as $L^{1/3}$ in 2D; in 3D, between $L^{1/3}$ and the scale of the
fluctuations along lines, which is larger (S6, S7). The thickness says how fast the mean density falls on the frozen
side; the first string along the diagonal, which sets $\ell$ there, lies a few sites beyond $u_0$ on the liquid side
($\langle\ell\rangle=12.3$ sites at $L=128$, against $u_0L=8.9$) and fluctuates over about two sites (``Cubic ice''). The thickness $D^*_1$ ($D^*_2$) is the distance, in sites of the
diagonal, from $u_c$ to the point of the frozen side where $\rho$ has fallen to $1/e$ ($1/e^2$) of its value at $u_c$,
with $e\simeq2.718$. The position $u_c$ is not known on cubic ice: it is
fitted with the head of the profile, on the sites where $0.03\le\rho\le0.2$, in four forms with free $u_c$: a power law
$\rho\propto d^\gamma$, $\sqrt d$ plus $d^{3/2}$, $d$ plus $d^2$, and $d$ plus $d\ln(1/d)$.

\textit{Calibration of the profile}---On square ice, with the exact edge, $D^*_1$ and $D^*_2$ grow as $L^{0.335(15)}$ and
$L^{0.336(11)}$, and the amplitude as $L^{-0.368(19)}$, for $1/3$ and $-1/3$. With $u_c$ fitted by the power-law or
$\sqrt d$ form, biased by 0.4 to 0.7 site, the exponents remain correct; with $d\ln(1/d)$, that of $D^*_2$ is
0.29(1), still far below 1. On the existence runs of cubic ice, $L=64$ to 160, three things appear.
Comparing the forms by their residuals is not reliable at these sizes: the best form changes with the window, in 2D as
in 3D, and in 2D $d\ln(1/d)$ even beats $\sqrt d$ in some windows. The amplitude cannot be measured on cubic ice: the edge
layer is about one site thick, while the fitted $u_c$ moves by 0.15 to 0.3 site from one size to the next. But the
effective exponent $\gamma$ of the head separates 3D from 2D.

\begin{figure}
\includegraphics[width=\columnwidth]{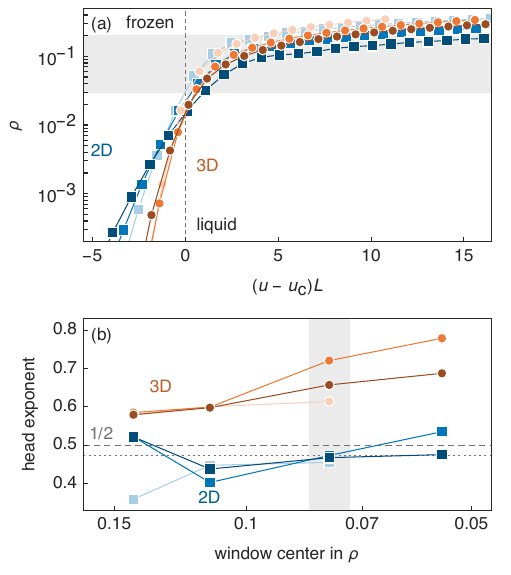}
\caption{\label{fig:edge}The edge of cubic ice along the body diagonal. (a) String density $\rho$ against the distance to
the edge in sites of the diagonal, $(u-u_c)L$, with $u_c$ from the form $d\ln(1/d)$ (criterion F2): cubic ice at $L=64$, 96,
and 128 (disks, light to dark; vertex-type runs), and square ice along its diagonal at $L=96$, 192, and 384 (squares);
gray band: the window of criterion F1. (b) Effective exponent $\gamma$ of the head, with free $u_c$, in four windows of
$\rho$, against the geometric mean of their bounds, which decreases to the right, toward the edge; gray: the window of F1;
dashed: 1/2; dotted: the threshold of F1.}
\end{figure}

\textit{A sharp edge with a steep head}---Two criteria judge the edge of the vertex-type runs at $L=64$, 96, and 128
(Criteria and checks): that $\gamma$ exceed its largest 2D value by five errors (F1), and that the thickness grow more
slowly than $L$ (F2). Both hold (Fig.~\ref{fig:edge}). F1 gives $\gamma=0.614(7)$, 0.721(6), and 0.657(2), each fit
using 4 to 8 sites, 0.14 to 0.25 above the threshold of 0.473. In errors, these margins are 19 to 108, but the scatter
between sizes is larger: it is set by the sites that fall in the window, not by noise, and the existence runs reproduce it
size by size (0.635, 0.729, 0.653). Over the eight runs, $\gamma$ goes from 0.61 to 0.73, with a standard deviation of
0.04, and the smallest value still exceeds the threshold by 3.4 times that scatter; in every window of
Fig.~\ref{fig:edge}(b), every 3D value (0.58 to 0.78) exceeds every 2D one (0.36 to 0.54). F2 gives
$D^*_2=0.92$, 1.00, and 1.17 sites, growing as $L^{0.35(9)}$ (its error is scaled by $\sqrt{\chi^2/\mathrm{ndf}}=3.5$, as by the Particle Data Group,
S1; this scaling absorbs the scatter between sizes, that of the fitted $u_c$ included): the edge is about one site thick at these sizes.
Reported without criterion, $\gamma$ grows as the window approaches the edge at every size, from 0.58 in
$0.08\le\rho\le0.25$ to 0.69 and 0.78 in $0.02\le\rho\le0.15$ at $L=128$ and 96, as expected for $d\ln(1/d)$, whose
logarithmic slope $1-1/\ln(d_0/d)$ rises toward 1; in 2D, it stays near 1/2, between 0.36 and 0.54, without trend. The
thickness exponents of all variants lie between 0.18 and 0.48, compatible with $1/3$ as with the larger exponents of the
fluctuations.

\textit{What the head does not decide}---F1 does not distinguish $d\ln(1/d)$ from a power law between 1/2 and 1; the
amplitude would, but it cannot be measured on cubic ice at these sizes. Nor does the effective exponent connect the
profile to Eq.~(3) quantitatively. The profile that local equilibrium predicts at the sites of the diagonal, $X_B(g(u-u_0))$
with the linear $\mu_{\rm loc}$ of S4, corrected by the finite-size shift measured on sawtooth tori at the same field
gradient, misses the 2D exponents of the head by $+0.21$, $+0.05$, and $+0.04$ at $L=96$, 192, and 384, against a tolerance
of 0.03: the method is not calibrated in 2D, and its 3D values are not used.

\begin{figure}
\includegraphics[width=\columnwidth]{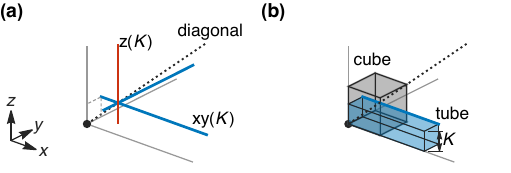}
\caption{\label{fig:probes}Probes anchored at a corner (schematic, not to scale), in the frame of the corner (dot). (a)
The lines of S5: the body diagonal (dotted), the lines xy$(K)$, $(x,K,K)$ and $(K,y,K)$ (blue), and the line z$(K)$,
$(K,K,z)$ (red), each at a distance $K$ from the two faces it runs along (gray dashes, for $(x,K,K)$). (b) The boxes of
S6: the corner cube $[0,b]^3$, paired with the diagonal, and the tube $[0,x]\times[0,K]^2$, paired with the line
$(x,K,K)$ that runs along its outer edge (thick line).
For a box, as for a line, $\ell$ counts sites: it is $b+1$, or $x+1$, for the largest box entirely in the state of
the corner.}
\end{figure}

\textit{Lines from the corners}---Along a line from a corner, $\ell$ is the number of consecutive vertices in the state
of the corner, sample by sample and without connectivity: the local height of the frozen region. It is recorded from
every corner along the body diagonal, and along lines parallel to the axes at a distance $K$ from the two faces that meet
there: $(x,K,K)$ and $(K,y,K)$ for the class xy$(K)$, and $(K,K,z)$ for z$(K)$, with $K=L/32$ and $L/16$
[Fig.~\ref{fig:probes}(a)]; in 2D, along the diagonal and $(x,K)$. Moments come from the histogram summed over corners, the lines of a class, and replicas, and errors
by jackknife over replicas. The standard deviation $\sigma_\ell$ carries Sheppard's correction for the unit grouping of
$\ell$~\cite{Sheppard1897}; skewness and kurtosis come from the raw moments (on the lines of cubic ice, correcting them would raise the skewness by
at most 5\% from $L=64$ on, and by up to 10\% at $L=32$ and 48).

\textit{Calibration on square ice}---Along the diagonal of square ice, $\ell$ is also the side of the largest frozen
square at the corner, whose law is known: $(Ly_c-\ell-1)/(\lambda L^{1/3})$ tends
to the Tracy-Widom distribution of the Gaussian unitary ensemble (GUE)~\cite{TracyWidom1994}, with $y_c=1-\sqrt3/2$ and
$\lambda=2^{-4/3}3^{-1/6}$ (Appendix~A). Colomo and Pronko derived this limit (Theorem 7 of Ref.~\cite{ColomoPronko2026}; details in
Ref.~\cite{ColomoPronko2024})
from a determinant formula for the probability that a corner square is frozen, which they had first conjectured in the language of the six-vertex
model~\cite{ColomoPronko2024}; Li proved this formula~\cite{Li2026}, and Liao, independently, this formula and the
convergence of the strings to the Airy line ensemble near the arctic curve~\cite{Liao2026}, both in recent preprints.
Monte Carlo sampling of square ice by coupling from the past, for sizes up to 2046, had already found this law along
the diagonal, with the scale that Kardar-Parisi-Zhang scaling theory predicts, equal to $\lambda$~\cite{PrahoferSpohn2024}; Monte Carlo
simulations of the six-vertex model with domain-wall boundaries for $0\le\Delta<1$, square ice included, had also given
strong evidence for it~\cite{LybergKorepinViti2023}. Its mean, $-1.771$, and standard deviation, 0.902~\cite{Bornemann2010}, give
$\langle\ell\rangle-y_cL\simeq0.5853\,L^{1/3}$ and $\sigma_\ell\simeq0.2980\,L^{1/3}$; its skewness, 0.224, gives $\ell$
the skewness $-0.224$, since $\ell$ is measured from the corner. These theorems state that the law of $\ell$ converges,
not its moments; the width and skewness quoted are those of the limit law, which the moments of $\ell$ are expected to
approach, as the measured width does (below). The chain recovers this law. On the square ice runs of S2,
$L=12$ to 384, $\sigma_\ell/(0.2980\,L^{1/3})$ is 0.986 at $L=12$, then 0.998 to 1.004 from 24 to 384, each with an error
below 0.006, and the skewness of $\ell$ goes from $-0.08$ at $L=12$ to $-0.195(8)$ at 384. Without Sheppard's
correction, the width exceeds the exact law by $1+1/(24\sigma_\ell^2)$, as for a grouped variable, 5\% at $L=24$; the
correction removes this excess. The mean, fitted as
$aL^{1/3}+b$ for $L\ge48$, gives $a=0.5805(49)(168)$ for 0.5853: the statistical error propagates the jackknife
errors of the points, not rescaled since $\chi^2/\mathrm{ndf}=0.16$ (S1), and the systematic error is the largest change over fitting windows ($L\ge24$, $L\ge96$, a correction in
$L^{-1/3}$). The ratio $a/(\sigma_\ell/L^{1/3})$, which does not depend on $\lambda$, is $1.945(17)(56)$, for
1.964 in the GUE and far from 0.952 in the orthogonal ensemble~\cite{TracyWidom1996,Bornemann2010}. In the runs for lines, $L=64$ to
256, $\sigma_\ell/(0.2980\,L^{1/3})=1.004(3)$ and 1.000(3) at $L=128$ and 256, the exponent of $\sigma_\ell$ is 0.326(3),
and the skewness of $\ell$ is $-0.19$ at 256 (Table~I); along xy$(L/16)$, which meets the arctic curve obliquely, it is
$-0.14$, which may be why it converges more slowly.

\textit{Cubic ice}---On the runs for lines at $L=64$, 96, and 128, the criteria (Criteria and checks) find the
fluctuations of the edge not Tracy-Widom, and growing faster than $L^{1/3}$. On the two main lines, the diagonal and xy$(L/16)$, the skewness of $\ell$
is positive, 68 to 265 errors above zero, the sign opposite to that measured on square ice by the same chain
(Table~\ref{tab:lines}). The exponents of $\sigma_\ell$, fitted from $L=64$ to 128, are 0.457(2) on the diagonal and
0.440(5) on xy$(L/16)$, 64 and 21 statistical errors above 1/3 (about 60 and 7 times the spread between pairs of sizes); between successive sizes, they are 0.456 and 0.458 on the diagonal,
0.446 and 0.431 on xy$(L/16)$. These are effective exponents at these sizes, not asymptotic ones. Reported without
verdict, the five classes of lines give exponents from 0.440 to 0.485, below 1/2, and on the two main lines neither
$L^{1/2}$ nor $(L/\ln L)^{1/2}$, with one free amplitude, fits ($\chi^2$ from 140 to 1241 for 2 degrees of freedom). Over
the five classes, the skewness is 0.35 to 0.57 and the excess kurtosis 0.16 to 0.42. For independent strings met with a
density growing linearly beyond the edge, the mean number of strings within $w$ vertices beyond it would grow as $w^2$,
and $P(\ell\ge\ell_0+w)=\exp(-cw^2)$, where $\ell_0$ is the position of the edge on the line: the Rayleigh law, with the
skewness 0.631 and the excess kurtosis 0.245 (S6, ``What the model predicts''). The measured skewness has its sign and is
smaller; the comparison with the model computed on the measured profile is in S6.

\begin{table*}
\caption{\label{tab:lines}Frozen length $\ell$ along the two main lines of cubic ice (runs for lines): mean and standard
deviation in sites, the latter with Sheppard's correction, skewness $\gamma_1$, and excess kurtosis $\gamma_2$, from the
raw moments; errors on $\sigma_\ell$: 0.001 to 0.002 along the diagonal and 0.009 to 0.011 along xy$(L/16)$; at most
0.008 on $\gamma_1$ and 0.025 on $\gamma_2$.}
\begin{ruledtabular}
\begin{tabular}{rllllllll}
 & \multicolumn{4}{l}{Diagonal} & \multicolumn{4}{l}{xy$(L/16)$} \\
$L$ & $\langle\ell\rangle$ & $\sigma_\ell$ & $\gamma_1$ & $\gamma_2$ & $\langle\ell\rangle$ & $\sigma_\ell$ & $\gamma_1$ &
$\gamma_2$ \\
\colrule
64 & 6.73 & 1.546 & 0.546 & 0.35 & 10.24 & 4.429 & 0.469 & 0.34 \\
96 & 9.55 & 1.860 & 0.558 & 0.38 & 14.76 & 5.308 & 0.452 & 0.36 \\
128 & 12.29 & 2.122 & 0.569 & 0.38 & 19.06 & 6.009 & 0.455 & 0.38 \\
\end{tabular}
\end{ruledtabular}
\end{table*}

\begin{figure}
\includegraphics[width=\columnwidth]{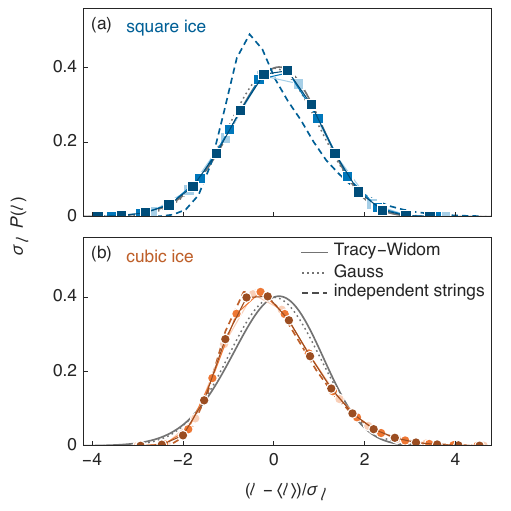}
\caption{\label{fig:laws}Reduced laws of the frozen length $\ell$ along the diagonal, $\sigma_\ell P(\ell)$ against
$(\ell-\langle\ell\rangle)/\sigma_\ell$, each law reduced by its own mean and standard deviation. (a) Square ice,
$L=64$, 128, and 256 (light to dark). (b) Cubic ice, $L=64$, 96, and 128. Gray: Tracy-Widom (GUE) law of $-\ell$, exact for
square ice; dotted: Gauss; dashed, in the color of the series: the independent-string model of S6 at the largest size,
reduced in the same way.}
\end{figure}

\textit{Reduced laws}---Figure~\ref{fig:laws} compares the laws of $\ell$ along the diagonal, each reduced by its own
mean and standard deviation. On square ice, they fall close to the Tracy-Widom law of $-\ell$ at all three sizes, while the law that
independent strings would give, computed from the mean string density of the same line (S6), leans the other way. On cubic
ice, the measured laws have the long tail at large $\ell$, and the independent-string law has their sign and nearly their
shape.

\textit{Not a lattice effect}---The positive sign is neither an effect of the discreteness of $\ell$ nor of the
corners and faces of the box. On sawtooth tori (S7), whose planar edges lie in a field known exactly, the 2D and 3D edges at $L=64$ have the same
mean $\ell$, 19.19 and 19.21 sites, and nearly the same width, 1.96 and 2.02 sites, yet skewnesses of $-0.239$ in 2D and
$+0.205$ in 3D. At these widths of about two sites, the discreteness of $\ell$ thus does not set the sign.

\subsection{Criteria and checks}

\textit{The edge}---F1 requires $\gamma$, in the window $0.03\le\rho\le0.2$, to exceed by $5\sigma$ its largest value on
square ice at $L=96$ to 384, 0.473: the head is not the square root of the 2D class. F2 requires the exponent of $D^*_2$,
with $u_c$ from the form $d\ln(1/d)$, to lie $5\sigma$ below 1: the edge is sharp. They judge the vertex-type runs at
$L=64$, 96, and 128 (S2, ``How the tests were run''); applied to the existence runs at $L=64$ to 160, on which they were
set, they give the same outcome, a consistency check rather than a test.

\textit{The lines}---At $L=32$ and 48, the skewness of $\ell$ is already positive on all five classes of lines,
$+0.33$ to $+0.62$, at a width in sites comparable to that of 2D (on the diagonal, 1.14 and 1.37 sites, against 1.21 for
square ice at $L=64$). The criteria take the diagonal and xy$(L/16)$ as main lines and judge $L=64$ to 128 (S2): H1
confirms on new sizes a sign seen at small ones. H0 requires, on square ice, the diagonal to lie within 2\% of the exact
width at $L=128$ and 256, the skewness of $\ell$ to be negative at $5\sigma$, and the exponent of $\sigma_\ell$ to lie in
$[0.30,0.37]$; judged on 2D runs already read when the criteria were set, it checks the chain rather than tests it. H1
requires the skewness of $\ell$ to be positive at $5\sigma$ at $L=64$, 96, and 128 on both main lines. H2 requires the
exponent of $\sigma_\ell$ to exceed 1/3 by $5\sigma$ over the three sizes, and by $3\sigma$ between the two largest.
Quality controls: eight replicas, the same version of the code as the 2D calibration runs, at least 100 autocorrelation
times of the mean $\ell$ per replica, and drift between halves and differences between corners within four errors. All
criteria and controls hold.

\section{Independent strings}
\label{sec:strings}

How far are the edge fluctuations from those of strings that ignore each other? In 2D, far: the measured width lies 47\%
below that of independent strings, near the 48.5\% expected in the limit, and the skewness has the opposite sign, because ordered
strings shorten the tail of long frozen lengths. In 3D, the width lies 10\% to 36\% below, depending on the probe; this
gap, $\delta=\sigma_\ell/\sigma_{\rm ind}-1$, shrinks with $L$ for every probe. Along lines, the skewness has the sign of independent strings; on every probe of Table~\ref{tab:probes}, it lies below
that of the model, in the direction in which ordering moves it in 2D, and on the diagonal, the corner cube, and the tubes
much less than in 2D.
Boxes, which see every string of their region, deviate less than lines, and their widths grow near $L^{1/3}$. All
comparisons of this section are descriptive.

\textit{The model}---Relative to the polarized state of a corner, the reversed arrows of a configuration form the
strings of S1, paths that are monotonic in the frame of the corner; at a vertex, they come in pairs, one in and one out.
For the corner $(0,0,0)$, a string steps along $-x$, $-y$, or $+z$, so neither $x-z$ nor $y-z$ ever increases, and their
sum decreases at every step: it meets the diagonal, where both vanish, at most once, and a line parallel to an axis,
along which its other two coordinates are fixed, in one run of consecutive vertices. On lines as in boxes, each
segment met is thus one string. The mean number of string passages at the $k$th vertex of a line, $n_k$, is half the
mean number of reversed free arrows there, read on the mean profile. A string that runs along the line between vertices
$k$ and $k+1$ passes both; $f_k$, the mean reversal of that edge, removes the double count. Thus
$\Lambda(D)=\sum_{k<D}n_k-\sum_{k<D-1}f_k$ is the mean number of string segments that meet the first $D$ vertices of the
line, whatever the pairing of strings at shared vertices (Appendix~B). If these segments were independent and each unlikely, their number
would follow a Poisson law, and $P(\ell\ge D)=\exp[-\Lambda(D)]$ [Eq.~(4)]: the whole law of $\ell$, without free
parameters, on the same runs. Two checks of the computation hold: along the diagonal, the profile is that of S5 to
rounding, and the bounds $1-\sum_{k<D}n_k\le P(\ell\ge D)\le1-n_{D-1}/3$, a vertex being crossed by at most three
strings, hold within the errors.

\textit{What the model predicts}---Near the edge, let the mean number of string passages grow as $n_k\simeq
n_0(j/L)^\beta$ at the $j$th vertex beyond the edge, as for a string density $\rho\propto(d/L)^\beta$ at a depth of $d$
sites. Within $w$ vertices
beyond the edge, the line then meets $\Lambda(w)\simeq n_0\,w^{1+\beta}/[(1+\beta)L^\beta]$ strings, and $\ell$
fluctuates over the depth where $\Lambda\simeq1$: $\sigma_\ell\propto L^{\beta/(1+\beta)}$. The square root of 2D,
$\beta=1/2$, gives $L^{1/3}$, the very exponent of the Tracy-Widom edge. In 3D, local equilibrium with Eq.~(3) gives
$\rho\propto(d/L)\ln(L/d)$ near the edge (S4); then $\Lambda(w)\propto(w^2/L)\ln(L/w)$, and
$\sigma_\ell\propto(L/\ln L)^{1/2}$, whose effective exponent approaches 1/2 only logarithmically. The argument assumes
independent strings, a mean profile of the form $(d/L)^\beta$ down to the depth $w$, and a line that crosses the edge at
a finite angle, as the diagonal does by symmetry. In 2D, the lines $(x,L/16)$ and $(x,L/32)$ also cross the exact arctic curve at finite angles, $29^\circ$ and
$19^\circ$, but few strings meet them: relative to the corner, every string runs from one far face to the other, so that
the line $(x,K)$ is touched, in every configuration, by exactly the $K+1$ strings that meet the face $x=L$ at a height of
at most $K$. $\Lambda$ then reaches only about $(K+1)/2$ at half an axis, and the law of the model keeps weight there
(Checks): with so few strings, their count is far from Poisson. This may be why the exponent drawn from the profile fails
on these lines; it is not checked, and at $K=L/32$ the exponent does not improve as the neglected tail shrinks. In 3D, a
string can pass beside a line, and no such count is fixed; why the model misses the exponent near the faces is not
checked (``Square ice'' and ``The growth of the width''). The argument also
separates two scales of the edge. The depth $w$ where a line meets its first string, $(L/\ln L)^{1/2}$ in 3D, is larger
than the Airy length over which a single string rounds the mean profile (S5); the two coincide only in 2D,
where $\beta=1/2$. The same expression sets the shape of the law: $P(\ell\ge\ell_0+w)=\exp[-\Lambda(w)]$, where $\ell_0$
is the position of the edge on the line, is a Weibull law of shape $1+\beta$. Its skewness is positive as long as the
shape is below 3.60, that is, as long as $\Lambda$ grows more slowly than $w^{3.6}$; for $\beta=1$, it is the Rayleigh
law, with skewness 0.631 and excess kurtosis 0.245. The long tail lies at large $\ell$: a long frozen length only
requires the strings expected beyond the edge to be absent, each independently. For a box anchored at the corner,
$\Lambda$ can grow faster, and the model itself gives the corner cube a negative skewness (Table~\ref{tab:probes}): the
sign of the skewness tells ordered from independent strings only along lines.

\textit{Why ordering reverses the skewness}---On square ice, $s=(Ly_c-\ell-1)/(\lambda L^{1/3})$ follows the
Tracy-Widom law (Appendix~A): $s$ is the scaled position of the first string, the analog of the largest eigenvalue, and it
grows as the string advances into the frozen corner. Its two tails have different origins~\cite{MajumdarSchehr2014}.
For the first string to advance by $s>0$, it moves alone, at the cost of one particle pulled out of a Coulomb gas: a long
tail, at small $\ell$. For it to retreat by $|s|$, it must push back all the strings behind it, which it cannot cross: a
short tail, at large $\ell$. Independent strings with the same mean density, the Airy density $\sqrt{|s|}/\pi$ behind the
edge, have the same tail at small $\ell$ to leading order; at large $\ell$, they need only be absent, at the cost
$\exp(-\Lambda)$ with $\Lambda\simeq\int_0^{|s|}\sqrt x\,dx/\pi$. The three tails are
\[
\underbrace{e^{-\frac43s^{3/2}}}_{\text{advance, both}},\qquad
\underbrace{e^{-|s|^3/12}}_{\text{retreat, ordered}},\qquad
\underbrace{e^{-\frac2{3\pi}|s|^{3/2}}}_{\text{retreat, independent}}.
\]
Hence the skewness $+0.224$ of the Tracy-Widom law, and $-0.224$ for $\ell$, measured from the corner; for independent
strings, the long tail moves to large $\ell$, and the skewness becomes positive. The two observations of 2D below, the
reversed skewness and the halved width, thus have one cause: ordering shortens the tail of long frozen lengths.

\textit{Square ice}---The model fails, as it should for ordered strings: along the diagonal, $\sigma_\ell$ lies 46.6\%,
46.5\%, and 46.6\% below it at $L=64$, 128, and 256, and its skewness, $+0.92$ to $+0.94$, has the sign opposite to the
measured one, $-0.16$ to $-0.19$ [Fig.~\ref{fig:laws}(a)]. Yet the model gives the exponent of $\sigma_\ell$, 0.325(2)
against 0.326(3): the ordering reduces the width by a nearly fixed factor. This agreement is
expected: the 2D edge has a single scale, $\lambda L^{1/3}$, which sets both the mean profile near the edge and the
fluctuations, so that any width drawn from either grows as $L^{1/3}$. Square ice thus checks the model, but cannot tell
whether the exponent comes from the profile or from the correlations of the strings, since its square-root profile and
its ordering both follow from their non-crossing; along the diagonal, the two exponents differ only in 3D (``The growth
of the width'').
The agreement of the exponents holds along the diagonal only. On the lines xy$(K)$, even in 2D, the exponent drawn from the profile is lower and
unstable (0.23 against 0.33 at $K=L/16$, 0.10 against 0.35 at $L/32$), and the gap drifts, from $-56\%$ and $-65\%$
at $L=64$ to $-49\%$ and $-51\%$ at 256, depending at small sizes on the truncation of the model (Checks).
The factor has an expected limit. If the mean density of strings near the edge tends to that of the Airy edge,
$K_{\rm Ai}(x,x)={\rm Ai}'(x)^2-x\,{\rm Ai}(x)^2$, as the convergence of the strings to the Airy line
ensemble~\cite{Liao2026} suggests without implying it, the model tends to the largest point of a Poisson process with this
density, whose distribution is $\exp[-\int_s^\infty K_{\rm Ai}(x,x)\,dx]$, with
\[
\int_s^\infty K_{\rm Ai}(x,x)\,dx=\tfrac13\bigl[2s^2{\rm Ai}^2-2s\,{\rm Ai}'^2-{\rm Ai}\,{\rm Ai}'\bigr],
\]
with Ai and Ai$'$ taken at $s$, and a standard deviation of 1.750, against 0.902 for the Tracy-Widom law~\cite{TracyWidom1994}, which lies 48.46\%
below; the skewness of its $\ell$, $+0.960$, lies near that of the model, which drifts away from it as the model's width
does, from $+0.94$ to $+0.92$ between $L=64$ and 256. Along the diagonal of square ice, the gap stays 1.9 points from
$-48.46\%$: the width of the model falls from 2.4\% to 3.5\% below this expected limit, by steps that shrink, while the
measured width reaches its own limit. At these sizes, the model thus does not approach this limit, and the data cannot
tell whether it will. Discretizing the Airy density on the sites of the diagonal, a site counting the strings that cross
it, changes the width of the Poisson process, with the same correction, by less than $10^{-4}$ from $L=48$ on: the drift
does not come from grouping $\ell$ into integers but from the measured mean profile, whose shape near the edge still
differs from the Airy density, and it is not explained. The measured width approaches the Tracy-Widom one from above,
1.1\% above it at $L=64$, 0.4\% at 128, and less than 0.1\% at 256, so that at these sizes $1+\delta$ is biased toward
independence by about 4\% (3.6\% to 3.8\%) along the diagonal.
On 2D sawtooth tori, the gap goes from $-43.3\%$ to $-45.8\%$ between $L=64$ and 1024 (S7). On each probe, the data give
the direction of the drift, not its limit; along the diagonal, the 2D gap does not drift from $L=64$ to 256, whereas the 3D one shrinks
(``Cubic ice'').

\textit{Cubic ice}---Along the lines of S5, the model does not reproduce the measurement: at $L=128$, $\sigma_\ell$ lies 14.4\% below it on the diagonal and 18.1\% on xy$(L/16)$, and the
measured skewness is smaller, 0.57 against 0.74 and 0.46 against 0.81. But the gap is three times smaller than in 2D, and
it shrinks with $L$, where it is constant in 2D from $L=64$ to 256: $-21.4\%$, $-18.9\%$, $-17.8\%$, $-15.7\%$, and $-14.4\%$ along the
diagonal from $L=32$ to 128 (runs for lines). The reduced law of the model has the sign and nearly the shape of the
measured one, with a larger skewness [Fig.~\ref{fig:laws}(b)].

\textit{The growth of the width}---Since $\sigma_\ell=\sigma_{\rm ind}(1+\delta)$, with $\delta=\sigma_\ell/\sigma_{\rm
ind}-1$, the exponent of $\sigma_\ell$ is that of $\sigma_{\rm ind}$, carried by the profile, plus that of $1+\delta$.
Along the diagonal, from $L=64$ to 128, the model gives 0.398(1) of the measured 0.457(2), and the shrinking gap the
rest, $\ln(0.856/0.822)/\ln2=0.058$. The 0.40 is carried by the measured profile, not predicted from the class: through
$\beta/(1+\beta)$, the effective head exponents of S5, 0.61 to 0.73, give 0.38 to 0.42, and F1 does not tell $d\ln(1/d)$
from such a power law. On tori, where the field is exact, the profile predicted from Eq.~(3) carries an exponent 0.010
to 0.012 higher than the measured one (S7). If $\sigma_\ell/\sigma_{\rm ind}$ tends to any nonzero limit, whether the gap
vanishes or not, the measured exponent must come down to that of the model, itself rising slowly. The share of $1+\delta$ falls only slowly, however: if the gap of the tori keeps decreasing as from $L=192$ to 256
(S7), its share of the exponent is still 0.04 from 256 to 512 and 0.02 from 4096 to 8192.
A flat measured exponent is thus expected whether the gap vanishes or not, and the data do not tell the two apart: the
effective exponents of $\sigma_\ell$ are 0.456 from 64 to 96 and 0.458 from 96 to 128, against 0.395 and 0.403 for the
model, and on sawtooth tori the effective exponent no longer decreases from $L=128$ to 256 (S7). Off the diagonal, the model misses
the exponent in 3D as in 2D: 0.36 against 0.44 on xy$(L/16)$, and 0.17 to 0.27 against 0.46 to 0.49 on the lines closer
to the faces.

\textit{Thick probes}---A line meets only the strings that touch it. It might thus be a thinned probe of an ordered
edge, which would push any statistics toward that of independent events: randomly removing eigenvalues turns the
Tracy-Widom law continuously into the Weibull law of uncorrelated variables~\cite{Bohigas2009,BothnerBuckingham2018}. Boxes anchored at the corner test this. The
down-set of a corner contains the site $(a,b,c)$ if and only if the whole box $[0,a]\times[0,b]\times[0,c]$, in the frame
of the corner, is in its state; its membership, recorded sample by sample, gives the frozen depth of any box anchored at
the corner. A box is frozen only if no string enters it: it sees every string of its region. We pair the corner cube
$[0,b]^3$ with the diagonal. We call tubes the boxes of square section $K\times K$ that extend from the corner along one
axis, $[0,x]\times[0,K]^2$ along $x$ (and its image along $y$) and $[0,K]^2\times[0,z]$ along $z$; the edge of a tube
farthest from the two faces it lies against, $(x,K,K)$ or $(K,K,z)$, is the line xy$(K)$ or z$(K)$ of S5, with which it is
paired [Fig.~\ref{fig:probes}(b)]. The boxes are read in the vertex-type runs, $L=48$ to 128. The model extends exactly to boxes: a string is a
monotonic path in the frame of the corner, so it cuts a box in a single segment, and the number of segments in a box
equals the number of string passages through its vertices minus the number of string steps along its inner edges, for
every configuration. In 2D, a string that touches the strip $[0,x]\times[0,K]$ crosses the line $(x,K)$, and one that
touches the square crosses its diagonal: boxes and lines coincide configuration by configuration, as the computation
confirms.

\textit{What thick probes show}---Thick probes see no more order than the line (Table~\ref{tab:probes}). At $L=128$,
the gap to independent strings is $-9.7\%$ for the cube against $-14.4\%$ for the diagonal, and $-10.6\%$ and $-11.0\%$
for the tubes against $-18.1\%$ and $-22.6\%$ for the lines xy$(L/16)$ and z$(L/16)$; every gap shrinks with $L$, and in
2D it is $-46.6\%$ for the square, as for the diagonal, from $L=64$ to 256. On these probes, the face layers carry less than
0.5\% of the string passages at the mean depth. The widths of boxes grow more slowly than those of lines: from $L=48$ to
128, the cube grows from 0.59 to 0.80 site, about as $L^{0.31}$, and the tubes as $L^{0.26}$ to $L^{0.32}$, against
$L^{0.44}$ to $L^{0.485}$ for lines (S5), at $L=128$ the tubes being 1.8 to 4.1 sites wide; the model gives them lower exponents still, 0.29 for the cube and 0.15 to 0.24
for the tubes. A box stays frozen until the first string reaches any of its sites, an extreme over a surface rather than
along a line. The cube is narrower than a site, where Sheppard's correction and the skewness feel the discreteness of
$\ell$ (gap $-8.7\%$ without the correction); its growth, about $L^{0.31}$, is measured below one site, where square ice gives
0.35, or 0.31 without the correction, for 1/3 (from $L=12$ to 24). Its skewness is negative, as the model's is, but more so, $-0.37$
against $-0.20$. For every probe of Table~\ref{tab:probes}, thin
or thick, the measured skewness lies below that of the model: by 0.17 on the diagonal and the cube, 0.22 to 0.23 on the
tubes at $L/16$ and 0.33 to 0.39 at $L/32$, against 1.1 on the diagonal of square ice; on the other lines, where the
model misses even the exponent, the shift is larger, 0.35 to 1.8. It goes the way ordering moves the skewness in 2D, by
shortening the tail at large $\ell$. The line is thus not a thinned probe of an
ordered edge. The positive skewness and the growth faster than $L^{1/3}$ are properties of lines; the gap to independent
strings, much smaller than in 2D for every probe and shrinking with $L$, is common to all.

\textit{A thinned ordered edge}---Keeping each point of the Airy edge with probability $p$, independently, gives a
determinantal process of kernel $pK_{\rm Ai}$, whose largest point lies below $s$ with probability $\det(1-pK_{\rm Ai})$
on $(s,\infty)$~\cite{Bohigas2009,BothnerBuckingham2018}; independent points of the same density give
$\exp[-p\int_s^\infty K_{\rm Ai}(x,x)\,dx]$. Thinning brings the two laws together: the width gap is $-48.5\%$ at $p=1$,
$-35\%$ at 0.8, $-20\%$ at 0.5, and $-11\%$ at 0.3, and the skewness of $\ell$ is already positive at $p=0.9$, $+0.39$. A
thinned ordered edge can thus show a small gap and a positive skewness, but not with the shapes measured here: at gaps of
11\% to 18\%, those of the diagonal, the tubes, and xy$(L/16)$, its skewness lies within 0.03 of that of independent
points, and closer still at smaller gaps, whereas the measured skewness lies 0.17 below the model on the diagonal and
the cube, 0.22 to 0.39 on the tubes and 0.35 on xy$(L/16)$; a shift of 0.17 would require $p\simeq0.7$, hence a gap near
$-30\%$, not $-14\%$. A line does not select strings at random, and the 3D profile is not that of Airy: the
comparison is indicative. Nor does a width below one site hide ordering: on square ice at $L=12$ and 24, where
$\sigma_\ell=0.67$ and 0.86 site, as for the cube, the corner square lies 48.9\% and 47.9\% below independent strings
(45.6\% and 45.9\% without Sheppard's correction). From these widths on, however, the gap of square ice also
shrinks, to 47.3\% at $L=48$ ($\sigma_\ell=1.08$ sites) and 46.5\% from $L=192$ on; $\sigma_\ell/\sigma_{\rm ind}$ rises
by 2.0\% from $L=12$ to 24, and that of the cube by 2.6\% from $L=48$ to 128: the level of the cube's gap, not its
trend, sets it apart from square ice. The
cube is measured up to $L=128$ only. Like the rest of this section, these comparisons are descriptive (S2): they were made after the measured values were
known.

\begin{table}
\caption{\label{tab:probes}Lines and boxes anchored at the corner of cubic ice [Fig.~\ref{fig:probes}; cube: $[0,b]^3$;
tube: the box of section $K\times K$ whose outer edge is the line]: gap $\delta=\sigma_\ell/\sigma_{\rm ind}-1$ (\%) of
the width to independent strings, by size, and skewness at $L=128$, measured and of the model. Lines: runs for lines
($L=64$ to 128); boxes: vertex-type runs ($L=48$ to 128); errors 0.1 to 0.3 on the gaps. Widths carry Sheppard's
correction, measurement and model alike. Square ice: the square and the diagonal coincide, $-46.6\%$, $-46.5\%$, and
$-46.6\%$ at $L=64$, 128, and 256.}
\begin{ruledtabular}
\begin{tabular}{lrrrrr}
 & \multicolumn{4}{c}{Gap $\delta$ (\%)} & Skewness \\
Probe & 48 & 64 & 96 & 128 & at 128 \\
\colrule
Diagonal & & $-17.8$ & $-15.7$ & $-14.4$ & 0.57 / 0.74 \\
Cube & $-12.0$ & $-10.8$ & $-10.0$ & $-9.7$ & $-0.37$ / $-0.20$ \\
xy$(L/16)$, line & & $-22.6$ & $-19.7$ & $-18.1$ & 0.46 / 0.81 \\
xy$(L/16)$, tube & $-17.3$ & $-14.5$ & $-12.1$ & $-10.6$ & $-0.35$ / $-0.11$ \\
z$(L/16)$, line & & $-33.2$ & $-26.2$ & $-22.6$ & 0.54 / 1.07 \\
z$(L/16)$, tube & $-17.7$ & $-14.3$ & $-12.0$ & $-11.0$ & $-0.32$ / $-0.09$ \\
xy$(L/32)$, line & & $-36.0$ & $-30.8$ & $-27.3$ & 0.36 / 1.11 \\
xy$(L/32)$, tube & & $-21.4$ & $-17.4$ & $-15.1$ & $-0.25$ / 0.14 \\
z$(L/32)$, line & & $-48.2$ & $-42.4$ & $-35.9$ & 0.44 / 2.21 \\
z$(L/32)$, tube & & $-19.9$ & $-16.1$ & $-13.9$ & $-0.33$ / $-0.01$ \\
\end{tabular}
\end{ruledtabular}
\end{table}

\subsection{Checks}

\textit{Truncation}---The model sums $\Lambda$ along the line, or through the box, up to half an axis. Where its law
still has weight there, as for the 2D lines and strips at $L=64$ to 256, the gap depends on this truncation; at
$L=128$, the neglected tail is below $10^{-17}$ for the main 3D probes (at most $3\times10^{-4}$ at $L=48$, and
$10^{-2}$ for the z$(L/32)$ tube at $L=64$).

\section{Tori with a sawtooth field}
\label{sec:tori}

What does a planar edge in a field known exactly show? Where strings are dense, local equilibrium in 3D: its criterion
C1 holds. The verdict of the tori is nevertheless ``not established'', since C0 requires the same test to hold in 2D:
there, the string density deviates from the bulk by less than 0.3\% at all the sizes of the test, $L=256$ to 1024, but at one value of $\mu$ this
deviation grows with $L$, which the test forbids (The verdict). Along lines, the
Tracy-Widom law in 2D and a positive skewness in 3D. In 3D, a gap $\delta$ to independent strings that still shrinks at $L=256$,
while the effective exponent of $\sigma_\ell$ does not come down. The profile that local equilibrium and Eq.~(3) predict
reproduces how the exponent carried by the profile rises with $L$, but not its level.

\textit{The tori}---In cubic ice, the field is set by the boundaries and known only through local equilibrium. On
sawtooth tori, it is imposed, on all three axes (S1, Fig.~\ref{fig:field}): the local log-weight per string step,
$\mu=\mu_0(4\tilde u/L-1)$ for $0\le\tilde u\le L/2$, vanishes at $\tilde u=L/4$ and $3L/4$, where two planar edges,
without corners, separate a polarized region centered at $\tilde u=0$ from a liquid region centered at $\tilde u=L/2$.
Along lines parallel to $x$ from the center of the polarized region, on both sides, $\ell_\pm$ is the number of
consecutive vertices in the state $(+,+,+)$. The runs cover $L\times L\times L$ tori with $L=32$ to 256 and $L\times L$
tori with $L=64$ to 1024, sampled with the two-defect worm (S1).

\textit{Local equilibrium and the law of $\ell$}---At $\mu=0.2$, 0.28, and 0.4, the string density at the depth where
the exact field gives $\mu$ is compared with that of the bulk, $R_\mu=\rho/\rho_{\rm bulk}-1$. In 3D, at $\mu=0.2$,
$R_\mu$ goes from $-4.3\%$ at $L=64$ to $-0.45\%$ at 192; in 2D, at the sizes of the criterion, $L=256$ to 1024, $|R_\mu|$ stays below 0.3\%. As in the box, this tests local
equilibrium where strings are dense, $\mu\ge0.2$, not at the edge itself. Closer to the edge, $R_\mu$, plotted against
$\Delta/L^{1/3}$ with $\Delta=\tilde u-L/4$, falls on one curve at all sizes in 3D, apart from the first site past the threshold, with a dip of about $-8\%$ to
$-9\%$ near $\Delta\simeq0.3$ to $0.5\,L^{1/3}$, and stays below 1\% beyond 1.2 to $1.4\,L^{1/3}$: the density departs
from local equilibrium only in a layer of the Airy width, a vanishing fraction of $L$. In 2D, it does not fall on one
curve: its dip deepens from $-21\%$ at $L=64$ to $-30\%$ at 1024, and it then oscillates, up to $+6\%$, until 1.8 to
$2.6\,L^{1/3}$. In 2D, the planar edge follows the Tracy-Widom
law: at $L=512$ and 1024, the skewness of $\ell$ is $-0.221$ and $-0.222$, for $-0.224$, its excess kurtosis 0.083 and
0.081, for 0.093, and the effective exponent of $\sigma_\ell$ is 0.337. In 3D, the skewness of $\ell$ is positive at every
size and grows, from 0.10 at $L=32$ to 0.37 at 256 (Table~\ref{tab:tori}). At $L=64$, the 2D and 3D edges have the same
mean $\ell$ and nearly the same width, yet opposite skewnesses (S5).

\textit{Independent strings}---On tori, the reference state is $(+,+,+)$, $n(\tilde u)=3\rho_v(\tilde u)$ in 3D and
$2\rho_v(\tilde u)$ in 2D, where $\rho_v(\tilde u)$ is the fraction of reversed arrows at the vertices with the same
$\tilde u$, and $f$ is the reversal of the $x$ edges. The model fails at every 2D size. In 3D, the gap fades: $\delta$
goes from $-22.9\%$ at $L=32$ to $-16.2\%$ at 192, a fit $L^{-\kappa}$ giving
$\kappa=0.204(10)$, against $-0.019(2)$ in 2D (Table~\ref{tab:tori}), and at $L=256$, $\delta=-15.14\%$, an
effective $\kappa$ of $0.234(1)$ from 192 to 256. The power law itself is rejected: its $\chi^2/\mathrm{ndf}$ is
about 3500, and the errors on the fitted $\kappa$, $0.204(10)$ up to $L=192$ and $0.210(8)$ up to 256, are
scaled by $\sqrt{\chi^2/\mathrm{ndf}}$ (S1). That the gap fades rests instead on the effective
values, all positive: they rise from 0.11 to 0.23 and level off. A decay as
$1/\ln L$, which a marginal repulsion suggests, keeps $\delta\ln L$ between $-0.84$ and $-0.86$ from $L=48$ to 256, but fits
worse than the power law, with a $\chi^2$ about ten times larger, and would make the effective $\kappa$ fall, while it
rises. The data do not decide whether the gap vanishes or tends to a finite limit.

\begin{figure*}
\includegraphics[width=\textwidth]{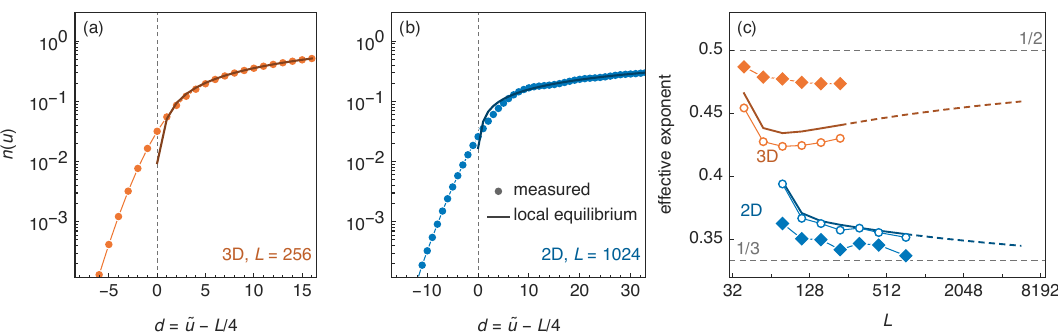}
\caption{\label{fig:torus}Sawtooth tori. (a) Mean number of string passages per vertex, $n(\tilde u)$, against the
distance $d=\tilde u-L/4$ to the edge (liquid for $d>0$), 3D torus at $L=256$: measured (disks) and predicted by local
equilibrium with the bulk equation of state (line), which vanishes for $d<0$. (b) The same for the 2D torus at $L=1024$.
(c) Effective exponents between successive sizes, against their geometric mean: $\sigma_\ell$ measured (diamonds);
$\sigma_{\rm ind}$ of independent strings carried by the measured profile (open circles) and by the predicted one (lines,
dashed beyond the largest measured size); dashed gray: 1/2 and 1/3.}
\end{figure*}

\textit{The exponent stops coming down}---The effective exponent of $\sigma_\ell$ falls from 0.487 ($32\to48$) to 0.4735
($128\to192$), and no further: it is $0.474(1)$ from 192 to 256, where the earlier trend would give 0.469
(Table~\ref{tab:tori}). The exact decomposition $\sigma_\ell=\sigma_{\rm
ind}(1+\delta)$ accounts for this: the effective exponent of $\sigma_\ell$ is that of $\sigma_{\rm ind}$, carried by the
profile, plus that of $1+\delta$, $0.4268+0.0467$ from 128 to 192 and $0.4301+0.0435$ from 192 to 256
[Fig.~\ref{fig:torus}(c)]. The share of the gap falls, while that of the profile rises, and the sum stays flat.

\begin{table*}
\caption{\label{tab:tori}Sawtooth tori: width of $\ell$ (sites, with Sheppard's correction), measured and for independent
strings carried by the measured profile; their gap $\delta$; skewness of $\ell$, measured and of the model; and
effective exponents between the size of the row and the previous one, of $\sigma_\ell$ and of $\sigma_{\rm ind}$ carried
by the measured profile and by the profile predicted from Eq.~(3). Errors, from a jackknife over replicas that recomputes measurement and model together, at most: in 3D, 0.0005 on the
widths, 0.01 points on $\delta$, and about 0.001 on the effective exponents of $\sigma_\ell$; in 2D, 0.006, 0.06 points,
and 0.0055; 0.003 on the skewness; 0.0005 (3D) and 0.0019 (2D) on those of $\sigma_{\rm ind}$, 0.0004 on the predicted
ones.}
\begin{ruledtabular}
\begin{tabular}{rllllllll}
 & \multicolumn{2}{l}{Width} & & \multicolumn{2}{l}{Skewness} & \multicolumn{3}{l}{Effective exponent} \\
$L$ & $\sigma_\ell$ & $\sigma_{\rm ind}$ & $\delta$ (\%) & measured & model & $\sigma_\ell$ & $\sigma_{\rm ind}$ &
predicted \\
\colrule
\multicolumn{9}{l}{3D, $L\times L\times L$} \\
32 & 1.444 & 1.872 & $-22.87$ & 0.100 & 0.421 & & & \\
48 & 1.759 & 2.250 & $-21.84$ & 0.164 & 0.520 & 0.4869 & 0.4542 & 0.4659 \\
64 & 2.019 & 2.545 & $-20.68$ & 0.205 & 0.544 & 0.4788 & 0.4275 & 0.4383 \\
96 & 2.450 & 3.022 & $-18.94$ & 0.259 & 0.557 & 0.4773 & 0.4238 & 0.4343 \\
128 & 2.808 & 3.414 & $-17.76$ & 0.294 & 0.564 & 0.4745 & 0.4244 & 0.4356 \\
192 & 3.402 & 4.059 & $-16.19$ & 0.339 & 0.573 & 0.4735 & 0.4268 & 0.4380 \\
256 & 3.899 & 4.594 & $-15.14$ & 0.369 & 0.579 & 0.4736 & 0.4301 & 0.4406 \\
\colrule
\multicolumn{9}{l}{2D, $L\times L$} \\
64 & 1.958 & 3.453 & $-43.27$ & $-0.239$ & 0.704 & & & \\
96 & 2.269 & 4.050 & $-43.99$ & $-0.232$ & 0.796 & 0.3625 & 0.3939 & 0.3960 \\
128 & 2.509 & 4.501 & $-44.25$ & $-0.232$ & 0.818 & 0.3504 & 0.3667 & 0.3708 \\
192 & 2.891 & 5.214 & $-44.54$ & $-0.226$ & 0.833 & 0.3497 & 0.3625 & 0.3645 \\
256 & 3.190 & 5.778 & $-44.79$ & $-0.225$ & 0.843 & 0.3417 & 0.3572 & 0.3615 \\
384 & 3.671 & 6.683 & $-45.06$ & $-0.223$ & 0.853 & 0.3466 & 0.3588 & 0.3591 \\
512 & 4.055 & 7.402 & $-45.22$ & $-0.221$ & 0.860 & 0.3454 & 0.3554 & 0.3569 \\
1024 & 5.122 & 9.445 & $-45.77$ & $-0.222$ & 0.877 & 0.3369 & 0.3515 & 0.3541 \\
\end{tabular}
\end{ruledtabular}
\end{table*}

\textit{The profile predicted by the bulk equation of state}---If local equilibrium holds up to the edge, each edge of a
torus carries the string density of the bulk at its own field, $\rho_e=X_B(\mu_0-2h_e)$, zero on the ordered side; $X_B$
interpolates the points of the bulk tori as in S4, with, below $\mu=0.0125$, Eq.~(3) fitted on W1, $\rho=0.459\mu
[1+0.293\ln(1/\mu)]$ (in 2D, $\rho=0.450\sqrt\mu+0.180\mu$). The independent-string model then gives $\sigma_{\rm ind}$
with no free parameter. The prediction is calibrated in 2D: on the $L\times L$ tori, which reach $L=1024$, the effective
exponents of $\sigma_{\rm ind}$, predicted and measured, agree within 0.005 at every pair of sizes. In 3D, the predicted exponent lies 0.010 to 0.012 above the measured one at every pair
[Fig.~\ref{fig:torus}(c)], and the predicted $\sigma_{\rm ind}$ lies 7.1\% to 5.0\% below the measured one from $L=32$
to 256. Figure~\ref{fig:torus}(a) suggests why: local equilibrium gives a sharp edge, while the measured profile has a
tail on the frozen side, which widens $\sigma_{\rm ind}$ by a share that falls with $L$ and so lowers the measured
exponent. The rise of the exponent, on the other hand, is reproduced without free parameters: $+0.0050(4)$ predicted and
$+0.0057(5)$ measured, from $96\to128$ to $192\to256$. The logarithm of Eq.~(3) accounts for the slope, not for the level.
Extrapolated, the prediction reaches 0.459 from $L=4096$ to 8192.

\textit{The tail on the frozen side}---Where local equilibrium predicts no string, the measured profile keeps a tail
[Figs.~\ref{fig:torus}(a) and \ref{fig:torus}(b)], at a mean depth of 1.2 to 1.7 sites in 3D and 1.5 to 2.8 in 2D. Its
decay lengths grow as $L^{0.42}$ to $L^{0.45}$ in 3D, as $\sigma_{\rm ind}$ does; in 2D, the
same lengths grow as $L^{0.38}$ to $L^{0.39}$, where the exact exponent is 1/3. At one to three sites, they do not
separate the Airy length $L^{1/3}$ of a single string from the scale of the fluctuations (S5, S6), and the edge is about
one site thick at the sizes measured. Whether the tail narrows relative to $\sigma_{\rm ind}$ cannot be told at one to
three sites: its mean depth, counted from the edge $\tilde u=L/4$, shrinks relative to $\sigma_{\rm ind}$, but counted
from the first site on the frozen side, one site deeper, it grows.

\textit{A scenario}---What follows is a speculation, consistent with the data but not established by them. If the gap
to independent strings vanishes, the edge fluctuations tend to those of independent strings carried by the profile of
local equilibrium and Eq.~(3), hence $\sigma_\ell\propto(L/\ln c'L)^{1/2}$ (S6), with $c'\approx32$ from the
extrapolated prediction, and an effective exponent $1/2-1/(2\ln c'L)$, 0.444 near $L=220$ and 0.459 near $L=5800$. The
measured exponents, 0.457 on cubic ice and 0.474 on tori, would then be crossover values: in $\sigma_\ell=\sigma_{\rm
ind}(1+\delta)$, the exponent of $\sigma_{\rm ind}$, carried by the profile, rises slowly toward 1/2, while that of
$1+\delta$ fades. A nonzero limit of the gap $\delta$, set by the tail on the frozen side or by a residual repulsion, is
not excluded at $L\le256$.

\subsection{Criteria and checks}

\textit{Criteria}---The tori are judged by three criteria (S2, ``How the tests were run''). C0 calibrates the chain on 2D tori: at $L=512$ and 1024, the skewness of $-\ell$ and the excess kurtosis must
match Tracy-Widom, 0.2241 and 0.0934, within three errors or 0.01 and 0.02, and the effective exponent of $\sigma_\ell$
must lie in $[0.30,0.37]$; the test of C1 must also hold in 2D. C1 tests local equilibrium: $|R_\mu|$ must not grow with
$L$, within two errors, over $L=64$, 96, 128, and 192 in 3D (256, 384, 512, and 1024 in 2D), and must be below 2\%, 1\%, and
0.5\% at the largest of these sizes for $\mu=0.2$, 0.28, and 0.4; in 3D, it must also stay below 0.5\% from $L=96$ on,
deeper in the liquid ($\mu=0.57$, 0.8, and 0.95). C2
requires the skewness of $\ell$ to be positive at $5\sigma$ at $L=128$ and 192, the sign opposite to Tracy-Widom.

\textit{The verdict}---C1 and C2 hold in 3D, and the Tracy-Widom part of C0 holds in 2D, but not its 2D test of C1:
$|R_\mu|$ stays below 0.3\% at all sizes, yet at $\mu=0.2$ it grows from 0.05\% at $L=384$ to 0.29\% at 512, beyond two
errors. The rule of non-growth does not suit 2D: there, $R_\mu$ oscillates near the edge with the discrete string
density, like the Airy density, and at fixed $\mu$ it crosses these oscillations as $L$ grows. The verdict of the tori is
therefore ``not established'', and it is kept. Judged on the thresholds of $|R_\mu|$ at all sizes alone, a rule set
after the data, the 2D test would hold and the verdict would read ``3D edge in local equilibrium, with non-Tracy-Widom
fluctuations''; this verdict is reported next to the original one, never in its place. This is the only rule set after
the data, in this work, that would alter a verdict; the Letter uses neither verdict, only descriptive comparisons on the
tori.

\section*{Data availability}

The data and code that support these findings are openly available~\cite{DepotDonnees,DepotCode}. In them,
cubic ice is called OvoidIce, square ice
LockedIce, the free box FreeIce, the banded boxes BandIce, and the tori TorusIce.

\bibliography{supplement}